%% file: mainLong.tex
\documentclass[envcountsame,orivec,runningheads,a4paper
]{llncs}

\usepackage[utf8]{inputenc}

\usepackage{microtype}
\usepackage{marginnote}
\usepackage{mathtools,amsthm}
\usepackage{multirow}
\usepackage[hidelinks,bookmarks,bookmarksopen,bookmarksdepth=2]{hyperref}

\hypersetup{colorlinks=false}

\newtheorem{lemmaAppendix}{Lemma} 
\newtheorem{theoremAppendix}{Theorem} 
\newtheorem{propositionAppendix}{Proposition} 

\newcommand{\macrospath}{.} 

\input{\macrospath/macros}
\input{local-macros} 

\author{Beniamino Accattoli\inst{1}\and Giulio Guerrieri\inst{2}}

\authorrunning{B.~Accattoli\and G.~Guerrieri}

\institute{INRIA, UMR 7161, LIX, \'Ecole Polytechnique, \email{\href{mailto:beniamino.accattoli@inria.fr}{beniamino.accattoli@inria.fr}} \and
Università di Bologna, Dipartimento di Informatica---Scienza e Ignegneria (DISI), Bologna, Italy,
\email{\href{mailto:giulio.guerrieri@unibo.it}{giulio.guerrieri@unibo.it}}
}

\title{Types of Fireballs (Extended Version)}

\begin{document}

\maketitle

\begin{abstract}
\input{00_-_Abstract}
\end{abstract}

\input{01_-_Introduction}

\paragraph{(No) Proofs.} All proofs have been moved to Appendix \ref{sect:proofs}. This paper is a long version of an accepted paper to appear in the proceedings of APLAS 2018.  

\input{02_-_The_Rise_of_Fireballs}

\input{03_-_The_Fall_of_Fireballs}

\input{04_-_Fireballs_Reloaded_the_Split_Fireball_Calculus}
\input{05_-_Multi_Types_for_Split}

\input{06_-_Correctness}
\input{07_-_Completeness}
\input{08_-_Tight_Types_and_Exact_Bounds_New}
\input{09_-_Conclusions}

\phantomsection
\addcontentsline{toc}{section}{References}
\bibliographystyle{splncs04}
\bibliography{\macrospath/biblio}

\newpage
\appendix

\input{\proofspath/proofs}

\end{document}

%% file: macros.tex
\usepackage[utf8]{inputenc}

\usepackage{amssymb}
\usepackage{stmaryrd}
\usepackage{bussproofs}
\usepackage{mathtools}
\usepackage{bbold}

\usepackage{ifthen}
\usepackage{xspace}
\usepackage{fancybox}
\usepackage{hhline}

\usepackage{xcolor}

\newcommand{\sem}[1]{[\![#1]\!]}

\newcommand{\ignore}[1]{}

\newcommand{\myinput}[1]{\ifthenelse{\boolean{withimages}}{\input{#1}}{}}

\newcommand{\reflemma}[1]{Lemma~\ref{l:#1}}
\newcommand{\reflemmap}[2]{Lemma~\ref{l:#1}.\ref{p:#1-#2}}

\newcommand{\reflemmasps}[3]{Lemmas~\ref{l:#1}.\ref{p:#1-#2}-\ref{p:#1-#3}} 

\newcommand{\refpoint}[1]{Point~\ref{p:#1}}

\newcommand{\refthm}[1]{Thm.~\ref{thm:#1}}
\newcommand{\refthms}[2]{Thm.~\ref{thm:#1} and \ref{thm:#2}} 

\newcommand{\refprop}[1]{Prop.~\ref{prop:#1}}
\newcommand{\refpropp}[2]{Prop.~\ref{prop:#1}.\ref{p:#1-#2}} 
\newcommand{\refsect}[1]{Sect.~\ref{sect:#1}}

\newcommand{\reffig}[1]{Fig.~\ref{fig:#1}}
\newcommand{\refcoro}[1]{Cor.~\ref{coro:#1}}
\newcommand{\refcor}[1]{Cor.\,\ref{coro:#1}}

\newcommand{\refrmk}[1]{Remark~\ref{rmk:#1}} 
\newcommand{\refex}[1]{Ex.~\ref{ex:#1}}

\newcommand{\ie}{\textit{i.e.}\xspace}
\newcommand{\eg}{\textit{e.g.}\xspace}
\newcommand{\ih}{\textit{i.h.}\xspace}

\newcommand{\defeq}{\coloneqq} 
\newcommand{\eqdef}{\eqqcolon} 
\newcommand{\grameq}{\Coloneqq} 

\newcommand{\nat}{\mathbb{N}}
\newcommand{\size}[1]{|#1|}
\newcommand{\sizeap}[1]{\size{#1}_@}
\renewcommand{\sizeap}[1]{\size{#1}}

\newcommand{\vsym}{\mathsf{v}}

\renewcommand{\l}{\lambda}
\newcommand{\isub}[2]{\{#1/#2\}}
\renewcommand{\isub}[2]{\{#1{\shortleftarrow}#2\}}
\newcommand{\esub}[2]{[#1/#2]}
\renewcommand{\esub}[2]{[#1{\shortleftarrow}#2]}
\newcommand{\fv}[1]{{\tt fv}(#1)}

\newcommand{\rootRew}[1]{\mapsto_{#1}}
\newcommand{\Rew}[1]{\rightarrow_{#1}}

\newcommand{\rtobv}{\rootRew{\betav}} 

\newcommand{\rtoin}{\rootRew{\inert}}

\newcommand{\tob}{\Rew{\beta}}

\newcommand{\betav}{{\beta_v}} 

\newcommand{\tobv}{\Rew{\betav}} 

\newcommand{\betain}{\beta_{\isym}} 
\newcommand{\inert}{\betain} 
\newcommand{\toin}{\Rew{\inert}}

\newcommand{\isym}{i}

\newcommand{\fsym}{f}

\newcommand{\subsym}{{\mathsf{sub}}}

\newcommand{\vmsym}{\mathsf{shuf}} 
\newcommand{\shuf}{\vmsym} 

\newcommand{\shufeqext}{\shufeqext} 

\newcommand{\tm}{t}
\newcommand{\tmtwo}{u}
\newcommand{\tmthree}{s}
\newcommand{\tmfour}{r}
\newcommand{\tmfive}{q}
\newcommand{\tmsix}{p}

\newcommand{\tmp}{\tm'}
\newcommand{\tmtwop}{\tmtwo'}
\newcommand{\tmthreep}{\tmthree'}

\newcommand{\var}{x}
\newcommand{\vartwo}{y}
\newcommand{\varthree}{z}

\newcommand{\val}{v}
\newcommand{\valtwo}{\val'}
\newcommand{\valthree}{\val''}

\newcommand{\ctxholep}[1]{\langle #1\rangle}
\newcommand{\ctxhole}{\ctxholep{\cdot}}

\newcommand{\ctx}{C}

\newcommand{\ctxp}[1]{\ctx\ctxholep{#1}}

\newcommand{\arbctxp}[1]{\arbctxp{#1}}
\newcommand{\arbctxtwop}[1]{\arbctxtwop{#1}}

\newcommand{\evctx}{\genevctx}

\newcommand{\evctxp}[1]{\evctx\ctxholep{#1}}

\newcommand{\emptyenv}{\epsilon}
\newcommand{\genv}{E}
\newcommand{\genvtwo}{E'}

\newcommand{\cons}{:}

\newcommand{\deriv}{d}
\newcommand{\derivtwo}{e}

\newcommand{\itm}{i}
\newcommand{\itmtwo}{\itm'}

 \newcommand{\gconst}{i} 
\newcommand{\gconsttwo}{\gconst'}
\newcommand{\gconstthree}{\gconst''}

\newcommand{\fire}{f}
\newcommand{\firetwo}{\fire'}
\newcommand{\firethree}{\fire''}

\newcommand{\betaf}{\beta_{\!\fsym}} 

\newcommand{\rtof}{\rootRew{\betaf}}
\newcommand{\tof}{\Rew{\betaf}}

\newcommand{\vsub}{{\vsym\subsym}} 

\newcommand{\unfsym}{\rotatebox[origin=c]{-90}{$\rightarrow$}}
\newcommand{\unf}[1]{#1\unfsym\,}

\newcommand{\la}[1]{\lambda #1.}

\newcommand{\myproof}[1]{
\ifthenelse{\boolean{omitproofs}}{\begin{IEEEproof} Proof available but omitted for readability. \end{IEEEproof}}{#1}}

\newcommand{\gregoire}{Gr{\'{e}}goire\xspace}

\newcommand{\withproofs}[1]{\ifthenelse{\boolean{withproofs}}{#1}{}}

\newcommand{\withoutproofs}[1]{\ifthenelse{\boolean{withproofs}}{}{#1}}

\newcommand{\NoteProof}[1]{\withproofs{\marginpar{\scriptsize \ \ Proof p.\,{\pageref{#1}}}}} 
\newcommand{\NoteState}[1]{\withproofs{\marginpar{\scriptsize \ \ See p.~{\pageref{#1}}}}} 

\newcommand{\vsubcalc}{\lambda_\vsub}

\newcommand{\firecalc}{\lambda_\mathsf{fire}}
\newcommand{\sfirecalc}{\mathsf{Split }\lambda_\mathsf{fire}}
\newcommand{\shufcalc}{\lambda_\shuf}

\newcommand{\plotcalc}{\lambda_\mathsf{Plot}}

\newcommand{\quiet}{inert\xspace}

\newcommand{\doubt}[1]{}

\newcommand{\lambdamucalc}{\lambdabar\tilde{\mu}}

\newcommand{\lambdabar}{\bar{\lambda}}

\newcounter{numberone}
\newcounter{numberoneroman}
\newcounter{numberonealph}

\newcommand{\cbn}{CbN\xspace}
\newcommand{\cbv}{CbV\xspace}
\newcommand{\ocbv}{Open \cbv}
\newcommand{\ccbv}{Closed \cbv}

\newcommand{\prog}{p}
\newcommand{\progtwo}{q}
\newcommand{\progthree}{r}
\newcommand{\progp}{\prog'}

\newcommand{\expr}{e}

\newcommand{\mset}[1]{[#1]}
\newcommand{\emptymset}{\zero}

\newcommand{\ptype}{M}
\newcommand{\ptypetwo}{N}
\newcommand{\ptypethree}{O}
\newcommand{\zero}{\mathbf{0}}

\newcommand{\ntype}{L}
\newcommand{\ntypetwo}{\ntype'}

\newcommand{\iptype}{M^i}
\newcommand{\iptypetwo}{N^i}

\newcommand{\intype}{L^i}

\newcommand{\typctx}{\Gamma}
\newcommand{\typctxtwo}{\Delta}
\newcommand{\typctxthree}{\Pi}

\newcommand{\tder}{\pi}
\newcommand{\tdertwo}{\sigma}
\newcommand{\tderthree}{\rho}

\newcommand{\ruleAp}{@}
\newcommand{\ruleAx}{\mathsf{ax}}

\newcommand{\appendOp}{@}
\newcommand{\hastype}{\!:\!}

\newboolean{withproofs}
\setboolean{withproofs}{false}

\newcommand{\domain}[1]{\mathsf{dom}(#1)}
\newcommand{\typelist}[1]{\underline{#1}}

\newcommand{\splitc}{split\xspace}
\newcommand{\Splitc}{Split\xspace}

%% file: local-macros.tex
\newcommand{\concl}[4]{#1 \vartriangleright #2 \vdash #3 \hastype #4}

\newcommand{\Pair}[2]{#1 \multimap #2}
\newcommand{\Dom}[1]{\mathsf{dom}(#1)}
\newcommand{\Fv}[1]{\mathsf{fv}(#1)}
\renewcommand{\fv}[1]{\Fv{#1}}
\newcommand{\Ax}{\mathsf{ax}}

\newcommand{\EsCoerc}{\mathsf{es}_\emptyenv}
\newcommand{\EsAppend}{\mathsf{es}_\appendOp}

\renewcommand{\sem}[2]{\llbracket #1 \rrbracket_{#2}}

\renewcommand{\vmsym}{\mathsf{sh}}

\newcommand{\valu}{value}
\newcommand{\nonempty}{nonempty\xspace}
\newcommand{\empt}{empty\xspace}

\DeclarePairedDelimiter\abs{\lvert}{\rvert}%

\renewcommand{\size}[1]{\abs{#1}}

\renewcommand{\cons}{\hastype}

\renewcommand{\evctx}{C}

\renewcommand{\NoteProof}[1]{
\marginnote{
\scriptsize{Proof p.\,{\pageref{#1}}}}}
\renewcommand{\NoteState}[1]{
\marginnote{
\scriptsize{See p.\,{\pageref{#1}}}}}

%% file: 00_-_Abstract.tex
The good properties of Plotkin's call-by-value lambda-calculus crucially rely on the restriction to weak evaluation and closed terms. Open call-by-value is the more general setting where evaluation is weak but terms may be open. 
Such an extension is delicate and the literature contains a number of proposals. 
Recently, 
we provided operational and implementative studies of these proposals, showing that they are equivalent 
with respect to termination, and also 
at the \mbox{level of time cost models.} 

This paper explores the denotational semantics of open call-by-value, adapting de Carvalho's analysis of call-by-name via multi types (aka non-idempotent intersection types). Our type system characterises normalisation and thus provides an adequate relational semantics. Moreover, type derivations carry quantitative information about the cost of evaluation: their size bounds the number of evaluation steps and the size of the normal form, and we also characterise derivations giving exact bounds. 

The study crucially relies on a new, refined presentation of the fireball calculus, the simplest proposal for open call-by-value, that is more apt to denotational investigations.

%% file: 01_-_Introduction.tex
\section{Introduction}
\label{sect:intro}

The core of functional programming languages and proof assistants is usually modelled as a variation over the $\l$-calculus. Even when one forgets about type systems, there are in fact many $\l$-calculi rather than a single $\l$-calculus, depending on whether evaluation is weak or strong (that is, only outside or also inside abstractions), call-by-name (\cbn for short), call-by-value (\cbv),\footnotemark
\footnotetext{In CbV, 
	function's arguments are evaluated before being passed to the function, so 
	$\beta$-redexes can fire only when their arguments are values, i.e. abstractions or variables.}
or call-by-need, whether terms are closed or may be open, not to speak of extensions with continuations, pattern matching, fix-points, linearity constraints, and so on. 

\paragraph{Benchmark for $\l$-calculi.} A natural question is \emph{what is a good $\l$-calculus}? It is of course impossible to give an absolute answer, because different settings value different properties. It is nonetheless possible to collect requirements that seem desirable in order to have an abstract framework that is also useful in practice. We can isolate at least six principles to be satisfied by a good $\l$-calculus:
\begin{enumerate}
  \item \emph{Rewriting}: there should be a small-step operational semantics having nice rewriting properties. Typically, the calculus should be non-deterministic but confluent, and a deterministic evaluation strategy should emerge naturally from some good rewriting property (factorisation\,/\,standardisation theorem, or the diamond property). The \emph{strategy emerging from the calculus} principle guarantees that the chosen evaluation is not ad-hoc. 
  \item \emph{Logic}: typed versions of the calculus should be in Curry-Howard correspondences with some proof systems, providing logical intuitions and guiding principles for the features of the calculus and the study of its properties.
  \item \emph{Implementation}: there should be a good understanding of how to decompose evaluation in micro-steps, that is, at the level of abstract machines, in order to guide the design of languages or proof assistants based on the calculus.
  \item \emph{Cost model}: the number of steps of the deterministic evaluation strategy should be a reasonable time cost model,\footnotemark
  \footnotetext{Here \emph{reasonable} is a technical word meaning that the cost model is polynomially equivalent to the one of Turing machines.
  } so that cost analyses of $\l$-terms are possible, and independent of implementative choices. 
  \item \emph{Denotations}: there should be denotational semantics, that is, syntax-free mathematical interpretations of the calculus that are invariant by evaluation and that reflect some of its properties. Well-behaved denotations guarantee that the calculus is somewhat independent from its own syntax, which is a further guarantee that it is not ad-hoc. 
  
  \item \emph{Equality}: contextual equivalence can be characterised by some form of bisimilarity, showing that there is a robust notion of program equivalence. Program equivalence is indeed essential for studying program transformations and optimisations at work in compilers.
\end{enumerate}

Finally, there is a sort of meta-principle: the more principles are connected, the better. For instance, it is desirable that evaluation in the calculus corresponds to cut-elimination in some logical interpretation of the calculus. 
Denotations are usually at least required to be \emph{adequate} with respect to the rewriting: the denotation of a term is non-degenerated if and only if its evaluation terminates. 
Additionally, denotations are \emph{fully abstract} if they reflect contextual equivalence. And implementations have to work within an overhead that respects the intended cost semantics. Ideally, all principles are satisfied and perfectly interconnected.

Of course, some specific cases may drop some requirements---for instance, a probabilistic $\l$-calculus would not be confluent---some properties may also be strengthened---for instance, equality may be characterised via a separation theorem akin to Bohm's---and other principles may be added---categorical semantics, graphical representations, etc.

What is usually considered \emph{the} $\l$-calculus, is, in our terminology, the strong 
\cbn $\l$-calculus with (possibly) open terms, and all points of the benchmark have been studied for it. 
Plotkin's original formulation of 
\cbv \cite{DBLP:journals/tcs/Plotkin75}, conceived for weak evaluation and closed terms and here referred to as \emph{\ccbv}, also boldly passes the benchmark. 
Unfortunately Plotkin's setting fails the benchmark as soon as it is extended to open terms, which is required when using \cbv for implementing proof assistants, see \gregoire and Leroy's \cite{DBLP:conf/icfp/GregoireL02}. Typically, denotations are no longer adequate, as first noticed by Paolini and Ronchi Della Rocca \cite{parametricBook}, and there is a mismatch between evaluation in the calculus and cut-elimination in its linear logic interpretation, as shown by Accattoli \cite{DBLP:journals/tcs/Accattoli15}. 
The failure can be observed also at other levels not covered by our benchmark, 
\eg the incompleteness of CPS translations, already noticed by Plotkin himself \cite{DBLP:journals/tcs/Plotkin75}.

\paragraph{Benchmarking open call-by-value.} The problematic interaction of \cbv and open terms is well known, and the fault is usually given to the rewriting---the operational semantics has to be changed somehow. 
The literature contains a number of proposals for extensions of \cbv out of the closed world, some of which were introduced to solve the incompleteness of CPS translations. 
In 
\cite{DBLP:conf/aplas/AccattoliG16}, 
we provided a comparative study of four extensions of \ccbv\ (with weak evaluation on possibly open terms), showing that they have equivalent rewriting theories (namely, they are equivalent from the point of view of termination), 
they are all adequate with respect to denotations, and 
they share the same time cost models---these proposals have then to be considered as different incarnations of a more abstract framework, 
which we call \emph{open call-by-value} (\ocbv). 
Together with Sacerdoti Coen, we provided also a theory of implementations respecting the cost semantics \cite{fireballs,AccattoliGuerrieri17b}, and a precise linear logic interpretation \cite{DBLP:journals/tcs/Accattoli15}. 
Thus, \ocbv passes the first five points of the benchmark.

This paper deepens the analysis of the fifth point, by refining the denotational un\-der\-standing of \ocbv with a quantitative relationship with the rewriting and the cost model. 
We connect the size of type derivations for a term with its evaluation via rewriting, and the size of elements in its denotation with the size of its normal form, in a model coming from the linear logic interpretation of \cbv and presented as a type system: 
Ehrhard's relational semantics for~
\cbv~\cite{DBLP:conf/csl/Ehrhard12}.

The last point of the benchmark---contextual equivalence for \ocbv---was shown by Lassen to be a difficult question \cite{DBLP:conf/lics/Lassen05}, and it is left to future work.

\paragraph{Multi types.} Intersection types are one of the standard tools to study $\l$-calculi, mainly used to characterise termination properties---classical references are Coppo and Dezani \cite{DBLP:journals/aml/CoppoD78,DBLP:journals/ndjfl/CoppoD80}, Pottinger \cite{Pottinger80}, and Krivine \cite{Kri}.  In contrast to other type systems, they do not provide a logical interpretation, at least not as smoothly as for simple or polymorphic types---see Ronchi Della Rocca and Roversi's \cite{DBLP:conf/csl/RoccaR01} or Bono, Venneri, and Bettini's \cite{DBLP:journals/tcs/BonoVB08} for details. They are better understood, in fact, as syntactic presentations of denotational semantics: they are invariant under evaluation and type all and only the terminating terms, thus naturally providing an adequate denotational model.

Intersection types are a flexible tool that can
be formulated in various ways. A flavour that emerged 
in the last 10 years is that of \emph{non-idempotent} intersection
types, where the
intersection $A \cap A$ is not equivalent to $A$. They were first considered by Gardner \cite{DBLP:conf/tacs/Gardner94}, and then Kfoury \cite{DBLP:journals/logcom/Kfoury00}, Neergaard and Mairson \cite{DBLP:conf/icfp/NeergaardM04}, and de Carvalho \cite{Carvalho07,deCarvalho18} provided a first wave of works abut them---a survey can be
found in Bucciarelli, Kesner, and Ventura's~\cite{BKV17}. Non-idempotent intersections can be seen as multisets, which is why, to ease the
language, we prefer to call them \emph{multi types} rather than
\emph{non-idempotent intersection types}.

Multi types retain the denotational character of intersection types, and they actually refine it along two correlated lines. First, taking types with multiplicities gives rise to a \emph{quantitative} approach, that reflects 
resource consumption in the evaluation of terms. Second, such a quantitative feature turns out to coincide exactly with the one at work in linear logic. 
Some care is needed here: multi types do not correspond to linear logic formulas, rather to the relational denotational semantics of linear logic (two seminal references for such a semantic are Girard's~\cite{Girard88} and Bucciarelli and Ehrhard's \cite{DBLP:journals/apal/BucciarelliE01}; see also \cite{DBLP:conf/csl/Carvalho16,GuerrieriPellissierTortora16})---similarly to intersection types, they provide a denotational rather than a logical interpretation.

An insightful use of multi types is de Carvalho's connection between the size of types and the size of normal forms, and between the size of type derivations and evaluation lengths for the 
\cbn $\l$-calculus \cite{deCarvalho18}.

\paragraph{Types of fireballs.} This paper develops a denotational analysis of \ocbv akin to de Carvalho's. There are two main translations of the $\l$-calculus into linear logic, due to Girard \cite{DBLP:journals/tcs/Girard87}, the \cbn one, that underlies de Carvalho's study \cite{Carvalho07,deCarvalho18}, and the \cbv one, that is explored here. 
The literature contains denotational semantics of 
\cbv  and also studies of multi types for 
\cbv. The distinguishing feature of our study is the use of multi types to provide  bounds on the number of evaluation steps and on the size of normal forms, which has never been done before for 
\cbv, and 
moreover we do it for the open case---the result for the closed case, refining Ehrhard's study \cite{DBLP:conf/csl/Ehrhard12}, follows as a special case. 
Besides, we provide a characterisation of types and type derivations that provide \emph{exact} bounds, similarly to de Carvalho \cite{Carvalho07,deCarvalho18}, Bernadet and Lengrand \cite{Bernadet-Lengrand2013}, and de Carvalho, Pagani, and Tortora de Falco \cite{DBLP:journals/tcs/CarvalhoPF11}, and along the lines of a very recent work by Accattoli, Graham-Lengrand, and Kesner \cite{AccattoliLengrandKesner18}, but using a slightly different approach.

Extracting exact bounds from the multi types system is however only half of the story. The other, subtler half is about tuning up the presentation of \ocbv as to accommodate as many points of the benchmark as possible. Our quantitative denotational inquire via multi types requires the following properties:
\begin{enumerate}
	  \setcounter{enumi}{-1}
	\item \emph{Compositionality}:\label{p:compositionality} if two terms have the same type assignments, then 
		the terms obtained by plugging them in the same context do so.
  \item \emph{Invariance under evaluation}:\label{p:invariance} type assignments have to be stable by evaluation.  
  \item \emph{Adequacy}:\label{p:adequacy} a term is typable if and only if it terminates.
  \item \emph{Elegant normal forms}\label{p:nfs}: normal forms have a simple structure, so that the technical development is simple and intuitive.
  \item \emph{Number of steps}:\label{p:steps} type derivations have to provide the number of steps to evaluate to normal forms, and this number must be a reasonable cost model.
  \item \emph{Matching of sizes}:\label{p:sizes} the size of normal forms has to be bounded by the size of their types.
\end{enumerate}
While property \ref{p:compositionality} is not problematic (type systems/denotational models are conceived to satisfy it),
it turns out that none of the incarnations of \ocbv 
we studied in \cite{DBLP:conf/aplas/AccattoliG16} (namely, Paolini and Ronchi Della Rocca's \emph{fireball calculus} $\firecalc$ \cite{DBLP:journals/ita/PaoliniR99,parametricBook,DBLP:conf/icfp/GregoireL02,fireballs}, Accattoli and Paolini's \emph{value substitution calculus} $\vsubcalc$ \cite{DBLP:conf/flops/AccattoliP12,DBLP:journals/tcs/Accattoli15}, and Carraro and Guerrieri's \emph{shuffling calculus} $\shufcalc$ \cite{DBLP:conf/fossacs/CarraroG14,GuerrieriPR15,Guerrieri15,DBLP:journals/lmcs/GuerrieriPR17,Guerrieri18})\footnotemark
\footnotetext{In \cite{DBLP:conf/aplas/AccattoliG16} a fourth incarnation, the \emph{value sequent calculus} (a fragment of Curien and Herbelin's $\lambdamucalc$ \cite{DBLP:conf/icfp/CurienH00}), is 
proved	isomorphic to a fragment of $\vsubcalc$, which then subsumes~it.}
satisfies all 
the properties \ref{p:invariance}--\ref{p:sizes} at the same time: 
$\firecalc$ lacks property \ref{p:invariance} (as shown here in \refsect{fireball}); 
$\vsubcalc$ lacks property \ref{p:nfs} (the inelegant characterisation of normal forms is in \cite{DBLP:conf/flops/AccattoliP12}); and 
$\shufcalc$, which in \cite{DBLP:conf/fossacs/CarraroG14} is shown to satisfy \ref{p:invariance}, \ref{p:adequacy}, and partially \ref{p:nfs}, lacks properties
\ref{p:steps} (the number of steps does not seem to be  
a reasonable cost model, see \cite{DBLP:conf/aplas/AccattoliG16}) and \ref{p:sizes} (see  the end of \refsect{correctness} in this paper).

We then introduce the \emph{\splitc fireball calculus}, that is a minor variant of the fireball calculus $\firecalc$, isomorphic to it but integrating some 
features of the value substitution calculus $\vsubcalc$, and satisfying all the requirements for our study. 
Thus, the denotational study follows smooth and natural, fully validating the design and the benchmark.


To sum up, our study adds new ingredients to the understanding of \ocbv, by providing a simple and quantitative denotational analysis via an adaptation of de Carvalho's approach \cite{Carvalho07,deCarvalho18}. 
The main features of our study are:
\begin{enumerate}
\item \emph{Split fireball calculus}: a new incarnation of \ocbv more apt to denotational studies, and conservative with respect to the other properties of the setting.
\item \emph{Quantitative characterisation of termination}: proofs that typable terms are exactly the normalising ones, and that types and type derivations provide bounds on the size of normal forms and on evaluation lengths.
\item \emph{Tight derivations and exact bounds}: a class of type derivations that provide 
the exact length of evaluations, and such that the types in the final judgements provide 
the exact size of normal forms.
\end{enumerate}

\paragraph{Related work.} Classical studies of the denotational semantics of \ccbv are due to Sieber \cite{DBLP:conf/fsttcs/Sieber90}, Fiore and Plotkin \cite{DBLP:conf/lics/FioreP94}, Honda and Yoshida \cite{DBLP:journals/tcs/HondaY99}, and Pravato, Ronchi Della Rocca and Roversi \cite{DBLP:journals/mscs/PravatoRR99}. 
A number of works rely on multi types or relational semantics to study property of programs and proofs. 
Among them, Ehrhard's \cite{DBLP:conf/csl/Ehrhard12}, Diaz-Caro, Manzonetto, and Pagani's \cite{DBLP:conf/lfcs/Diaz-CaroMP13}, Carraro and Guerrieri's \cite{DBLP:conf/fossacs/CarraroG14}, Ehrhard and Guerrieri's \cite{DBLP:conf/ppdp/EhrhardG16}, and Guerrieri's \cite{Guerrieri18} deal with \cbv, while de Carvalho's \cite{Carvalho07,deCarvalho18}, Bernadet and Lengrand's \cite{Bernadet-Lengrand2013}, de Carvalho, Pagani, and Tortora de Falco's \cite{DBLP:journals/tcs/CarvalhoPF11}, Accattoli, Graham-Lengrand, and Kesner's \cite{AccattoliLengrandKesner18} provide exact bounds. Further related work about multi types is by Bucciarelli, Ehrhard, and Manzonetto 
\cite{DBLP:journals/apal/BucciarelliEM12}, de Carvalho and Tortora de Falco \cite{DBLP:journals/iandc/CarvalhoF16}, Kesner and Vial \cite{DBLP:conf/rta/KesnerV17}, and Mazza, Pellissier, and Vial \cite{DBLP:journals/pacmpl/MazzaPV18}---this list is not exhaustive.


%% file: 02_-_The_Rise_of_Fireballs.tex
\section{The Rise of Fireballs}
\label{sect:fireball}
\input{image_fireball_calculus}
In this section we recall the fireball calculus $\firecalc$, the simplest presentation of \ocbv. For the issues of Plotkin's setting with respect to open terms and for alternative presentations of \ocbv, we refer the reader to 
our work \cite{DBLP:conf/aplas/AccattoliG16}.

The fireball calculus  was introduced without a name and studied first by Paolini and Ronchi Della Rocca in~\cite{DBLP:journals/ita/PaoliniR99,parametricBook}.
It has then been rediscovered by \gregoire and Leroy
in \cite{DBLP:conf/icfp/GregoireL02} to improve the implementation of Coq, and later by Accattoli and Sacerdoti Coen in \cite{fireballs} to study cost models, where it was also named. We present it following \cite{fireballs}, changing only  inessential, cosmetic details.

\paragraph{The fireball calculus.} The fireball calculus $\firecalc$ is defined in \reffig{fireball-calculus}.
The idea is that the values of the \cbv $\l$-calculus
\,---\,\ie~abstractions and variables\,---\,are generalised to \emph{fireballs}, by extending variables to more general \emph{inert terms}. 
Actually fireballs (noted $\fire, \firetwo, \dots$) and inert terms (noted $\gconst, \gconsttwo, \dots$) are defined by mutual induction (in \reffig{fireball-calculus}). 
For instance, $\var$ and $\la\var\vartwo$ are fireballs as values, while $\vartwo(\la\var\var)$, $\var\vartwo$, and $(\varthree(\la\var\var))(\varthree\varthree) (\la\vartwo(\varthree\vartwo))$ are fireballs as inert terms. 

The main feature of inert terms is that they are open, normal, and that when plugged in a context they cannot create a redex, hence the name. Essentially, they are the \emph{neutral terms} of \ocbv. In \gregoire and Leroy's presentation \cite{DBLP:conf/icfp/GregoireL02}, inert terms are called \emph{accumulators} and fireballs are simply called values.

Terms are always identified up to $\alpha$-equivalence and the set of 
free variables of a term $\tm$ is denoted by $\fv{\tm}$. We use $\tm\isub\var\tmtwo$ for the term obtained by the capture-avoiding substitution of $\tmtwo$ for each free occurrence of $\var$ in $\tm$.

Variables are, morally, both values and inert terms. In \cite{fireballs} they were considered as inert terms, while here, for minor technical reasons we prefer to consider them as values and not as inert terms---the change is inessential.

\paragraph{Evaluation rules.} Evaluation is given by \emph{call-by-fireball} $\beta$-reduction $\tof$: the $\beta$-rule can fire, \emph{lighting} the argument, 
 only if the argument  
 is a fireball (\emph{fireball} is a catchier version of \emph{fire-able term}).
 We actually distinguish two sub-rules: one that \emph{lights} values, noted $\tobv$,
 and one that \emph{lights} inert terms, noted $\toin$ (see \reffig{fireball-calculus}). 
 Note that evaluation is \emph{weak}: it does not reduce under abstractions.
 
 We endow the calculus with the 
 (deterministic) right-to-left evaluation strategy, defined via right evaluation contexts $\ctx$---note the production $\ctx \fire$, forcing the right-to-left order. A more general calculus is defined in \cite{DBLP:conf/aplas/AccattoliG16}, for which 
 the right-to-left strategy is shown to be complete. 
 The left-to-right strategy, often adopted in the literature on \ccbv, is also complete, but in the open case the right-to-left one has stronger invariants that lead to simpler abstract machines (see \cite{AccattoliGuerrieri17b}), which is why we adopt it here. 
 We omit details about the rewriting theory of the fireball calculus because 
our focus here is on 
 denotational semantics.

\paragraph{Properties.} A famous key property of \ccbv (whose evaluation is exactly $\tobv$) is \emph{harmony}: given a closed term $\tm$, either it diverges or it evaluates to an abstraction, \ie~$\tm$ is $\betav$-normal if and only if $\tm$ is an abstraction. 
The fireball calculus $\firecalc$ satisfies an analogous property in the (more general) \emph{open} setting by replacing abstractions with fireballs (\refpropp{distinctive-fireball}{open-harmony}). 
Moreover, the fireball calculus is a \emph{conservative extension} of \ccbv: on closed terms it collapses on \ccbv (\refpropp{distinctive-fireball}{conservative}). 
No other presentation of \ocbv has these good properties.

\newcounter{prop:distinctive-fireball} 
\addtocounter{prop:distinctive-fireball}{\value{proposition}}
\begin{proposition}[Distinctive properties of $\firecalc$]
  \label{prop:distinctive-fireball}
  Let $\tm$ be a term.
\NoteProof{propappendix:distinctive-fireball}
  \begin{enumerate}
    \item\label{p:distinctive-fireball-open-harmony} \emph{Open harmony}:  $\tm$ is $\betaf$-normal if and only if $\tm$ is a fireball.
    \item\label{p:distinctive-fireball-conservative} \emph{Conservative open extension}: $\tm \tof \tmtwo$ if and only if $\tm \tobv \tmtwo$, when $\tm$ is closed.
  \end{enumerate}
\end{proposition}

\begin{example}
\label{ex:torf}  
  Let $\tm \defeq (\la{\varthree}{\varthree(\vartwo\varthree)})\la{\var}{\var}$.
  Then, $\tm \tof (\la{\var}{\var})(\vartwo \, \la{\var}{\var}) \tof \vartwo \, \la{\var}{\var}$, where the final term $\vartwo \, \la{\var}{\var}$ is a fireball (and $\betaf$-normal).
\end{example}

The key property of inert terms is summarised by the following proposition:
substitution of inert terms does not create or erase $\betaf$-redexes, and hence can always be avoided. It plays a role in \refsect{split-calculus}.

\newcounter{prop:inerts-and-creations}
\addtocounter{prop:inerts-and-creations}{\value{proposition}}
\begin{proposition}[Inert substitutions and evaluation commute]
\label{prop:inerts-and-creations}
\NoteProof{propappendix:inerts-and-creations}
  Let $\tm, \tmtwo$ be terms, $\gconst$ be an inert term. 
  Then, $\tm \tof \tmtwo$ if and only if $\tm \isub\var\gconst \tof \tmtwo \isub\var\gconst$.
\end{proposition}

With general terms (or even fireballs) instead of inert ones, evaluation and substitution do not commute, in the sense that both directions of \refprop{inerts-and-creations} do not hold.
Direction $\Leftarrow$ is false because substitution can create $\betaf$-redexes, as in $(\var \vartwo) \isub \var { \la\varthree \varthree } = (\la\varthree \varthree) \vartwo$;
direction $\Rightarrow$ is false because substitution can erase $\betaf$-redexes, as in $((\la\var\varthree)(\var \var)) \isub \var {\delta} = (\la\var \varthree)(\delta\delta)$ where $\delta \defeq \la\vartwo \vartwo\vartwo$.\footnotemark
\footnotetext{As well-known, 
\refprop{inerts-and-creations} with ordinary (\ie~\cbn
) $\beta$-reduction $\tob$ instead of $\tof$ and general terms instead of inert ones holds only \mbox{in direction $\Rightarrow$.}}

%% file: image_fireball_calculus.tex
\begin{figure}[t]
  \centering
  \scalebox{0.9}{
\begin{tabular}{c}
  
   $
    \begin{array}{r@{\hspace{.5cm}}rll}
	    	    \textsc{Terms} & \tm,\tmtwo,\tmthree,\tmfour &\grameq& \var \mid  \la\var\tm \mid \tm\tmtwo\\
    
    \textsc{Values} & \val,\valtwo,\valthree & \grameq & \var \mid \la\var\tm\\
    
    \textsc{Fireballs} & \fire, \firetwo, \firethree & \grameq & \val \mid \gconst\\
	    
	    \textsc{Inert terms} & \gconst,\gconsttwo, \gconstthree & \grameq &  \var \fire_1\ldots \fire_n\ \ \ \ n> 0\\
    
	    \textsc{Right evaluation contexts} & \evctx  & \grameq & \ctxhole\mid \tm\evctx \mid \evctx\fire \\\\	
\end{array}$
\\
	    $\begin{array}{@{\hspace{.8cm}}r@{\hspace{.3cm}}c@{\hspace{.3cm}}l@{\hspace{.8cm}}rll}
      \multicolumn{3}{c}{\textsc{Rule at top level}} & \multicolumn{3}{c}{\textsc{Contextual closure}} \\
	    (\la\var\tm)\val & \rtobv & \tm\isub\var{\val} &
	    \evctxp \tm & \tobv & \evctxp \tmtwo \textrm{~~~if } \tm \rtobv     \tmtwo \\
	    (\l\var.\tm)\gconst & \rtoin & \tm\isub\var\gconst &
	    \evctxp \tm & \toin & \evctxp \tmtwo \quad\textup{if } \tm \rtoin \tmtwo \\\\ 

	    \multicolumn{3}{c}{\textsc{Reduction}} & \tof & \defeq & \!\!\!\!\tobv \!\cup \toin
    \end{array}
    $
    \end{tabular}
 }
  \caption{\label{fig:fireball-calculus} The fireball calculus $\firecalc$.}
\end{figure}

%% file: 03_-_The_Fall_of_Fireballs.tex
\section{The Fall of Fireballs}
\label{sect:fall-fireballs}
Here we introduce Ehrhard's multi type system for \cbv \cite{DBLP:conf/csl/Ehrhard12} and show that---with respect to it---the fireball calculus $\firecalc$ fails the denotational test of the benchmark sketched in \refsect{intro}.
This is an issue of $\firecalc$: 
to our knowledge, all denotational models that are adequate for (some variant of) \cbv are not invariant under the evaluation rules of 
$\firecalc$, because of the rule $\toin$ substituting inert terms\footnotemark
\footnotetext{Clearly, any denotational model for the \cbn $\l$-calculus is invariant under $\betaf$-reduction (since $\tof \, \subseteq \, \tob$),
but there is no hope that it could be adequate for the fireball calculus. 
Indeed, such a model would identify the interpretations of $(\la{\var}{\vartwo})\Omega$ (where $\Omega$ is a diverging term and $\var \neq \vartwo$) and $\vartwo$, but in a \cbv setting these two terms have a completely different behaviour: 
$\vartwo$ is normal, whereas $(\la{\var}{\vartwo})\Omega$ cannot normalise.}. 

In the next sections we shall use this type system, while the failure  is not required for the main results of the paper, and may be skipped on a first reading.

\paragraph{Relational semantics.} We analyse the failure considering a concrete and well-known denotational model for \cbv: \emph{relational semantics}.
For Plotkin's original \cbv $\l$-calculus, it has been introduced by Ehrhard \cite{DBLP:conf/csl/Ehrhard12}. 
More generally, relational semantics provides a sort of canonical model of linear logic \cite{DBLP:journals/tcs/Girard87,DBLP:journals/apal/BucciarelliE01,GuerrieriPellissierTortora16,DBLP:conf/csl/Carvalho16}, and Ehrhard's model is the one obtained by representing the \cbv $\l$-calculus into linear logic, and then interpreting it according to the relational semantics. It is also strongly related to other denotational models for \cbv based on linear logic such as Scott domains and coherent semantics \cite{DBLP:conf/csl/Ehrhard12,DBLP:journals/mscs/PravatoRR99}, and it has a well-studied \cbn counterpart \cite{Carvalho07,deCarvalho18,DBLP:journals/apal/BucciarelliEM12,DBLP:journals/mscs/PaoliniPR17,DBLP:conf/lics/Ong17,DBLP:journals/pacmpl/MazzaPV18,AccattoliLengrandKesner18}.

Relational semantics for \cbv admits a nice syntactic presentation as a \emph{multi type system} (aka non-idempotent intersection types), introduced right next.
This type system, 
first studied by Ehrhard in \cite{DBLP:conf/csl/Ehrhard12}, 
is nothing but the \cbv version of de Carvalho's System $\mathsf{R}$ for \cbn $\lambda$-calculus \cite{Carvalho07,deCarvalho18}.

\paragraph{Multi types.} Multi types 
and linear types are defined by mutual induction:
 \begin{align*}
   &\textsc{\small Linear types} & \ntype,\ntypetwo &\grameq \Pair{\ptype}{\ptypetwo} \\[-.2\baselineskip]
   &\textsc{\small Multi types} & \ptype,\ptypetwo &\grameq \mset{\ntype_1, \dots, \ntype_n} \quad \text{(with } n \in \nat\text{)}
 \end{align*}
where $[\ntype_1, \dots, \ntype_n]$ is our notation for multisets. Note the absence of base types: their role is played by the \emph{empty multiset} $[\ ]$ 
(obtained for $n = 0$), that we rather note $\zero$ and refer to as \emph{the empty (multi) type}.
A multi type $[\ntype_1, \dots, \ntype_n]$ has to be intended as a conjunction $\ntype_1 \land \dots \land \ntype_n$ of linear types $\ntype_1, \dots, \ntype_n$, for a commutative and associative conjunction connective $\land$ that is not idempotent (morally a tensor $\otimes$) and whose neutral element is $\emptymset$.

The intuition is that a linear type corresponds to a single use of a term $\tm$, and that $\tm$ is typed with a multiset $\ptype$ of $n$ linear types if it is going to be used (at most) $n$ times. The  meaning of \emph{using a term} is not easy to define precisely. Roughly, it means that if $\tm$ is part of a larger term $\tmtwo$, then (at most) $n$ copies of $\tm$ shall end up in evaluation position during the evaluation of $\tmtwo$. More precisely, the $n$ copies shall end up in evaluation positions where \mbox{they are applied to some terms.}

The derivation rules for the multi types system are in \reffig{typesPlotkin}---
they are exactly the same as in \cite{DBLP:conf/csl/Ehrhard12}.
In this system, \emph{judgements} have the shape $\typctx \vdash \tm \hastype \ptype$ where $\tm$ is a term, $\ptype$ is a multi type and $\typctx$ is a \emph{type context}, that is, a total function from variables to multi types such that  the set $\domain{\typctx} \defeq \{\var \mid \typctx(\var) \neq \emptymset\}$ is finite. Note that terms are always assigned a multi type, and never a linear type---this is dual to what happens in de Carvalho's System $\mathsf{R}$ for \cbn \cite{Carvalho07,deCarvalho18}.

The application rule has a multiplicative formulation (in linear logic terminology), as it collects the type contexts of the two premises. 
The involved operation is the \emph{sum of type contexts} $\typctx \uplus \typctxtwo$, that is defined as $(\typctx \uplus \typctxtwo)(\var) \defeq \typctx(\var) \uplus \typctxtwo(\var)$, where the $\uplus$ in the RHS stands for the multiset sum. 
A type context $\typctx$ such that $\domain{\typctx} \subseteq \{\var_1, \dots, \var_n\}$ with $\var_i \neq \var_j$ and $\typctx(x_i) = \ptype_i$ for all $1 \leq i \neq j \leq n$ is often written as $\typctx = \var_1 \hastype \ptype_1,\dots, \var_n \hastype \ptype_n$. 
Note that the sum of type contexts $\uplus$ is commutative and associative, and its neutral element is the type context $\typctx$ such that $\domain{\typctx} = \emptyset$, which is called the \emph{empty type context} (all types in $\typctx$ are $\emptymset$).
The notation $\concl{\tder}{\typctx}{\tm}{\ptype}$ means that $\tder$ is a \emph{type derivation} $\tder$ (\ie a tree constructed using the rules in \reffig{typesPlotkin}) with conclusion the judgement $\typctx \vdash \tm \hastype \ptype$.

\input{typesPlotkin}

\paragraph{Intuitions: the empty type $\emptymset$.} Before digging into technical details let us provide some intuitions. A key type specific to the \cbv setting is the empty multiset $\emptymset$, also known as the empty (multi) type. 
The idea 
is that $\emptymset$ is the type of terms that can be erased. To understand its role in \cbv, 
we first recall its role in \cbn.

In the \cbn multi type system \cite{Carvalho07,deCarvalho18,AccattoliLengrandKesner18} every term, even a diverging one, is typable with $\emptymset$. On the one hand, this is correct, because in \cbn every term can be erased, and erased terms can also be divergent, because they are never evaluated. On the other hand, adequacy is formulated with respect to non-empty types: a term terminates if and only if it is typable with a non-empty type.

In \cbv, instead, terms have to be evaluated before being erased. And, of course, their evaluation has to terminate. Therefore, terminating terms and erasable terms coincide. Since the multi type system is meant to characterise terminating terms, in \cbv a term is typable
if and only if it is typable with $\emptymset$, as we shall prove in \refsect{tight}. 
Then the empty type is not a degenerate type, as in \cbn, it rather is \emph{the} type, characterising (adequate) typability altogether.

Note that, in particular, in a typing judgement $\typctx \vdash \expr \hastype \ptype$ the type context $\typctx$ may give the empty type to a variable $\var$ occurring in $\expr$, as for instance in the axiom $\var \hastype \emptymset \vdash \var \hastype \emptymset$---this may seem very strange to people familiar with \cbn multi types. We hope that instead, according to the provided intuition that $\emptymset$ is the type of termination, it would rather seem natural.

\paragraph{The model.} The idea to build the denotational model from the type system is that the interpretation (or 
semantics) of a term is simply the set of its type assignments, \ie the set of its derivable types together with their type contexts. 
More precisely, let $\tm$ be a term and $\var_1, \dots, \var_n$ (with $n \geq 0$) be pairwise distinct variables.
If $\Fv{\tm} \subseteq \{\var_1, \dots, \var_n\}$, we say that the list $\vec{\var} = (\var_1, \dots, \var_n)$ is \emph{suitable for} $\tm$.
If $\vec{\var} = (\var_1, \dots, \var_n)$ is suitable for $\tm$, the (\emph{relational}) \emph{semantics} 
\emph{of} $\tm$ \emph{for} $\vec{\var}$ is
\begin{equation*}
    \sem{\tm}{\vec{\var}} \defeq \{((\ptype_1,\dots, \ptype_n),\ptypetwo) \mid \exists \, \concl{\pi}{\var_1 \colon\! \ptype_1, \dots, \var_n \colon\! \ptype_n}{\tm}{\ptypetwo}\} \,.
\end{equation*}
%
%
Ehrhard proved that this is a denotational model for Plotkin's \cbv $\lambda$-calculus \cite[p.~267]{DBLP:conf/csl/Ehrhard12}, in the sense that the semantics of a \mbox{term is invariant under $\betav$-reduction.}

\begin{theorem}[Invariance for $\tobv$, \cite{DBLP:conf/csl/Ehrhard12}]
\label{thm:invariance-Plotkin}
  Let $\tm$ and $\tmtwo$ be two terms and $\vec{\var} = (\var_1, \dots, \var_n)$ be a suitable list of variables for $\tm$ and $\tmtwo$.
  If $\tm \tobv \tmtwo$ then $\sem{\tm}{\vec{\var}} = \sem{\tmtwo}{\vec{\var}}$.
\end{theorem}

Note that terms are not assumed to be closed. Unfortunately, relational semantics is not a denotational model of the fireball calculus $\firecalc$: \refthm{invariance-Plotkin} does not hold if we replace $\tobv$ with $\toin$ (and hence with $\tof$), 
as we show in the following example---the reader can skip it on a first reading.

\begin{example}[On a second reading: non-invariance of multi types in the fireball calculus]
\label{ex:counterexample}
Consider the fireball step $(\la{\varthree}{\vartwo})(\var\var) \tof \vartwo$, where the inert sub-term $\var\var$ is erased. Let us construct the interpretations of the terms $(\la{\varthree}{\vartwo})(\var\var) $ and $ \vartwo$. 
All type derivations for $\var\var$ are as follows ($\ptype$ and $\ptypetwo$ are arbitrary multi types):
{\small
\begin{equation*}
    \tder_{\ptype,\ptypetwo} =
    \AxiomC{}
    \RightLabel{\footnotesize$\Ax$}
    \UnaryInfC{$\var \hastype \mset{\Pair{\ptype}{\ptypetwo}} \vdash \var \hastype \mset{\Pair{\ptype}{\ptypetwo}}$}
    \AxiomC{}
    \RightLabel{\footnotesize$\Ax$}
    \UnaryInfC{$\var \hastype \ptype \vdash \var \hastype \ptype$}
    \RightLabel{\footnotesize$\ruleAp$}
    \BinaryInfC{$\var \hastype \mset{\Pair{\ptype}{\ptypetwo}} \uplus \ptype \vdash \var\var \hastype \ptypetwo$}
    \DisplayProof
\end{equation*}
}

Hence, all type derivations for $(\la{\varthree}{\vartwo})(\var\var)$ and $\vartwo$ have the following forms:

{\small
\begin{equation*}
\def\ScoreOverhang{1pt}
    \AxiomC{}
    \RightLabel{\footnotesize$\Ax$}
    \UnaryInfC{$\vartwo \hastype \ptypetwo \vdash \vartwo \hastype \ptypetwo$}
    \RightLabel{\footnotesize$\lambda$}
    \UnaryInfC{$\vartwo \hastype \ptypetwo \vdash \la{\varthree}{\vartwo} \hastype \mset{\Pair{\emptymset}{\ptypetwo}}$}
    \AxiomC{$\ \vdots\,\tder_{\ptype,\emptymset}$}
    \noLine
    \UnaryInfC{$\var \hastype \mset{\Pair{\ptype}{\emptymset}} \uplus \ptype \vdash \var\var \hastype \emptymset$}
    \RightLabel{\footnotesize$\ruleAp$}
    \BinaryInfC{$\var \hastype \mset{\Pair{\ptype}{\emptymset}} \uplus \ptype, \vartwo \hastype \ptypetwo \vdash (\la{\varthree}{\vartwo})(\var\var) \hastype \ptypetwo$}
    \DisplayProof
    \qquad
    \AxiomC{}
    \RightLabel{\footnotesize$\Ax$}
    \UnaryInfC{$\var \hastype \emptymset, \vartwo \hastype \ptypetwo \vdash \vartwo \hastype \ptypetwo$}
    \DisplayProof
\end{equation*}
}

Therefore,
\begin{align*}
\sem{(\la{\varthree}{\vartwo})(\var\var)}{\var,\vartwo} & =  \{((\mset{\Pair{\ptype}{\emptymset}} \uplus \ptype, \ptypetwo),\ptypetwo) \mid \ptype, \ptypetwo \text{ multi types}\}\\
\sem{\vartwo}{\var,\vartwo} &= \{ ((\emptymset, \ptypetwo),\ptypetwo) \mid \ptypetwo \text{ multi type}\}
\end{align*}

To sum up, in the fireball calculus $(\la{\varthree}{\vartwo})(\var\var) \tof \vartwo$, but $\sem{(\la{\varthree}{\vartwo})(\var\var)}{\var,\vartwo} \not\subseteq \sem{\vartwo}{\var,\vartwo}$ as $((\mset{\Pair{\emptymset}{\emptymset}}, \emptymset),\emptymset) \in \sem{(\la{\varthree}{\vartwo})(\var\var)}{\var,\vartwo} \smallsetminus \sem{\vartwo}{\var,\vartwo}$, and $\sem{\vartwo}{\var,\vartwo} \not\subseteq \sem{(\la{\varthree}{\vartwo})(\var\var)}{\var,\vartwo}$ because $((\emptymset, \emptymset),\emptymset) \in \sem{\vartwo}{\var,\vartwo} \smallsetminus \sem{(\la{\varthree}{\vartwo})(\var\var)}{\var,\vartwo}$.
\end{example}

An analogous problem affects the reduction step $(\la{\varthree}{\varthree\varthree})(\var\var) \tof (\var\var)(\var\var)$, where the inert term  $\var\var$ is instead duplicated.
In general, all counterexamples to the invariance of the relational semantics under $\betaf$-reduction are due to $\betain$-reduction, when the argument of the fired $\betaf$-redex is an inert term that is erased or duplicated.
Intuitively, to fix this issue, we should modify the syntax and operational semantics of 
$\firecalc$ in such a way that the $\betain$-step destroys the $\beta$-redex without  erasing nor duplicating its inert argument: \refprop{inerts-and-creations} guarantees that this modification is harmless.
This new presentation of 
$\firecalc$ is in the next section.

\begin{remark}[On a second reading: additional remarks about relational semantics]
\label{rmk:adequate}
\begin{enumerate}
  \item Relational semantics is invariant for Plotkin's \cbv even in presence of \emph{open} terms, but \emph{it no longer is an adequate model}: 
  the term $(\la{\vartwo}{\delta})(\var\var)\delta$ (where $\delta \defeq \la{\varthree}{\varthree\varthree}$) has an empty semantics (\ie is not typable in the multi type system of \reffig{typesPlotkin}) but it is $\betav$-normal. 
  Note that, instead, it diverges in $\firecalc$ because a $\betain$-step ``unblocks" it: $(\la{\vartwo}{\delta})(\var\var)\delta \toin \delta\delta \tobv \delta\delta \tobv \dots$
  \item Even though it is not a denotational model for the fireball calculus, relational semantics is \emph{adequate} for it, in the sense that a term is typable in the multi types system of \reffig{typesPlotkin} if and only if it $\betaf$-normalises.
  This follows from two results involving the shuffling calculus, an extension of Plotkin's \cbv that is another presentation of \ocbv: 
  \begin{itemize}
    \item the adequacy of the relational semantics for the shuffling calculus \cite{DBLP:conf/fossacs/CarraroG14,Guerrieri18};
    \item the equivalence of the fireball calculus $\firecalc$ and shuffling calculus $\shufcalc$ from the termination point of view, \ie a term normalises in one calculus if and only if it normalises in the other one \cite{DBLP:conf/aplas/AccattoliG16}.
  \end{itemize}
\end{enumerate}
Unfortunately, the shuffling calculus $\shufcalc$ has issues with respect to the quantitative aspects of the semantics (it is unknown whether its number of steps is a reasonable cost model \cite{DBLP:conf/aplas/AccattoliG16}; the size of $\shufcalc$-normal forms is not bounded by the size of their types, as we 
show in \refex{normal-sizes-shuffling}), 
which instead better fit the fireball calculus
$\firecalc$. 
This is why in the next section we slightly modify 
$\firecalc$, rather than~switching~to~$\shufcalc$. 
\end{remark}

%% file: typesPlotkin.tex
\begin{figure}[!t]
  \begin{center}
  \begin{tabular}{c@{\hspace{1.2cm}}c}
      \AxiomC{}
      \RightLabel{\footnotesize$\Ax$}
      \UnaryInfC{$\var \hastype \ptype \vdash \var \hastype \ptype$}
      \DisplayProof
&
      \AxiomC{$\typctx \vdash \tm \hastype \mset{ \Pair\ptype\ptypetwo }$}
      \AxiomC{$\typctxtwo \vdash \tmtwo \hastype \ptype$}
      \RightLabel{\footnotesize$\ruleAp$}
      \BinaryInfC{$\typctx \uplus \typctxtwo \vdash \tm\tmtwo \hastype \ptypetwo$}
      \DisplayProof
      \\\\
  \multicolumn{2}{c}{
      \AxiomC{$\typctx_1, \var \hastype \ptype_1 \vdash \tm \hastype \ptypetwo_1$}
      \AxiomC{$\overset{n \in \nat}{\dots}$}
      \AxiomC{$\typctx_n, \var \hastype \ptype_n \vdash \tm \hastype \ptypetwo_n$}
      \RightLabel{\footnotesize$\mathsf{\lambda}$}
      \TrinaryInfC{$\typctx_1 \uplus \dots \uplus \typctx_n \vdash \la{\var}{\tm} \hastype \mset{ \Pair{\ptype_1}{\ptypetwo_1}, \dots, \Pair{\ptype_n}{\ptypetwo_n} }$}
      \DisplayProof}
      \end{tabular}
            \end{center}
\caption{Multi types system for Plotkin's \cbv $\lambda$-calculus \cite{DBLP:conf/csl/Ehrhard12}.  }
  \label{fig:typesPlotkin}
\end{figure}

%% file: 04_-_Fireballs_Reloaded_the_Split_Fireball_Calculus.tex
\section{\texorpdfstring{Fireballs Reloaded: the \Splitc Fireball Calculus $\sfirecalc$}{Fireballs Reloaded: the \Splitc Fireball Calculus}}
\label{sect:split-calculus}
\input{image_split_fireball_calculus}
This section presents the \emph{split fireball calculus} $\sfirecalc$, that is the refinement of the fireball calculus $\firecalc$ correcting the issue explained in the previous section (\refex{counterexample}), namely the non-invariance of type assignments by evaluation.

The calculus $\sfirecalc$ is defined in \reffig{split-fireball-calculus}. 
The underlying idea is simple: the problem with the fireball calculus is the substitution of inert terms, as discussed in \refex{counterexample};
but some form of $\betain$-step is needed to get the adequacy of relational semantics in presence of open terms, as shown in \refrmk{adequate}. 
Inspired by \refprop{inerts-and-creations}, the solution is to keep trace of the inert terms involved in 
$\betain$-steps in an auxiliary environment, without substituting them in the body of the abstraction. 
Therefore, we introduce the syntactic category of \emph{programs} $\prog$, that are terms with an \emph{environment} $\genv$, which in turn is 
a list of 
explicit (\ie delayed) substitutions paring variables and inert terms. 
We use \emph{expressions} $\expr,\expr',\ldots$ to refer to the union of terms and programs.
Note the new form of the rewriting rule $\toin$, that does not substitute the inert term and rather adds an entry to the environment.
Apart from storing inert terms, the environment does not play any active role in $\betaf$-reduction for $\sfirecalc$.
Even though $\tof$ is a binary relation on programs, we use `\emph{normal expression}' to refer to either a normal (with respect to $\tof$) program or a term $\tm$ such that the program $(\tm, \genv)$ is normal (for any environment $\genv$).

The good properties of the fireball calculus are retained. 
Harmony in $\sfirecalc$ 
takes the following form (for arbitrary fireball $\fire$ and environment $\genv$):

\newcounter{prop:harmony-split}
\addtocounter{prop:harmony-split}{\value{proposition}}
\begin{proposition}[Harmony]
\label{prop:harmony-split}
\NoteProof{propappendix:harmony-split}
 A program $\prog$ is normal if and only if $\prog = (\fire, \genv)$.
\end{proposition}
\noindent So, an expression is normal iff it is a fireball $\fire$ or a program of the form $(\fire,\genv)$.

Conservativity with respect to the closed case is also immediate, because in the closed case the rule $\toin$ never fires and so the environment is always empty.

\paragraph{On a second reading: no open size explosion.} Let us mention that avoiding the substitution of inert terms is also natural at the implementation\,/\,cost model level, as substituting them causes \emph{open size explosion}, an issue studied at length in previous work on the fireball calculus \cite{fireballs,AccattoliGuerrieri17b}. Avoiding the substitution of inert terms altogether is in fact what is done by the other incarnations of \ocbv, as well as by abstract machines. 
The split fireball calculus $\sfirecalc$ can in fact be seen as adding the environment of abstract machines but without having to deal with the intricacies of decomposed evaluation rules. It can also be seen as the (open fragment of) Accattoli and Paolini's \emph{value substitution calculus} \cite{DBLP:conf/flops/AccattoliP12}, where indeed inert terms are never substituted. In particular, it is possible to prove that the normal forms of the split fireball calculus are isomorphic to those of the value substitution up to its structural equivalence (see \cite{DBLP:conf/aplas/AccattoliG16} for the definitions).

\paragraph{On a second reading: relationship with the fireball calculus.} The split and the (plain) fireball calculus are isomorphic at the rewriting level. To state the relationship we need the concept of \emph{program unfolding} $\unf{(\tm,\genv)}$, that is, the term obtained by substituting the inert terms in the environment $\genv$ into the main~term~$\tm$: 
\begin{align*}
	\unf{(\tm,\emptyenv)} & \defeq \tm 
  &    \unf{(\tm,\esub\vartwo\itm\cons \genv)} & \defeq  \unf{(\tm\isub\var\itm,\genv)}
\end{align*}

From the commutation of evaluation and substitution of inert terms in the fireball calculus (\refprop{inerts-and-creations}), it follows that normal programs (in $\sfirecalc$) unfold to normal terms (in $\firecalc$), that is, fireballs. 
Conversely, every fireball can be seen as a normal program with respect to the empty environment.

For evaluation, the same commutation property easily gives the following strong bisimulation between the split $\sfirecalc$ and the plain $\firecalc$ fireball calculi.

\newcounter{prop:strong-bisimulation}
\addtocounter{prop:strong-bisimulation}{\value{proposition}}
\begin{proposition}[Strong bisimulation]
\label{prop:strong-bisimulation}
\NoteProof{propappendix:strong-bisimulation}
    Let $\prog$ be a program (in $\sfirecalc$).
    \begin{enumerate}
    \item \emph{Split to plain}: if $\prog \tof \progtwo$ then  $\unf\prog \tof \unf\progtwo$.
    \item \emph{Plain to split}: if $\unf\prog \tof \tmtwo$ then there exists $\progtwo$ such that $\prog \tof \progtwo$ and $\unf\progtwo = \tmtwo$.
    \end{enumerate}
\end{proposition}

It is then immediate that termination in the two calculi coincide, as well as the number of steps to reach a normal form. Said differently, the split fireball calculus can be seen as an \emph{isomorphic} refinement of the fireball calculus.

%% file: image_split_fireball_calculus.tex
%

\begin{figure}[t]
  \centering
  \scalebox{0.9}{
\begin{tabular}{c}
%
       $\begin{array}{r@{\hspace{.5cm}}rcl}
	    	    \textsc{Terms}, \textsc{Values}, \textsc{Fireballs}, & \multicolumn{3}{c}{\multirow{2}{*}{\textup{as for the fireball calculus $\firecalc$}}} \\
	    	    \textsc{Inert terms},\ \textsc{Right ev.~contexts}  \\[.2cm]
	    	    \textsc{Environments} & \genv	& \grameq & \emptyenv \mid \esub{\var}{\itm} \cons \genv\\
\textsc{Programs} & \prog & \grameq & (\tm, \genv)\\[.2cm]
    \multirow{2}{*}{\textsc{Rules}} & (\evctxp{(\la\var \tm)~\val}, \genv) & \tobv & (\evctxp{\tm \isub\var\val},\genv) \\
     & (\evctxp{(\la\var \tm)~\itm}, \genv) & \toin & (\evctxp{\tm)}, \esub\var\itm \cons \genv) \\[.1cm]
	    \textsc{Reduction} & \tof & \defeq & \tobv \!\cup \toin
\end{array}$

    \end{tabular}
 }
  \caption{\label{fig:split-fireball-calculus} The split fireball calculus $\sfirecalc$.}
\end{figure}

%% file: 05_-_Multi_Types_for_Split.tex
\section{\texorpdfstring{Multi Types for $\sfirecalc$}{Multi Types for the split fireball calculus}} 
\label{sect:types-for-split}

The multi type system for the \splitc fireball calculus $\sfirecalc$ is the natural extension to terms with environments of the multi type system for Plotkin's \cbv $\lambda$-calculus seen in \refsect{fireball}.
Multi and linear types are the same.
The only novelty is that now judgements type expressions, not only terms, hence we add two new rules for the two cases of environment, $\EsCoerc$ and $\EsAppend$, see \reffig{types}. Rule $\EsCoerc$ is trivial, it is simply the coercion of a term to a program with an empty environment. Rule $\EsAppend$ uses the \emph{append} operation $\genv \appendOp \esub\var\itm$ that appends an entry $\esub\var\itm$ to the end of an environment $\genv$, formally defined as follows:
\begin{align*}
 \emptyenv\appendOp \esub\var\itm & \defeq \esub\var\itm	    &
 (\esub\vartwo\itmtwo\cons \genv) \appendOp \esub\var\itm & \defeq  \esub\vartwo\itmtwo\cons (\genv \appendOp \esub\var\itm)
\end{align*}

We keep all the notations already used for multi types in \refsect{fall-fireballs}. 

\input{typesNew}

\paragraph{Sizes, and basic properties of typing.} 
For our quantitative analyses, we need the notions of size for terms, programs and type derivations.

The \emph{size $\size\tm$ of a term} $\tm$ is the number of its applications not under the scope of an abstraction.
The \emph{size $\size{(\tm, \genv)}$ of a program} $(\tm, \genv)$ is the size of $\tm$ plus the size of the (inert) terms in the environment $\genv$. Formally, they are defined as follows:
    \begin{align*}
    \sizeap{\val} &\defeq 0 & \sizeap{\tm\tmtwo} &\defeq \sizeap{\tm} + \sizeap{\tmtwo} + 1 & \qquad
    \sizeap{(\tm, \emptyenv)} &\defeq \sizeap{\tm} & \sizeap{(\tm, \genv  \appendOp \esub{\var}{\itm})} &\defeq \sizeap{(\tm,\genv)} +  \sizeap{\itm}
    \end{align*}
    
  The \emph{size $\size{\tder}$ of a type derivation} $\tder$ is the number of its $\ruleAp$ rules.

The proofs of the next basic properties of type derivations are straightforward.

\newcounter{l:free}
\addtocounter{l:free}{\value{lemma}}
\begin{lemma}[Free variables in typing]
\label{l:free}
  If $\concl{\tder}{\typctx}{\expr}{\ptype}$ then $\Dom{\typctx} \subseteq \Fv{\expr}$.
\end{lemma}



%
%

The next lemma collects some basic properties of type derivations for values.

\begin{lemma}[Typing of \valu s]
\label{l:value-typing} 
 Let $\concl{\tder}{\typctx}{\val}{\ptype}$ be a type derivation for a value $\val$.
 Then,
 \begin{enumerate}
 \item \label{p:value-typing-empty}
 \emph{Empty multiset implies null size}: 
 if $\ptype = \emptymset$ \mbox{then $\domain{\typctx} = \emptyset$ and $\size{\tder} = 0 = \sizeap{\val}$.}

 \item \label{p:value-typing-dec}
 \emph{Multiset splitting}: if 
 $\ptype = \ptypetwo \uplus \ptypethree$, then there are two type contexts $\typctxtwo$ and $\typctxthree$ and two type derivations $\concl{\tdertwo}{\typctxtwo}{\val}{\ptypetwo}$ and $\concl{\tderthree}{\typctxthree}{\val}{\ptypethree}$ such that $\typctx = \typctxtwo \uplus \typctxthree$ and $\size{\tder} = \size \tdertwo + \size \tderthree$.


\item \label{p:value-typing-judg}
\emph{Empty judgement}: there is a type derivation $\concl{\tdertwo}{\ }{\val}{\emptymset}$.

 \item \label{p:value-typing-merg}
 \emph{Multiset merging}: for any two type derivations $\concl{\tder}{\typctx}{\val}{\ptype}$ and $\concl{\tdertwo}{\typctxtwo}{\val}{\ptypetwo}$ there is a type derivation 
 $\concl{\tderthree}{\typctx \uplus \typctxtwo}{\val}{\ptype \uplus \ptypetwo}$ such that 
 $\size \tderthree = \size \tder + \size \tdertwo$.
 \end{enumerate}
\end{lemma}

The next two sections prove that the multi type system is correct (\refsect{correctness}) and complete (\refsect{completeness}) for termination in the split fireball calculus $\sfirecalc$, also providing bounds 
for the length $\size{\deriv}$ of a normalising evaluation $\deriv$
and for the size of normal forms. 
At the end of \refsect{completeness} we discuss the adequacy of the relational model induced by this multi type system, with respect to $\sfirecalc$. \refsect{tight} characterises types and type derivations that provide exact bounds.



%% file: typesNew.tex
\begin{figure}[!t]
  \centering
  
      \AxiomC{}
      \RightLabel{\footnotesize$\Ax$}
      \UnaryInfC{$\var \hastype \ptype \vdash \var \hastype \ptype$}
      \DisplayProof
      \qquad
      \AxiomC{$\typctx \vdash \tm \hastype \mset{ \Pair\ptype\ptypetwo }$}
      \AxiomC{$\typctxtwo \vdash \tmtwo \hastype \ptype$}
      \RightLabel{\footnotesize$\ruleAp$}
      \BinaryInfC{$\typctx \uplus \typctxtwo \vdash \tm\tmtwo \hastype \ptypetwo$}
      \DisplayProof
      \\[\baselineskip]
      \AxiomC{$\typctx_1, \var \hastype \ptype_1 \vdash \tm \hastype \ptypetwo_1$}
      \AxiomC{$\overset{n \in \nat}{\dots}$}
      \AxiomC{$\typctx_n, \var \hastype \ptype_n \vdash \tm \hastype \ptypetwo_n$}
      \RightLabel{\footnotesize$\mathsf{\lambda}$}
      \TrinaryInfC{$\typctx_1 \uplus \dots \uplus \typctx_n \vdash \la{\var}{\tm} \hastype \mset{ \Pair{\ptype_1}{\ptypetwo_1}, \dots, \Pair{\ptype_n}{\ptypetwo_n} }$}
      \DisplayProof
          \\[\baselineskip]
			
      \AxiomC{$\typctx \vdash \tm \hastype \ptype$}      
      \RightLabel{\footnotesize$\EsCoerc$}
      \UnaryInfC{$\typctx \vdash (\tm,\emptyenv) \hastype \ptype$}
      \DisplayProof
      \qquad
      \AxiomC{$\typctx, \var \hastype \ptype \vdash (\tm,\genv) \hastype \ptypetwo$}
      \AxiomC{$\typctxtwo \vdash \itm \hastype \ptype$}
      \RightLabel{\footnotesize$\EsAppend$}
      \BinaryInfC{$\typctx \uplus \typctxtwo \vdash (\tm,\genv \appendOp \esub\var\itm) \hastype \ptypetwo$}
      \DisplayProof
  \caption{Multi types system for the split fireball calculus.  }
  \label{fig:types}
\end{figure}

%% file: 06_-_Correctness.tex
\section{Correctness}
\label{sect:correctness}
Here we prove correctness (\refthm{correctness}) of multi types for $\sfirecalc$, refined with quantitative information: if a term is typable then it terminates, and the type derivation provides bounds for both the number of steps to normal form and the size of the normal form. After the correctness theorem we show that even types by themselves---without the derivation---bound the size of normal forms.

\paragraph{Correctness.} The proof technique is standard. 
Correctness is obtained 
from subject reduction 
(\refprop{quant-subject-reduction}) 
plus a 
property of 
typings of normal forms (\refprop{size-normal}).

\newcounter{prop:size-normal}
\addtocounter{prop:size-normal}{\value{proposition}}
\begin{proposition}[Type derivations bound the size of normal forms]
\label{prop:size-normal} 
\NoteProof{propappendix:size-normal}
Let $\concl{\tder}{\typctx}{\expr}{\ptype}$ be a type derivation for a normal 
expression $\expr$. \mbox{Then 
$\sizeap\expr \leq \sizeap\tder$.}
\end{proposition}

%

As it is standard in the study of type systems, subject reduction requires a substitution lemma for typed terms, here refined with quantitative information. 

\newcounter{l:substitution}
\addtocounter{l:substitution}{\value{lemma}}
\begin{lemma}[Substitution]
\label{l:substitution}
\NoteProof{lappendix:substitution}
  Let $\concl{\tder}{\typctx, \var \hastype \ptypetwo}{\tm}{\ptype}$ and $\concl{\tdertwo}{\typctxtwo}{\val}{\ptypetwo}$ (where $\val$ is a value). 
  Then there exists $\concl{\tderthree}{\typctx \uplus \typctxtwo}{\tm\isub\var\val}{\ptype}$ such that $\size{\tderthree} = \size{\tder} + \size{\tdertwo}$.
\end{lemma}

The key point of the next \emph{quantitative} subject reduction property is the fact that the size of the derivation decreases by \emph{exactly} 1 at each evaluation step.

\newcounter{prop:quant-subject-reduction}
\addtocounter{prop:quant-subject-reduction}{\value{proposition}}
\begin{proposition}[Quantitative subject reduction]
\label{prop:quant-subject-reduction}
\NoteProof{propappendix:quant-subject-reduction}
  Let $\prog$ and $\progp$ be programs and $\concl{\tder}{\typctx}{\prog}{\ptype}$ be a type derivation for $\prog$.
If $\prog \tof \progp$ then $\size{\tder} > 0$ and there exists a type derivation $\concl {\tder'} \typctx \progp \ptype$ 
such that $\size{\tder'} = \size{\tder} - 1$.
\end{proposition}

%
Correctness now 
follows as an easy induction on the size of the type derivation, which bounds both the length $\size{\deriv}$ of the---normalising---evaluation $\deriv$ (\ie the number of $\betaf$-steps in $\deriv$) by \refprop{quant-subject-reduction}, and the size of the normal form by \refprop{size-normal}.

\newcounter{thm:correctness}
\addtocounter{thm:correctness}{\value{theorem}}
\begin{theorem}[Correctness]
\label{thm:correctness}
\NoteProof{thmappendix:correctness}
Let $\concl{\tder}{\typctx}{\prog}{\ptype}$ be a type derivation. 
Then there exist a normal program $\progtwo$ and an evaluation $\deriv \colon \prog \tof^* \progtwo$ with $\size\deriv + \sizeap\progtwo \leq \sizeap\tder$. 
\end{theorem}

\paragraph{Types bound the size of normal forms.} In our multi type system, not only type derivations but also multi types provide quantitative information, in this case on the size of normal forms.

First, we need to define the size for multi types and type contexts, which is simply given by the number of occurrences of $\Pair{}{}$. Formally, the size of linear and multi types are defined by mutual induction by $\size{\Pair{\ptype}{\ptypetwo}} \defeq 1 + \size\ptype + \size\ptypetwo$ and $\size{\mset{\ntype_1, \dots, \ntype_n}} \defeq \sum_{i=1}^n \size{\ntype_i}$. 
Clearly, $\size{\ptype} \geq 0$ and $\size{\ptype} = 0$ if and only if $\ptype = \emptymset$. 

Given a type context $\typctx = \var_1 \hastype \ptype_1, \dots, \var_n \hastype \ptype_n$ we often consider the list of its types, 
noted  $\typelist\typctx \defeq (\ptype_1, \dots, \ptype_n)$.  Since any list of multi types $(\ptype_1, \dots, \ptype_n)$ can be seen as extracted from a type context $\typctx$, we 
use the notation $\typelist\typctx$ for lists of multi types.
The size of a list of multi types is given by $\size{(\ptype_1, \dots, \ptype_n)} \allowbreak\defeq \allowbreak\sum_{i=1}^n \size{\ptype_i}$. 
 Clearly, $\domain{\typctx} = \emptyset$ if and only if $\size{\typelist\typctx} = 0$.


The quantitative information is that the size of types bounds the size of normal forms. In the case of inert terms a stronger bound actually holds.

\newcounter{prop:types-bound-nfs}
\addtocounter{prop:types-bound-nfs}{\value{proposition}}
\begin{proposition}[Types bound the size of normal forms]
\label{prop:types-bound-nfs}
\NoteProof{propappendix:types-bound-nfs}
  Let $\expr$ be a normal expression.
  For any type derivation $\concl{\tder}{\typctx}{\expr}{\ptype}$, one has $\sizeap{\expr} \leq \size{(\typelist\typctx, \ptype)}$.
  If moreover $\expr$ is an inert term, then $\sizeap{\expr} + \size{\ptype}\leq \size{\typelist\typctx}$.
\end{proposition}

\begin{example}[On a second reading: types, normal forms, and $\shufcalc$]
\label{ex:normal-sizes-shuffling}
The fact that multi types bound the size of normal forms is a quite delicate result that holds in the \splitc fireball calculus $\sfirecalc$ but does not hold in other presentations of \ocbv, like the shuffling calculus $\shufcalc$ \cite{DBLP:conf/fossacs/CarraroG14,Guerrieri18}, as we now show---this is one of the reasons motivating the introduction of $\sfirecalc$.
Without going into the details of $\shufcalc$, consider $\tm \defeq ({\la{\varthree}{\varthree}})(\var\var)$: it is normal for 
$\shufcalc$ but it\,---\,or, more precisely, the program $\prog \defeq (\tm, \emptyenv)$\,---\,is not normal for $\sfirecalc$, indeed $\prog \tof^* (\varthree, \esub{\vartwo}{\var\var}) \eqdef \progtwo$ and $\progtwo$ is normal in $\sfirecalc$.
Concerning sizes, $\sizeap{\tm} = \sizeap{\prog} = 2$ and $\sizeap{\progtwo} = 1$.
Consider the following type derivation for $\tm$ (the type derivation $\tder_{\emptymset,\emptymset}$ is defined in \refex{counterexample}):

{\small
\begin{center}
\def\ScoreOverhang{1pt}
    \AxiomC{}
    \RightLabel{\footnotesize$\Ax$}
    \UnaryInfC{$\varthree \hastype \emptymset \vdash \varthree \hastype \emptymset$}
    \RightLabel{\footnotesize$\lambda$}
    \UnaryInfC{$\vdash \la{\varthree}{\varthree} \hastype \mset{\Pair{\emptymset}{\emptymset}}$}
    \AxiomC{$\ \vdots\,\tder_{\emptymset,\emptymset}$}
    \noLine
    \UnaryInfC{$\var \hastype \mset{\Pair{\emptymset}{\emptymset}} \vdash \var\var \hastype \emptymset$}
    \RightLabel{\footnotesize$\ruleAp$}
    \BinaryInfC{$\var \hastype \mset{\Pair{\emptymset}{\emptymset}} \vdash (\la{\varthree}{\varthree})(\var\var) \hastype \emptymset$}
    \DisplayProof
\end{center}
}

\noindent So, $\size{\tm} = 2 > 1 = \size{(\mset{\Pair{\emptymset}{\emptymset}},\emptymset)}$, which gives a counterexample to \refprop{types-bound-nfs} in 
$\shufcalc$.
\end{example}

%% file: 07_-_Completeness.tex
\section{Completeness}
\label{sect:completeness}
Here we prove completeness (\refthm{completeness}) of multi types for $\sfirecalc$, refined with quantitative information: if a term terminates then it is typable, and the quantitative information is the same as in the correctness theorem (\refthm{correctness} above). 
After that, we discuss the adequacy of the relational semantics induced by the multi type system, with respect to termination in $\sfirecalc$.

\paragraph{Completeness.} The proof technique, again, is standard. Completeness is obtained by a subject expansion property plus the fact that all normal forms are typable. 

\newcounter{prop:nfs-are-typable}
\addtocounter{prop:nfs-are-typable}{\value{proposition}}
\begin{proposition}[Normal forms are typable]
\label{prop:nfs-are-typable}
\NoteProof{propappendix:nfs-are-typable}
  \begin{enumerate}
    \item \emph{Normal expression:} for any normal expression $\expr$, there exists a type derivation $\concl{\tder}{\typctx}{\expr}{\ptype}$ for some type context $\typctx$ and some multi type $\ptype$.
    \item \emph{Inert term:} for any multi type $\ptypetwo$ and any inert term $\itm$, there exists a type derivation $\concl{\tdertwo}{\typctxtwo}{\itm}{\ptypetwo}$ for some type context $\typctxtwo$.
  \end{enumerate}
\end{proposition}

In the proof of \refprop{nfs-are-typable}, the stronger statement for inert terms is required, 
to type a normal expression that is a program with non-empty environment.

For quantitative subject expansion (\refprop{quant-subject-expansion}), which is dual to subject reduction (\refprop{quant-subject-reduction} above), we 
need an anti-substitution lemma that is the dual of the substitution one (\reflemma{substitution} above).

\newcounter{l:anti-substitution}
\addtocounter{l:anti-substitution}{\value{lemma}}
\begin{lemma}[Anti-substitution]
\label{l:anti-substitution}
\NoteProof{lappendix:anti-substitution}
  Let $\tm$ be a term, $\val$ be a value, and $\concl{\tder}{\typctx}{\tm\isub\var\val}{\ptype}$ be a type derivation. 
  Then there exist two type derivations $\concl{\tdertwo}{\typctxtwo, \var \hastype \ptypetwo}{\tm}{\ptype}$ and $\concl{\tderthree}{\typctxthree}{\val}{\ptypetwo}$ such that $\typctx = \typctxtwo \uplus \typctxthree$ and $\size{\tder} = \size{\tdertwo} + \size{\tderthree}$.
\end{lemma}

Subject expansion follows. Dually to subject reduction, the size of the type derivation grows by \emph{exactly} 1 along every expansion (\ie along every anti-$\betaf$-step).

\newcounter{prop:quant-subject-expansion}
\addtocounter{prop:quant-subject-expansion}{\value{proposition}}
\begin{proposition}[Quantitative subject expansion]
\label{prop:quant-subject-expansion}
\NoteProof{propappendix:quant-subject-expansion}
  Let $\prog$ and $\progp$ be programs and $\concl{\tder'}{\typctx}{\progp}{\ptype}$ be a type derivation for $\progp$.
If $\prog \tof \progp$ then there exists a type derivation $\concl {\tder} \typctx \prog \ptype$ for $\prog$ such that $\size{\tder'} = \size{\tder} - 1$.
\end{proposition}

\newcounter{thm:completeness}
\addtocounter{thm:completeness}{\value{theorem}}
\begin{theorem}[Completeness]
\label{thm:completeness}
\NoteProof{thmappendix:completeness}
Let $\deriv \colon \prog \tof^* \progtwo$ be a normalising evaluation. 
Then there is a type derivation $\concl{\tder}{\typctx}{\prog}{\ptype}$, and it satisfies $\size\deriv + \sizeap\progtwo \leq \sizeap\tder$.
\end{theorem}

\paragraph{Relational semantics.}
Subject reduction (\refprop{quant-subject-reduction}) and expansion (\refprop{quant-subject-expansion}) imply that the set of typing judgements of a term is invariant by evaluation, and so they provide a denotational model of the split fireball calculus (\refcoro{invariance} below).

The definitions seen in \refsect{fireball} of the interpretation $\sem{\tm}{\vec{\var}} $ of a term with respect to a list $\vec{\var}$ of suitable variables for $\tm$  extends to the split fireball calculus by simply replacing terms with programs, with no surprises.

\newcounter{cor:invariance}
\addtocounter{cor:invariance}{\value{definition}}
\begin{corollary}[Invariance]
\label{coro:invariance}
  Let $\prog$ and $\progtwo$ be two programs and $\vec{\var} = (\var_1, \dots, \var_n)$ be a suitable list of variables for $\prog$ and $\progtwo$.
  If $\prog \tof \progtwo$ then $\sem{\prog}{\vec{\var}} = \sem{\progtwo}{\vec{\var}}$.
\end{corollary}

From correctness (\refthm{correctness}) and completeness (\refthm{completeness}) it follows that the relational semantics is adequate for the split fireball calculus $\sfirecalc$.

\newcounter{cor:adequacy}
\addtocounter{cor:adequacy}{\value{definition}}
\begin{corollary}[Adequacy]
\label{coro:adequacy}
  Let $\prog$ be a program and $\vec{\var} = (\var_1, \dots, \var_n)$ be a suitable list of variables for $\prog$.
  The following are equivalent:
  \begin{enumerate}
    \item \emph{Termination}: the evaluation of $\prog$ terminates;
    \item \emph{Typability}: there is a type derivation $\concl{\tder}{\typctx}{\prog}{\ptype}$ for some $\typctx$ and $\ptype$;
    \item \emph{Non-empty denotation}: $\sem{\prog}{\vec{\var}} \neq \emptyset$.
  \end{enumerate}
\end{corollary}

Careful about the third point: it requires the interpretation to be non-empty---a program typable with the empty multiset $\emptymset$ has a non-empty interpretation. Actually, a term is typable if and only if it is typable with $\emptymset$, as we show next.

\begin{remark}
By \refpropp{distinctive-fireball}{conservative} and \ref{prop:strong-bisimulation}, (weak) evaluations in Plotkin's original \cbv $\lambda$-calculus $\plotcalc$, in the fireball calculus $\firecalc$ and in its split variant  $\sfirecalc$ coincide on closed terms.
 So, \refcor{adequacy} says  that 
 relational semantics is adequate also for $\plotcalc$ 
 \emph{restricted to closed terms} (but adequacy for $\plotcalc$ fails on open terms, see \refrmk{adequate}).
\end{remark}

%% file: 08_-_Tight_Types_and_Exact_Bounds_New.tex
\section{Tight Type Derivations and Exact Bounds}
\label{sect:tight}
In this section we study a class of minimal type derivations, called \emph{tight}, providing exact bounds for evaluation lengths and sizes of normal forms.

\paragraph{Typing values and inert terms.} Values can always be typed with $\zero$ 
in an empty type context (\reflemmap{value-typing}{judg}), by means of an axiom for variables or of a $\l$-rule with zero premises for abstractions. 
We are going to show that inert terms can also always be typed with $\zero$. 
There are differences, however. 
First, the type context in general is not empty. 
Second, the  derivations typing with $\zero$ have a more complex structure, having sub-derivations for inert terms whose right-hand type might not be $\zero$. 
It is then necessary, for inert terms, to consider a more general class of type derivations, that, as a special case, include derivations typing with $\zero$.

%
First of all, we define two class of types: 
\begin{align*}
 & \textsc{\small Inert linear types} & \intype & \grameq \Pair{\zero}{\iptypetwo} \\[-.3\baselineskip]
  &\textsc{\small Inert multi types} & \iptype,\iptypetwo & \grameq \mset{\intype_1, \dots, \intype_n} \quad \text{ (with } n \in  \nat\text{).}
\end{align*}
 
A type context $\typctx$ is \emph{inert} if it assigns only inert multi types to variables.

In particular, the empty multi type $\zero$ is inert (take $n = 0$), and hence the empty type context is inert.
Note that  inert multi types and inert multi contexts are closed under summation $\uplus$.

We also introduce two notions of type derivations, \emph{inert} and \emph{tight}. The tight ones are those we are actually interested in, but, as explained, for inert terms we need to consider a more general class of type derivations, the inert ones.
Formally, given
  an expression $\expr$, a type derivation $\concl{\tder}{\typctx}{\expr}{\ptype}$ is
  \begin{itemize}
    \item \emph{inert} if $\typctx$ is a inert type context and $\ptype$ is a inert multi type;
    \item \emph{tight} if $\tder$ is inert and $\ptype = \zero$;
    \item \emph{\nonempty} (resp.~\emph{empty}) if $\typctx$ is a non-empty (resp.~\empt) type context.
  \end{itemize}

Note that tightness and inertness of type derivations depend only on the judgement in their conclusions.
The general property is that inert terms admit a inert type derivation \emph{for every inert multi type $\iptype$}.

\newcounter{l:type-characterization-inert}
\addtocounter{l:type-characterization-inert}{\value{lemma}}
\begin{lemma}[Inert typing of inert terms]
\label{l:type-characterization-inert}
\NoteProof{lappendix:type-characterization-inert}
  Let $\itm$ be a inert term. 
  For any inert multi type $\iptype$ there exists a \nonempty inert type derivation $\concl{\tder}{\typctx}{\itm}{\iptype}$.  
\end{lemma}

\reflemma{type-characterization-inert} holds with respect to \emph{all} inert multi types, in particular $\zero$, so inert terms can be always typed with a \nonempty \emph{tight} derivation.
Since values can be always typed with an \empt tight derivation (\reflemmap{value-typing}{judg}), we can conclude:

\begin{corollary}[Fireballs are tightly typable]
\label{coro:tight-characterisation-nfs}
	For any fireball $\fire$ there exists a tight type derivation $\concl{\tder}{\typctx}{\fire}{\zero}$.
	Moreover, if $\fire$ is a inert term then $\tder$ is \nonempty, otherwise $\fire$ is a value and $\tder$ is \empt.
\end{corollary}

By harmony (\refprop{harmony-split}), 
it follows that any normal expression is tightly typable (\refprop{tight-characterization-nfs} below). 
\emph{Terminology}: a \emph{coerced value} is a program of the form $(\val, \emptyenv)$.

\newcounter{prop:tight-characterization-nfs}
\addtocounter{prop:tight-characterization-nfs}{\value{proposition}}
\begin{proposition}[Normal expressions are tightly typable]
	\label{prop:tight-characterization-nfs}
\NoteProof{propappendix:tight-characterization-nfs}
	Let $\expr$ be a normal expression. 
	Then there exists a tight derivation $\concl{\tder}{\typctx}{\expr}{\zero}$. 
	Moreover, 
	$\expr$ is a value or a coerced value if and only if $\tder$ is \empt.
\end{proposition}

\paragraph{Tight derivations and exact bounds.} 
The next step is to show that tight derivations are minimal and provide exact bounds. 
Again, we have to detour through inert derivations for inert terms. 
And we need a further property of inert terms: if the type context is inert then the right-hand type is also inert.

\newcounter{l:inert-ctx-imply-inert-type}
\addtocounter{l:inert-ctx-imply-inert-type}{\value{lemma}}
\begin{lemma}[Inert spreading on inert terms]
\label{l:inert-ctx-imply-inert-type}
\NoteProof{lappendix:inert-ctx-imply-inert-type}
	Let $\concl{\tder}{\typctx}{\itm}{\ptype}$ be a type de\-rivation for a inert term $\itm$. If $\typctx$ is a inert type context then $\ptype$ 
 	and $\tder$ are inert.
\end{lemma}

Next, we prove that inert derivations provide exact bounds for inert terms.

\newcounter{l:size-normal-inert}
\addtocounter{l:size-normal-inert}{\value{lemma}}
\begin{lemma}[Inert derivations are minimal and provide the exact size of inert terms]
\label{l:size-normal-inert} 
\NoteProof{lappendix:size-normal-inert}
Let $\concl{\tder}{\typctx}{\itm}{\iptype}$ be a inert type derivation for a inert term $\itm$. 
Then $\sizeap\itm = \sizeap\tder$ and $\size{\tder}$ is minimal among the type derivations of $\itm$.
\end{lemma}

We can now extend the characterisation of sizes to all normal expressions, via tight derivations, refining \refprop{size-normal}.

\newcounter{l:size-normal-tight}
\addtocounter{l:size-normal-tight}{\value{lemma}}
\begin{lemma}[Tight derivations are minimal and provide the exact size of normal forms]
	\label{l:size-normal-tight} 
\NoteProof{lappendix:size-normal-tight}
	Let $\concl{\tder}{\typctx}{\expr}{\emptymset}$ be a tight derivation and $\expr$ be a normal 
	expression. 
	Then $\sizeap\expr = \sizeap\tder$ and $\size{\tder}$ is minimal among the type derivations of $\expr$.
\end{lemma}

The bound on the size of normal forms using types rather than type derivations
(\refprop{types-bound-nfs}) can also be refined: \emph{tight} derivations end with judgements whose (inert) \emph{type contexts} provide the  \emph{exact} size of normal forms.

\newcounter{prop:tight-types-bound-nfs}
\addtocounter{prop:tight-types-bound-nfs}{\value{proposition}}
\begin{proposition}[
	Inert types and the exact size of normal forms]
\label{prop:tight-types-bound-nfs}
\NoteProof{propappendix:tight-types-bound-nfs}
  Let $\expr$ be a normal expression and $\concl{\tder}{\typctx}{\expr}{\zero}$ be a tight derivation. 
  Then $\sizeap{\expr} = \size{\typelist\typctx}$.
\end{proposition}

\paragraph{Tightness and general programs.} Via subject reduction and expansion, exact bounds can be extended to all normalisable programs. Tight derivations indeed induce refined correctness and completeness theorems replacing inequalities with equalities (see \refthms{tight-correctness}{tight-completeness} below and compare them with \refthms{correctness}{completeness} above, respectively): an \emph{exact} quantitative information relates the length $\size{\deriv}$ of evaluations, the size of normal forms and the size of \emph{tight} type derivations.

\newcounter{thm:tight-correctness}
\addtocounter{thm:tight-correctness}{\value{theorem}}
\begin{theorem}[Tight correctness]
\label{thm:tight-correctness}
\NoteProof{thmappendix:tight-correctness}
Let $\concl{\tder}{\typctx}{\prog}{\zero}$ be a tight type derivation. Then there 
is a normalising evaluation $\deriv \colon \prog \tof^* \progtwo$ with 
$\sizeap\tder = \size\deriv + \sizeap\progtwo = \size\deriv + \sizeap{\typelist\typctx}$.
    In particular, if $\domain{\typctx} = \emptyset$, then $\size{\tder} = \size{\deriv}$ and $\progtwo$ is a coerced value.
\end{theorem}

\newcounter{thm:tight-completeness}
\addtocounter{thm:tight-completeness}{\value{theorem}}
\begin{theorem}[Tight completeness]
\label{thm:tight-completeness}
\NoteProof{thmappendix:tight-completeness}
Let $\deriv \colon \prog \tof^* \progtwo$ be a normalising evaluation. 
Then there is a tight type derivation $\concl{\tder}{\typctx}{\prog}{\zero}$ with $\sizeap\tder = \size\deriv + \sizeap{\progtwo} = \size{\deriv} + \size{\typelist\typctx}$.
In particular, if $\progtwo$ is a coerced value, then $\size{\tder} = \size{\deriv}$ and $\domain{\typctx} = \emptyset$. 
\end{theorem}

\noindent Both theorems are proved analogously to their corresponding non-tight version (\refthms{correctness}{completeness}), the only difference is in the base case: here 
\reflemma{size-normal-tight} provides an equality on sizes for normal forms, instead of the inequality given by \refprop{size-normal} and used in the non-tight versions. 
The proof of tight completeness (\refthm{tight-completeness}) uses also 
that normal programs are \emph{tightly} typable (\refprop{tight-characterization-nfs}).

%% file: 09_-_Conclusions.tex
\section{Conclusions}
This paper studies multi types 
for \cbv\ weak evaluation. 
It recasts in \cbv de Carvalho's work for \cbn~\cite{Carvalho07,deCarvalho18}, building on a type system introduced by Ehrhard \cite{DBLP:conf/csl/Ehrhard12} for Plotkin's original \cbv $\lambda$-calculus $\plotcalc$ \cite{DBLP:journals/tcs/Plotkin75}. 
Multi types provide a denotational model that we show to be adequate for $\plotcalc$, but only when evaluating \emph{closed} terms; and for \ocbv \cite{DBLP:conf/aplas/AccattoliG16}, an extension of $\plotcalc$ where weak evaluation is on possibly \emph{open} terms.
More precisely, our main contributions are:

\begin{enumerate}
	\item 
	The formalism itself: we point out the issues with respect to subject reduction and expansion of the simplest presentation of \ocbv, the fireball calculus $\firecalc$, and introduce a refined calculus (isomorphic to $\firecalc$) that satisfies them.

	\item 
	The characterisation of termination both in a \emph{qualitative} and \emph{quantitative} way. 
	Qualitatively, typable terms and normalisable terms coincide. Quantitatively, types provide bounds on the size of normal forms, and type derivations bound the number of evaluation steps to normal form.
	
	\item 
	The identification of a class of type derivations that provide \emph{exact} bounds on evaluation lengths.
\end{enumerate}

%% file: proofs.tex
\section{Technical appendix: omitted proofs and lemmas}
\label{sect:proofs}

The enumeration of propositions, theorems, lemmas already stated in the body of the article is unchanged.

\input{\proofspath/proofs-fireball}

\input{\proofspath/proofs-ES_calculus}

\input{\proofspath/proofs-correctness}

\input{\proofspath/proofs-completeness}

\input{\proofspath/proofs-normal-forms-tightNew}

%% file: proofs-fireball.tex
\subsection{Omitted proofs and lemmas of \refsect{fireball}}

Note that any inert term $\var\fire_1\dots\fire_n$ (with $n > 0$) can be seen as term of the form $g\fire_n$ where $g$ is either a variable (if $n = 1$) or an inert term (if $n > 1$).

\setcounter{propositionAppendix}{\value{prop:distinctive-fireball}}
\begin{propositionAppendix}[Distinctive properties of $\firecalc$]
  \label{propappendix:distinctive-fireball}
\NoteState{prop:distinctive-fireball}
  Let $\tm$ be a term.
  \begin{enumerate}
    \item\label{pappendix:distinctive-fireball-open-harmony} \emph{Open harmony}:  $\tm$ is $\betaf$-normal if and only if $\tm$ is a fireball.
    \item\label{pappendix:distinctive-fireball-conservative} \emph{Conservative open extension}: $\tm \tof \tmtwo$ if and only if $\tm \tobv \tmtwo$, when $\tm$ is closed.
  \end{enumerate}
\end{propositionAppendix}

\begin{proof}\hfill
  \begin{enumerate}
    \item   
    \begin{description}
      \item [$\Rightarrow$:] Proof by induction on the term $\tm$.
      If $\tm$ is a value then $\tm$ is a fireball.
      
      Otherwise $\tm = \tmtwo\tmthree$ for some terms $\tmtwo$ and $\tmthree$.
      Since $\tm$ is $\betaf$-normal, then $\tmtwo$ and $ \tmthree$ are $\betaf$-normal, and either $\tmtwo$ is not an abstraction or $\tmthree$ is not a fireball.
      By induction hypothesis, $\tmtwo$ and $\tmthree$ are fireballs. 
      Summing up, $\tmtwo$ is either a variable or an inert term, and $\tmthree$ is a fireball, therefore $\tm = \tmtwo\tmthree$ is an inert term and hence a fireball.
      
      \item [$\Leftarrow$:] By hypothesis, $\tm$ is either a value or an inert term. 
      If $\tm$ is a value, then it is $\betaf$-normal since $\tof$ does not reduce under $\l$'s. 
      Otherwise $\tm$ is an inert term and we prove by induction on the definition of inert term that $\tm$ is $\betaf$-normal.  
      \begin{itemize}
	\item If $\tm = g(\la\var\tmtwo)$ where $g$ is a variable or an inert term then $g$ $\betaf$-normal (by \ih if $g$ is inert), and $\la\var\tmtwo$ is $\betaf$-normal as we have just shown; also, $g$ is not an abstraction, so $\tm$ is $\betaf$-normal.

	\item Finally, if $\tm = g\gconst$ where $g$ is a variable of an inert term, then $g$ is $\betaf$-normal (by \ih if $g$ is inert) and $\gconst$ is $\betaf$-normal by \ih, moreover $g$ is not an abstraction, hence $\tm$ is $\betaf$-normal.
      \end{itemize}

    \end{description}

    \item 
    \begin{description}
      \item [$\Rightarrow$:] By induction on the definition of $\tm \tof \tmtwo$. Cases:
      \begin{itemize}
        \item \emph{Step at the root}, \ie $\tm = (\la\var\tmthree)\fire \rtof \tmthree\isub\var{\fire} = \tmtwo$. Since $\tm$ is closed, then $\fire$ is closed and hence cannot be an inert term or a variable, so $\fire$ is a (closed) abstraction and thus $\tm = (\la\var\tmthree)\fire \rtobv \tmthree\isub\var{\fire} = \tmtwo$.
        
        \item \emph{Application left}, \ie $\tm = \tmthree\tmfour \tof \tmthreep\tmfour = \tmtwo$ with $\tmthree \tof \tmthreep$. Since $\tm$ is closed, $\tmthree$ is so and hence $\tmthree \tobv \tmthreep$ by \ih, thus $\tm = \tmthree\tmfour \tobv \tmthreep\tmfour = \tmtwo$.
        
        \item \emph{Application right}, \ie $\tm = \tmfour\tmthree \tof \tmfour\tmthreep = \tmtwo$ with $\tmthree \tof \tmthreep$. Since $\tm$ is closed, $\tmthree$ is so and hence $\tmthree \tobv \tmthreep$ by \ih, so $\tm = \tmfour\tmthree \tobv \tmfour\tmthreep = \tmtwo$.
      \end{itemize}

      \item [$\Leftarrow$:] We have $\tobv \, \subseteq \, \tof$ since every value is a fireball.
      \qedhere
    \end{description}

  \end{enumerate}
\end{proof}

\begin{lemma}[Fireballs are closed under substitution and anti-substitution of \quiet terms]
\label{l:inert-anti-sub-bis} 
  Let 
  $\tm$ be a term and $\itm$ be an inert term.
  \begin{enumerate}
    \item \label{p:inert-anti-sub-bis-abs} $\tm\isub\var\itm$ is an abstraction if and only if $\tm$ is an abstraction. 
    \item \label{p:inert-anti-sub-bis-inert} $\tm\isub\var\itm$ is an inert term or a variable if and only if $\tm$ is an inert term or a variable; in particular, if $\tm\isub\var\itm$ is a variable then $\tm$ is a variable; if $\tm$ is an inert term then $\tm\isub{\var}{\itm}$ is an inert term. 
    \item\label{p:inert-anti-sub-bis-fire} $\tm\isub\var\itm$ is a fireball if and only if $\tm$ is a fireball.
  \end{enumerate}
\end{lemma}

%
%

\begin{proof}\hfill
  \begin{enumerate}
    \item If $\tm\isub\var\gconst = \la\vartwo\tmthree$ then we can suppose without loss of generality that $\vartwo \notin \fv{\gconst} \cup \{\var\}$ and thus there is $\tmfour$ such that $\tmthree = \tmfour\isub\var\gconst$, hence  $\tm\isub\var\gconst = \la\vartwo(\tmfour\isub\var\gconst) = (\la\vartwo\tmfour)\isub\var\gconst$, therefore $\tm = \la\vartwo\tmfour$ is an abstraction.   

    Conversely, if $\tm = \la\vartwo\tmthree$ then we can suppose without loss of generality that $\vartwo \notin \fv{\gconst} \cup \{\var\}$ and thus $\tm\isub\var\gconst = \la\vartwo{(\tmthree\isub\var\gconst)}$ which is an abstraction.    
    
    \item ($\Rightarrow$): By induction on the 
    structure of $\tm\isub\var\gconst$. Cases:
    \begin{itemize}
      \item \emph{Variable}, \ie~$\tm\isub\var\gconst = \vartwo$, possibly with $\var = \vartwo$. Then $\tm = \var$ or $\tm = \vartwo$, and in both cases $\tm$ is a variable.
      \item \emph{Inert term}, \ie~$\tm\isub\var\gconst = g\fire$ where $g$ is a variable or an inert term. If $\tm$ is a variable, we are done. Otherwise it is an application $\tm = \tmtwo \tmthree$, and so $\tmtwo\isub\var\gconst = g$ and $\tmthree\isub\var\gconst = \fire$. By \ih, $\tmtwo$ is an inert term or a variable. Consider $\fire$. Two cases:
      
      \begin{enumerate}
	\item $\fire$ is an abstraction. Then by \refpoint{inert-anti-sub-bis-abs} $\tmthree$ is an abstraction.
	\item $\fire$ is an inert term or a variable. Then by \ih $\tmthree$ is an inert term or a variable.
      \end{enumerate}
      
      In both cases $\tmthree$ is a fireball, and so $\tm = \tmtwo \tmthree$ is an inert term.      
      \end{itemize}

      ($\Leftarrow$): By induction on the 
      structure of $\tm$. Cases:
    \begin{itemize}
      \item \emph{Variable}, \ie either $\tm = \var$ or $\tm = \vartwo$: in the first case  $\tm\isub\var\gconst = \gconst$, in the second case  $\tm\isub\var\gconst = \vartwo$; in both cases $\tm\isub\var\gconst$ is  an inert term or a variable.
      \item \emph{Inert term}, \ie $\tm = g \fire$ where $g$ is a variable or an inert term. Then $\tm\isub\var\gconst = g\isub\var\gconst \fire\isub\var\gconst$.
      By \ih, $g\isub\var\gconst$ is an inert term or a variable. Concerning $\fire$, there are two cases:
      \begin{enumerate}
	\item $\fire$ is an abstraction. Then by \refpoint{inert-anti-sub-bis-abs} $\fire\isub\var\gconst$ is an abstraction.
	\item $\fire$ is an inert term or a variable. Then by \ih $\fire\isub\var\gconst$ is an inert term or a variable.
      \end{enumerate}
      In both cases $\fire\isub\var\gconst$ is a fireball, and hence $\tm\isub\var\gconst = g\isub\var\gconst\fire\isub\var\gconst$ is an inert term.      
      \end{itemize}
    
    \item Immediate consequence of \reflemmasps{inert-anti-sub-bis}{abs}{inert}, since every fireball is either an abstraction or an inert term.
    \qedhere
  \end{enumerate}
\end{proof}

\begin{lemma}[Substitution of \quiet terms does not create $\betaf$-redexes]
\label{l:inert-anti-red-bis} 
  Let 
  $\tm, \tmtwo$~be terms and $\itm$ be an inert term.
  If $\tm\isub\var\itm \tof \tmtwo$ then $\tm \tof \tmthree$ for some term $\tmthree$ such that $\tmthree\isub\var\itm = \tmtwo$.
%
\end{lemma}

\begin{proof}
  By induction on the evaluation context closing the root redex. Cases:  
    \begin{itemize}
      \item \emph{Step at the root}:
      \begin{enumerate}
      \item \emph{Abstraction step}, \ie $\tm\isub\var\gconst \defeq (\la\vartwo\tmfour\isub\var\gconst) \tmfive\isub\var\gconst \allowbreak\rtof\allowbreak \tmfour\isub\var\gconst\isub\vartwo{\tmfive\isub\var\gconst} \allowbreak\eqdef \tmtwo$ where $\tmfive\isub\var\gconst$ is an abstraction.
      By \reflemmap{inert-anti-sub-bis}{abs}, $\tmfive$ is an abstraction, since $\tmfive\isub\var\gconst$ is an abstraction by hypothesis. Then $\tm = (\la\vartwo\tmfour) \tmfive \rtof \tmfour\isub\vartwo{\tmfive}$. So $\tmthree \defeq \tmfour\isub\var{\tmfive}$ verifies the statement, as $\tmthree\isub\var\gconst = (\tmfour\isub\vartwo{\tmfive})\isub\var\gconst = \tmfour\isub\var\gconst\isub\vartwo{\tmfive\isub\var\gconst} = \tmtwo$.     
      
      \item \emph{Inert or variable step}, identical to the abstraction subcase, just replace \emph{abstraction} with \emph{inert term} or \emph{variable}, and the use of \reflemmap{inert-anti-sub-bis}{abs} with the use of \reflemmap{inert-anti-sub-bis}{inert}.
      \end{enumerate}      
      
      \item \emph{Application left}, \ie $\tm = \tmfour\tmfive$ and reduction takes place in $\tmfour$:
      \begin{enumerate}
      \item \emph{Abstraction step}, \ie $\tm\isub\var\gconst \defeq \tmfour\isub\var\gconst \tmfive\isub\var\gconst \tof \tmsix \tmfive\isub\var\gconst \eqdef \tmtwo$. 
      By \ih there is a term $\tmthreep$ such that $\tmsix = \tmthreep\isub\var\gconst$ and $\tmfour \tof \tmthreep$. Then $\tmthree \defeq \tmthreep \tmfive$ satisfies the statement, as $\tmthree\isub\var\gconst = (\tmthreep \tmfive)\isub\var\gconst = \tmthreep\isub\var\gconst \tmfive\isub\var\gconst = \tmtwo$.
      
      \item \emph{Inert step}, identical to the abstraction subcase.
      \end{enumerate}

      \item \emph{Application Right}, \ie $\tm = \tmfour\tmfive$ and reduction takes place in $\tmfive$. Identical to the \emph{application left} case, just switch left and right.
      \qedhere
    \end{itemize}
      
\end{proof}

We can now prove that \emph{evaluation} and \emph{substitution of inert terms} commute.
Said differently, substitution of inert terms can always be avoided.

\setcounter{propositionAppendix}{\value{prop:inerts-and-creations}}
\begin{propositionAppendix}[Inert substitutions and evaluation commute]
\label{propappendix:inerts-and-creations}
\NoteState{prop:inerts-and-creations}
  Let $\tm, \tmtwo$ be terms and $\itm$ be an inert term. Then, $\tm \tof \tmtwo$ if and only if $\tm \isub\var\itm \tof \tmtwo \isub\var\itm$.
\end{propositionAppendix}

\begin{proof}
  The right-to-left direction is \reflemma{inert-anti-red-bis}.
  The left-to-right direction is proved by induction on the definition of $\tm \tof \tmtwo$. Cases:
  \begin{itemize}
    \item \emph{Step at the root}:
    \begin{enumerate}
      \item \emph{Abstraction step}, \ie~$\tm = (\la\vartwo\tmthree)\tmfour \rtobv \tmthree\isub\vartwo\tmfour = \tmtwo$ where $\tmfour$ is an abstraction.
      We can suppose without loss of generality that $\vartwo \notin \fv{\gconst} \cup \{\var\}$.
      By \reflemmap{inert-anti-sub-bis}{abs}, $\tmfour\isub\var\gconst$ is an abstraction.
      Therefore, $\tm\isub\var\gconst = (\la\vartwo\tmthree\isub\var\gconst)\tmfour\isub\var\gconst \rtobv\! \tmthree\isub\var\gconst\isub\vartwo{\tmfour\isub\var\gconst} \allowbreak= \tmthree\isub\vartwo\tmfour\isub\var\gconst \allowbreak=\allowbreak \tmtwo\isub\var\gconst$.
      \item \emph{Inert or variable step}, \ie~$\tm = (\la\vartwo\tmthree)g \rtof \tmthree\isub\vartwo{g} = \tmtwo$ where $g$ is an 
      inert term or a variable.
      We can suppose without loss of generality that $\vartwo \notin \fv{\gconst} \cup \{\var\}$.
      According to \reflemmap{inert-anti-sub-bis}{inert}, $\tmfour\isub\var\gconst$ is an inert term or a variable.
      So, $\tm\isub\var\gconst = (\la\vartwo\tmthree\isub\var\gconst)g\isub\var\gconst \rtof \tmthree\isub\var\gconst\isub\vartwo{g\isub\var\gconst} = \tmthree\isub\vartwo{g}\isub\var\gconst = \tmtwo\isub\var\gconst$.
    \end{enumerate}
    \item \emph{Application left}, \ie~$\tm = \tmthree\tmfour \tof \tmthreep\tmfour = \tmtwo$ with $\tmthree \tof \tmthreep$.
    By \ih, $\tmthree\isub\var\gconst \allowbreak\tof\allowbreak \tmthreep\isub\var\gconst$, so $\tm\isub\var\gconst = \tmthree\isub\var\gconst\tmfour\isub\var\gconst \tof \tmthreep\isub\var\gconst\tmfour\isub\var\gconst = \tmtwo\isub\var\gconst$.
    \item \emph{Application right}, \ie~$\tm = \tmfour\tmthree \tof \tmfour\tmthreep = \tmtwo$ with $\tmthree \tof \tmthreep$.
    Analogous to the previous case.
    \qedhere
  \end{itemize}
\end{proof}

%% file: proofs-ES_calculus.tex
\subsection{Omitted proofs and lemmas of \refsect{split-calculus}}

\setcounter{propositionAppendix}{\value{prop:harmony-split}}
\begin{propositionAppendix}[Harmony]
\label{propappendix:harmony-split}
\NoteState{prop:harmony-split}
 A program $\prog$ is normal if and only if $\prog = (\fire, \genv)$.
\end{propositionAppendix}

\begin{proof}
    \begin{description}
      \item [$\Rightarrow$:] 
      If $\prog = (\tm, \genv)$ is $\betaf$-normal in $\sfirecalc$ then $\tm$ cannot be of the form $\evctxp{(\la{\var}{\tm})\val}$ or $\evctxp{(\la{\var}{\tm})\itm}$, which means that $\tm$  is $\betaf$-normal in the fireball calculus $\firecalc$.
      By open harmony in $\firecalc$ (\refpropp{distinctive-fireball}{open-harmony}), $\tm$ is a fireball.
%
      
      \item [$\Leftarrow$:] 
      If $\prog = (\fire, \genv)$ then, by open harmony in the fireball calculus $\firecalc$ (\refpropp{distinctive-fireball}{open-harmony}), $\fire$ is $\betaf$-normal in $\firecalc$, which means that $\fire$ cannot be of the form $\evctxp{(\la{\var}{\tm})\val}$ or $\evctxp{(\la{\var}{\tm})\itm}$;
      therefore, $\prog$ is $\betaf$-normal in $\sfirecalc$.
%
        \qedhere
    \end{description}

\end{proof}

\begin{lemma}[Compositionality of evaluation]
\label{l:compositionality}
  Let $\tm$, $\tmp$ be terms, $\genv, \genvtwo$ be environments, and $\itm$ be an inert term. 
  If $(\tm,\genv) \tof (\tmp,\genvtwo)$, then $(\tm,\genv \appendOp\esub{\var}{\itm}) \tof (\tmp,\genvtwo \appendOp\esub{\var}{\itm})$. 
\end{lemma}

\begin{proof}
  By straightforward induction on the length of the environment $\genv$.
\end{proof}

\setcounter{propositionAppendix}{\value{prop:strong-bisimulation}}
\begin{propositionAppendix}[Strong bisimulation]
\label{propappendix:strong-bisimulation}
\NoteState{prop:strong-bisimulation}
    Let $\prog$ be a program (in $\sfirecalc$).
    \begin{enumerate}
    \item \emph{Split to plain}: if $\prog \tof \progtwo$ then  $\unf\prog \tof \unf\progtwo$.
    \item \emph{Plain to split}: if $\unf\prog \tof \tmtwo$ then there exists $\progtwo$ such that $\prog \tof \progtwo$ and $\unf\progtwo = \tmtwo$.
    \end{enumerate}
\end{propositionAppendix}

\begin{proof} Let $\prog \defeq (\tm, \genv)$, with $\genv \defeq \esub{\var_1}{\itm_1}\dots\esub{\var_n}{\itm_n}$ ($n \geq 0$).
  \begin{enumerate}
    \item Since $\prog \tof \progtwo$, by necessity $\tm = \evctxp{(\la{\vartwo}{\tmtwo})\fire}$.
    The proof is by induction on the definition of $\evctx$.
    We prove explicitly only the case where $\evctx = \ctxhole$, the other cases follow just by applying the \ih, since $\sfirecalc$ and $\firecalc$ have the same right evaluation contexts.
    We can suppose without loss of generality that $\vartwo \notin \bigcup_{i=1}^n (\fv{\fire_i} \cup \{\var_i\})$, so $\unf{\prog} = (\la{\vartwo}{\tmtwo\isub{\var_1}{\itm_1}\dots\isub{\var_n}{\itm_n}})\fire\isub{\var_1}{\itm_1}\dots\isub{\var_n}{\itm_n}$.
    For $\evctx = \ctxhole$ there are three cases:
    \begin{enumerate}
      \item $\fire = \val$ is an \emph{abstraction}:
      then, $\prog \tobv (\tmtwo\isub{\vartwo}{\val}, \genv) \eqdef \progtwo$.
      By \reflemmap{inert-anti-sub-bis}{abs}, $\val\isub{\var_1}{\itm_1}\dots\isub{\var_n}{\itm_n}$ is an abstraction, hence 
      \begin{equation*}
      \begin{split}
        \unf{\prog} \tobv \tmtwo\isub{\var_1}{\itm_1}\dots\isub{\var_n}{\itm_n}\isub{\vartwo}{\val\isub{\var_1}{\itm_1}\dots\isub{\var_n}{\itm_n}} \qquad\\
        = \tmtwo\isub{\vartwo}{\val}\isub{\var_1}{\itm_1}\dots\isub{\var_n}{\itm_n} = \unf{\progtwo}
      \end{split}
      \end{equation*}
      
      \item $\fire = \itm$ is an \emph{inert term}:
            then, $\prog \toin (\tmtwo, \esub{\vartwo}{\itm} \cons \genv) \eqdef \progtwo$.
      By \reflemmap{inert-anti-sub-bis}{inert}, $\itm\isub{\var_1}{\itm_1}\dots\isub{\var_n}{\itm_n}$ is a inert term, hence 
      \begin{equation*}
      \begin{split}
        \unf{\prog} \toin \tmtwo\isub{\var_1}{\itm_1}\dots\isub{\var_n}{\itm_n}\isub{\vartwo}{\itm\isub{\var_1}{\itm_1}\dots\isub{\var_n}{\itm_n}} \qquad\\
        = \tmtwo\isub{\vartwo}{\itm}\isub{\var_1}{\itm_1}\dots\isub{\var_n}{\itm_n} = \unf{\progtwo}
      \end{split}
      \end{equation*}

      \item $\fire = \varthree$ is a \emph{variable}:
      then, $\prog \tobv (\tmtwo\isub{\vartwo}{\varthree}, \genv) \eqdef \progtwo$.
      There are two subcases:
%
      \begin{itemize}
        \item either $\varthree \neq \var_j$ for all $1 \leq j \leq n$; thus, $\varthree\isub{\var_1}{\itm_1}\dots\isub{\var_n}{\itm_n}$ is a variable and hence  
      \begin{align*}
        \unf{\prog} &= (\la{\vartwo}{\tmtwo\isub{\var_1}{\itm_1}\dots\isub{\var_n}{\itm_n}})\varthree \tobv \tmtwo\isub{\var_1}{\itm_1}\dots\isub{\var_n}{\itm_n}\isub{\vartwo}{\varthree} \qquad\\
        &= \tmtwo\isub{\vartwo}{\varthree}\isub{\var_1}{\itm_1}\dots\isub{\var_n}{\itm_n} = \unf{\progtwo}
      \end{align*}
        \item or $\varthree = \var_j$ for some $1 \leq j \leq n$; thus, $\varthree\isub{\var_1}{\itm_1}\dots\isub{\var_n}{\itm_n}$ is a inert term by \reflemmap{inert-anti-sub-bis}{inert} and hence  
      \begin{equation*}
      \begin{split}
        \unf{\prog} \toin \tmtwo\isub{\var_1}{\itm_1}\dots\isub{\var_n}{\itm_n}\isub{\vartwo}{\varthree\isub{\var_1}{\itm_1}\dots\isub{\var_n}{\itm_n}} \qquad\\
        = \tmtwo\isub{\vartwo}{\varthree}\isub{\var_1}{\itm_1}\dots\isub{\var_n}{\itm_n} = \unf{\progtwo}\,.
      \end{split}
      \end{equation*}

      \end{itemize}
    \end{enumerate}

    \item Proof by induction on the length of the environment $\genv$.
    Cases:
    \begin{enumerate}
      \item \emph{Empty environment}, \ie $\prog = (\tm, \emptyenv)$.
      Then, $\unf{\prog} = \tm$. There are two subcases:
      \begin{itemize}
        \item \emph{$\betav$-step:} either $\unf{\prog} \tobv \tmtwo$, then $\tm \defeq \evctxp{(\la{\var}{\tmthree})\val} \tobv \evctxp{\tmthree \isub{\var}{\val}} \eqdef \tmtwo$ and so $\prog  \tobv (\tmtwo, \emptyenv) \eqdef \progtwo$, with $\unf{\progtwo} = \tmtwo$;
        \item \emph{$\betain$-step:} or $\unf{\prog} \toin \tmtwo$, then $\tm \defeq \evctxp{(\la{\var}{\tmthree})\itm} \toin \evctxp{\tmthree \isub{\var}{\itm}} \eqdef \tmtwo$ and we can suppose without loss of generality that $\var \notin \fv{\evctx}$;
        thus, $\prog \toin (\evctxp{\tmthree}, \esub{\var}{\itm}) \eqdef \progtwo$, with $\unf{\progtwo} = \evctxp{\tmthree }\isub{\var}{\itm} = \evctxp{\tmthree \isub{\var}{\itm}}  = \tmtwo$.
      \end{itemize}
      \item \emph{Non-empty environment}, \ie $\prog = (\tm, \genv \appendOp \esub{\var}{\itm})$.
      Then, $\unf{\prog} = \unf{(\tm, \genv)}\isub{\var}{\itm} \allowbreak\tof \tmtwo$. 
      By \reflemma{inert-anti-red-bis}, $\unf{(\tm,\genv)} \tof \tmthree$ for some term $\tmthree$ such that $\tmthree\isub{\var}{\itm} = \tmtwo$.
      By \ih, $(\tm,\genv) \tof (\tmthreep,\genvtwo)$ with $\unf{(\tmthreep,\genvtwo)} = \tmthree$.
      Let $\progtwo \defeq (\tmthreep, \genvtwo \appendOp \esub{\var}{\itm})$; 
      therefore, $\unf{\progtwo} = \unf{(\tmthreep,\genvtwo)}\isub{\var}{\itm} \tmthree\isub{\var}{\itm} = \tmtwo$ and $\prog \tof \progtwo$ by \reflemma{compositionality}.
    \qedhere
   \end{enumerate}

  \end{enumerate}
\end{proof}

%% file: proofs-correctness.tex
\subsection{Omitted proofs and lemmas of \refsect{correctness}}

\setcounter{propositionAppendix}{\value{prop:size-normal}}
\begin{propositionAppendix}[Type derivations bound the size of normal forms]
\label{propappendix:size-normal} 
\NoteState{prop:size-normal}
Let $\concl{\tder}{\typctx}{\expr}{\ptype}$ be a type derivation and $\expr$ be a normal 
expression. Then 
$\sizeap\expr \leq \sizeap\tder$.
\end{propositionAppendix}

\begin{proof}
By induction on $\expr$. 
\begin{itemize}
\item \emph{$\expr$ is a value}: $\sizeap\expr = 0 \leq \sizeap\tder$.

\item \emph{$\expr$ is a inert term}. Then $\expr = \var \fire_1 \ldots \fire_n$ and $\tder$ has the form:

{\small
  \begin{equation*}
    \AxiomC{}
    \RightLabel{\footnotesize$\ruleAx$}
    \UnaryInfC{$\var \hastype \mset{\Pair{\ptypetwo_1}{\mset{\Pair{\ldots}{\mset{\Pair{\ptypetwo_n}{\ptype}}}}}}  \vdash \var \hastype \mset{\Pair{\ptypetwo_1}{\mset{\Pair{\ldots}{\mset{\Pair{\ptypetwo_n}{\ptype}}}}}}$}
    \AxiomC{$\ \vdots\,\tdertwo_1$}
    \noLine
    \UnaryInfC{$\typctxtwo_1 \vdash \fire_1 \hastype \ptypetwo_1$}
    \RightLabel{\footnotesize$\ruleAp$}
    \BinaryInfC{$\var \hastype  \mset{\Pair{\ptypetwo_1}{\mset{\Pair{\ldots}{\mset{\Pair{\ptypetwo_n}{\ptype}}}}}}  \uplus \typctxtwo_1  \vdash \var \fire_1 \hastype  \mset{\Pair{\ptypetwo_2}{\mset{\Pair{\ldots}{\mset{\Pair{\ptypetwo_n}{\ptype}}}}}}$}
    \UnaryInfC{$\vdots$}
        \AxiomC{$\ \vdots\,\tdertwo_n$}
    \noLine
    \UnaryInfC{$\typctxtwo_n \vdash \fire_n \hastype \ptypetwo_n$}
    \RightLabel{\footnotesize$@$}
    \BinaryInfC{$\var \hastype \mset{\Pair{\ptypetwo_1}{\mset{\Pair{\ldots}{\mset{\Pair{\ptypetwo_n}{\ptype}}}}}}  \uplus (\biguplus_{i=1}^n\typctxtwo_i ) \vdash \var \fire_1 \ldots \fire_n \hastype \ptype$}
    \DisplayProof
  \end{equation*}
  }
  
  By \ih, $\sizeap{\fire_i} \leq \sizeap{\tdertwo_i}$, and so $\sizeap{\expr} = \sizeap{\var \fire_1 \ldots \fire_n} = n + \sum_{i=1}^n\sizeap{\fire_i} \leq_{\ih} n + \sum_{i=1}^n\sizeap{\tder_i} = \sizeap\tder$.

	\item \emph{$\expr$ is a program $\prog$}. By induction on the length $n$ of the environment. If $n=0$ then $\prog = (\fire, \emptyenv)$ and we apply the case for fireballs. Otherwise $n>0$ and $\prog = (\fire, \genv\appendOp\esub\var\itm)$. Then $\tder$ has the following form:
    \begin{equation*}
	\AxiomC{$\ \vdots\, \tdertwo$}
        \noLine
        \UnaryInfC{$\typctxthree, \var \hastype \ptypetwo \vdash (\fire, \genv) \hastype \ptype$}
        \AxiomC{$\ \vdots\, \tderthree$}
        \noLine
        \UnaryInfC{$\typctxtwo\vdash \itm \hastype \ptypetwo$}
         \RightLabel{\footnotesize$\EsAppend$}
         \BinaryInfC{$\typctxthree \uplus \typctxtwo \vdash (\fire, \genv \appendOp \esub\var\itm) \hastype \ptype$}
        \DisplayProof
      \end{equation*}
    with $\typctx = \typctxthree \uplus \typctxtwo$. 

    By \ih, $\sizeap{(\fire, \genv)} \leq \sizeap\tdertwo$ and $\sizeap \itm \leq \sizeap\tderthree$. 
    Then, $\sizeap \expr = \sizeap{(\fire, \genv \appendOp \esub\var\itm)} = \sizeap{(\fire, \genv)} + \sizeap \itm \leq \sizeap\tdertwo + \sizeap\tderthree = \sizeap\tder$.
    \qedhere
	\end{itemize}
\end{proof}

\setcounter{lemmaAppendix}{\value{l:substitution}}
\begin{lemmaAppendix}[Substitution]
\label{lappendix:substitution}
\NoteState{l:substitution}
  Let $\concl{\tder}{\typctx, \var \hastype \ptypetwo}{\tm}{\ptype}$ and $\concl{\tdertwo}{\typctxtwo}{\val}{\ptypetwo}$. Then there exists $\concl{\tderthree}{\typctx \uplus \typctxtwo}{\tm\isub\var\val}{\ptype}$ such that $\size{\tderthree} = \size{\tder} + \size{\tdertwo}$.
\end{lemmaAppendix}

\input{\proofspath/substitution}

\setcounter{propositionAppendix}{\value{prop:quant-subject-reduction}}
\begin{propositionAppendix}[Quantitative subject reduction]
\label{propappendix:quant-subject-reduction}
\NoteState{prop:quant-subject-reduction}
  Let $\prog$ and $\progp$ be programs and $\concl{\tder}{\typctx}{\prog}{\ptype}$ be a type derivation for $\prog$.
If $\prog \tof \progp$ then $\size{\tder} > 0$ and there exists a type derivation $\concl {\tder'} \typctx \progp \ptype$ 
such that $\size{\tder'} = \size{\tder} - 1$.
\end{propositionAppendix}

\input{\proofspath/subject-reduction}

\setcounter{theoremAppendix}{\value{thm:correctness}}
\begin{theoremAppendix}[Correctness]
\label{thmappendix:correctness}
\NoteProof{thm:correctness}
Let $\concl{\tder}{\typctx}{\prog}{\ptype}$ be a type derivation. Then there exists a normal form $\progtwo$ and an evaluation $\deriv: \prog \tof^* \progtwo$ with $\size\deriv + \sizeap\progtwo \leq \sizeap\tder$. 
\end{theoremAppendix}

\begin{proof}
By induction on $\sizeap\tder$. 
If $\prog$ is normal then the statement holds with $\progtwo \defeq \prog$ and $\deriv$ the empty evaluation ($\size{\deriv} = 0$), because $\sizeap\progtwo \leq \sizeap\tder$ by \refprop{size-normal}. 
Otherwise $\prog \tof \progthree$ and by quantitative subject reduction (\refprop{quant-subject-reduction}) there is $\concl{\tdertwo}{\typctx}{\progthree}{\ptype}$ such that $\sizeap\tdertwo = \sizeap\tder -1$. By \ih, there exists $\progtwo$ normal and an evaluation $\derivtwo: \progthree \tof^* \progtwo$ such that $\size\derivtwo + \sizeap\progtwo \leq \sizeap\tdertwo$. Then the evaluation $\deriv:\prog  \tof^* \progtwo$ obtained by prefixing $\derivtwo$ with the step $\prog \tof \progthree$ verifies $\size\deriv + \sizeap\progtwo = \size\derivtwo + 1 + \sizeap\progtwo \leq \sizeap\tdertwo + 1 = \sizeap\tder$.
\end{proof}

\setcounter{propositionAppendix}{\value{prop:types-bound-nfs}}
\begin{propositionAppendix}[Types bound the size of normal forms]
\label{propappendix:types-bound-nfs}
\NoteState{prop:types-bound-nfs}
  Let $\expr$ be a normal expression.
  For every type derivation $\concl{\tder}{\typctx}{\expr}{\ptype}$, one has $\sizeap{\expr} \leq \size{(\typelist\typctx, \ptype)}$.
  If moreover $\expr$ is an inert term, then $\sizeap{\expr} + \size{\ptype}\leq \size{\typelist\typctx}$.
\end{propositionAppendix}

\begin{proof}
There are two cases, either $\expr$ is a fireball or a normal program.
Note that $\sizeap{\expr} + \size{\ptype}\leq \size{\typelist\typctx}$ implies $\sizeap{\expr} \leq \size{(\typelist\typctx, \ptype)}$.

\begin{itemize}
\item \emph{$\expr$ is a fireball $\fire$}. By induction on the definition of fireballs. 
If $\fire$ is a value then $\sizeap{\expr} = 0 \leq \size{(\typelist\typctx, \ptype)}$. 
Otherwise $\fire$ is a inert term of the form $\var \fire_1 \ldots \fire_n$ ($n > 0$). 
Any type derivation $\concl{\tder}{\typctx}{\var \fire_1 \ldots \fire_n}{\ptype}$ is constructed as follows, where $\ptypetwo_1 \defeq \ptype$, $\ptypetwo_{i+1} \defeq \mset{\Pair{\ptype_{n-i+1}} {\ptypetwo_i}}$ ($1 \leq i \leq n$) and $\typctx = \var \hastype \ptypetwo_{n+1}  \uplus (\biguplus_{i=1}^n\typctxtwo_i)$:
  \begin{equation*}
    \tder \defeq 
    \AxiomC{}
    \RightLabel{\footnotesize$\ruleAx$}
    \UnaryInfC{$\var \hastype \ptypetwo_{n+1}  \vdash \var \hastype \ptypetwo_{n+1}$}
    \AxiomC{$\ \vdots\,\tdertwo_1$}
    \noLine
    \UnaryInfC{$\typctxtwo_1 \vdash \fire_1 \hastype \ptype_1$}
    \RightLabel{\footnotesize$\ruleAp$}
    \BinaryInfC{$\var \hastype  \ptypetwo_{n+1}  \uplus \typctxtwo_1  \vdash \var \fire_1 \hastype  \ptypetwo_n$}
    \UnaryInfC{$\vdots$}
        \AxiomC{$\ \vdots\,\tdertwo_n$}
    \noLine
    \UnaryInfC{$\typctxtwo_n \vdash \fire_n \hastype \ptype_n$}
    \RightLabel{\footnotesize$@$}
    \BinaryInfC{$\var \hastype \ptypetwo_{n+1}  \uplus (\biguplus_{i=1}^n\typctxtwo_i ) \vdash \var \fire_1 \ldots \fire_n \hastype \ptypetwo_1$}
    \DisplayProof
  \end{equation*}
    By \ih, $\sizeap{\fire_i} \leq \size{(\typelist\typctxtwo_i,\ptype_i)}$.
    Note that $\size{\ptypetwo_{n+1}} = n + \size{\ptype} + \sum_{i=1}^n \size{\ptype_i}$. 
    Therefore, $\sizeap{\var \fire_1\dots\fire_n} + \size{\ptypetwo_1} = n + \sum_i^n\sizeap{\fire_i} + \size{\ptypetwo_1} \leq n + \sum_i^n \size{(\typelist\typctxtwo_i,\ptype_i)} + \size{\ptypetwo_1} =\size{\typelist{\var \hastype \ptypetwo_{n+1}  \uplus (\biguplus_{i=1}^n\typctxtwo_i )}}$.

	\item \emph{$\expr$ is a program $\prog$}. 
	By induction on the length $n$ of the environment in $\prog$. 
	If $n=0$ then $\prog = (\fire, \emptyenv)$ and the last rule of any type derivation for $\prog$ is $\EsCoerc$ (which preserves types), thus we can apply the case for fireballs, since $\sizeap{(\fire, \emptyenv)} = \sizeap{\fire}$. 
	Otherwise $n>0$ and $\prog = (\fire, \genv\appendOp\esub\var\itm)$. 
	Any type derivation $\tder$ for $\prog$ is then constructed as follows, where $\typctx = \typctxthree \uplus \typctxtwo$:
	    \begin{equation*}
        \tder \defeq
	\AxiomC{$\ \vdots\, \tderthree$}
        \noLine
        \UnaryInfC{$\typctxthree, \var \hastype \ptypetwo \vdash (\fire, \genv) \hastype \ptype$}
        \AxiomC{$\ \vdots\, \tdertwo$}
        \noLine
        \UnaryInfC{$\typctxtwo\vdash \itm \hastype \ptypetwo$}
         \RightLabel{\footnotesize$\EsAppend$}
         \BinaryInfC{$\typctxthree \uplus \typctxtwo \vdash (\fire, \genv \appendOp \esub\var\itm) \hastype \ptype$}
        \DisplayProof
      \end{equation*}
      By \ih, $\sizeap{(\fire,\genv)} \leq 
      \size{(\typelist{\typctxthree, \var \hastype \ptypetwo},\ptype)}$ and $\sizeap{\itm} + \size{\ptypetwo} \leq \size{\typelist\typctxtwo}$. 
      Hence, $\sizeap{(\fire, \genv \appendOp \esub\var\itm)} = \sizeap{(\fire, \genv)} + \sizeap{\itm} \leq \size{(\typelist{\typctxthree, \var \hastype \ptypetwo},\ptype)} + \size{\typelist\typctxtwo} - \size{\ptypetwo} = \size{(\typelist{\typctxthree \uplus \typctxtwo},\ptype)}$.
      \qedhere
\end{itemize}
\end{proof}

\setcounter{propositionAppendix}{\value{prop:nfs-are-typable}}
\begin{propositionAppendix}[Normal forms are typable]
\label{propappendix:nfs-are-typable}
\NoteState{prop:nfs-are-typable}
  \begin{enumerate}
    \item \emph{Normal expression:} For any normal expression $\expr$, there exist a type derivation $\concl{\tder}{\typctx}{\expr}{\ptype}$ for some type context $\typctx$ and some multi type $\ptype$.
    \item \emph{Inter term:} For any multi type $\ptypetwo$ and any inert term $\itm$, there exist a type derivation $\concl{\tdertwo}{\typctxtwo}{\itm}{\ptypetwo}$ for some type context $\typctxtwo$.
  \end{enumerate}
\end{propositionAppendix}

\input{\proofspath/normal-forms-typable}

%% file: substitution.tex
\begin{proof}
  By induction on $\tm \in \Lambda$.
  \begin{enumerate}
  \item \emph{Variable}, two sub-cases:
  	\begin{enumerate}
		\item $\tm = \var$, then $\tm\isub\var\val = \var\isub\var\val = \val$ and the last rule of $\tder$ is $\ruleAx$ with $\ptypetwo = \ptype$ and $\typctx = \vartwo_1 \hastype \emptymset, \dots, \vartwo_n \hastype \emptymset$ ($\vartwo_i \neq \var$ for all $1 \leq i \leq n$) and $\size{\tder} = 0$. If $\typctx \uplus \typctxtwo = \typctxtwo$ then the statement holds by taking $\tderthree \defeq \tdertwo$. 
  
		\item $\tm = \vartwo \neq \var$, then $\tm\isub\var\val = \vartwo$ and the last rule of $\tder$ is $\ruleAx$ with $\ptypetwo = \emptymset$ (since $\var \neq \vartwo$), whence $\size{\tder} = 0$ and 
  $\typctx = \varthree_1 \hastype \emptymset, \dots, \varthree_k \hastype \emptymset, \vartwo \hastype \ptype$.
  By \reflemmap{value-typing}{empty}, from $\typctxtwo \vdash \val \hastype \emptymset$ it follows that $\size{\tdertwo} = 0$ and that all types in $\typctxtwo$ are $\emptymset$, therefore $\typctx \uplus \typctxtwo = \typctx$.
  Analogously to the previous case, $\tderthree$ is either $\tder$ itself or obtained by repeated weakenings of $\tder$ with the variables in $\Dom\typctxtwo \setminus \Dom\typctx$.
  \end{enumerate}
  
  \item \emph{Application}, \ie $\tm = \tmtwo\tmthree$. Then $\tm\isub\var\val = \tmtwo\isub\var\val\tmthree\isub\var\val$ and
  \begin{equation*}
      \tder = 
      \AxiomC{$\ \vdots\,\tder_1$}
      \noLine
      \UnaryInfC{$\typctx_1, \var \hastype \ptypetwo_1 \vdash \tmtwo \hastype [\Pair{\ptypethree}{\ptype}]$}
      \AxiomC{$\ \vdots\,\tder_2$}
      \noLine
      \UnaryInfC{$\typctx_2, \var \hastype \ptypetwo_2 \vdash \tmthree \hastype \ptypethree$}
      \RightLabel{\footnotesize$@$}
      \BinaryInfC{$\typctx, \var \hastype \ptypetwo \vdash \tm \hastype \ptype$}
      \DisplayProof
  \end{equation*}
  with $\size{\tder} = \size{\tder_1} + \size{\tder_2} + 1$, $\typctx = \typctx_1 \uplus \typctx_2$ and $\ptypetwo = \ptypetwo_1 \uplus \ptypetwo_2$.

  According to \reflemmap{value-typing}{dec} applied to $\concl{\tdertwo}{\typctxtwo}{\val}{\ptypetwo}$ and to the decomposition $\ptypetwo = \ptypetwo_1 \uplus \ptypetwo_2$, there are environments $\typctxtwo_1, \typctxtwo_2$ and derivations $\concl{\tdertwo_1}{\typctxtwo_1}{\val}{\ptypetwo_1}$ and $\concl{\tdertwo_2}{\typctxtwo_2}{\val}{\ptypetwo_2}$ such that $\typctxtwo = \typctxtwo_1 \uplus \typctxtwo_2$ and $\size{\tdertwo} = \size{\tdertwo_1} + \size{\tdertwo_2}$.

  By \ih, there are derivations $\tderthree_1$ and $\tderthree_2$ with conclusion $\typctx_1 \uplus \typctxtwo_1 \vdash \tmtwo\isub\var\val \hastype [\Pair{\ptypethree}{\ptype}]$ and $\typctx_2 \uplus \typctxtwo_2 \vdash \tmthree\isub\var\val \hastype \ptypethree$, respectively, such that $\size{\tderthree_1} = \size{\tder_1} + \size{\tdertwo_1}$ and $\size{\tderthree_2} = \size{\tder_2} + \size{\tdertwo_2}$.
  As $\typctx \uplus \typctxtwo = \typctx_1 \uplus \typctxtwo_1 \uplus \typctx_2 \uplus \typctxtwo_2$, there is a derivation 
  \begin{equation*}
      \tderthree \defeq 
      \AxiomC{$\ \vdots\,\tderthree_1$}
      \noLine
      \UnaryInfC{$\typctx_1 \uplus \typctxtwo_1 \vdash \tmtwo\isub\var\val \hastype \mset{\Pair{\ptypethree}{\ptype}}$}
      \AxiomC{$\ \vdots\,\tderthree_2$}
      \noLine
      \UnaryInfC{$\typctx_2 \uplus \typctxtwo_2 \vdash \tmthree\isub\var\val \hastype \ptypethree$}
      \RightLabel{\footnotesize$@$}
      \BinaryInfC{$\typctx \uplus \typctxtwo \vdash \tm\isub\var\val \hastype \ptype$}
      \DisplayProof
  \end{equation*}
  where $\size{\tderthree} = \size{\tderthree_1} + \size{\tderthree_2} + 1 =  \size{\tder_1} + \size{\tdertwo_1} + \size{\tder_2} + \size{\tdertwo_2} + 1 = \size{\tder} + \size{\tdertwo} + 1$.
  
  \item \emph{Abstraction}, \ie $\tm = \la\vartwo\tmtwo$. 
  We can suppose without loss of generality that $\vartwo \notin \Fv{\val} \cup \{\var\}
  $, therefore $\tm\isub\var\val = \la\vartwo\tmtwo\isub\var\val$ and there are $n \in \nat$, typing contexts $\typctx_1, \dots, \typctx_n$, multi types $\ptypetwo_1, \ptypethree_1, \ptype_1, \dots, \ptypetwo_n, \ptypethree_n, \ptype_n$ such that 
  \begin{equation*}
    \tder = 
    \AxiomC{$\ \vdots\,\tder_1$}
    \noLine
    \UnaryInfC{$\typctx_1, \vartwo \hastype \ptypethree_1, \var \hastype \ptypetwo_1 \vdash \tmtwo \hastype \ptype_1$}
    \AxiomC{$\overset{n}{\dots}$}
    \AxiomC{$\ \vdots\,\tder_n$}
    \noLine
    \UnaryInfC{$\typctx_n, \vartwo \hastype \ptypethree_n, \var \hastype \ptypetwo_n \vdash \tmtwo \hastype \ptype_n$}
    \RightLabel{\footnotesize$\lambda$}
    \TrinaryInfC{$\typctx, \var \hastype \ptypetwo \vdash \la\vartwo\tmtwo \hastype \ptype$}
    \DisplayProof
  \end{equation*}
  with $\typctx = \biguplus_{i=1}^n \typctx_i$, $\ptypetwo = \biguplus_{i = 1}^n \ptypetwo_i$, $\ptype = \biguplus_{i=1}^n [\Pair{\ptypethree_i}{\ptype_i}]$, and $\size{\tder} = \sum_{i=1}^n \size{\tder_i}$.
    Since $\vartwo \notin \fv{\val}$, then $\vartwo \notin \domain{\typctxtwo}$ by \reflemma{free}, so that $\concl{\tdertwo}{\typctxtwo, \vartwo \hastype \emptymset}{\val}{\ptypetwo}$.
  Now, we can decompose 
  $\tdertwo$ according to the partitioning $\ptypetwo = \biguplus_{i=1}^n \ptypetwo_i$ by repeatedly applying \reflemmap{value-typing}{dec}, and so for all $1 \leq i \leq n$ there are an environment $\typctxtwo_i$ and a derivation $\concl{\tdertwo_i}{\typctxtwo_i, \vartwo \hastype \emptymset}{\val}{\ptypetwo_i}$ such that $\typctxtwo = \biguplus_{i=1}^n \typctxtwo_i$ and $\size{\tdertwo'} = \sum_{i=1}^n \size{\tdertwo_i}$.
  By \ih, for all $1 \leq i \leq n$, there is a derivation $\concl{\tdertwo_i'}{\typctx_i \uplus \typctxtwo_i, \vartwo \hastype \ptypethree_i}{\tmtwo\isub\var\val}{\ptype_i}$ such that $\size{\tdertwo_i'} = \size{\tder_i} + \size{\tdertwo_i}$.	
  Since $\typctx \uplus \typctxtwo = \biguplus_{i=1}^n \typctx_i \uplus \typctxtwo_i$, there is a derivation
  \begin{equation*}
    \tderthree = 
    \AxiomC{$\ \vdots\,\tdertwo_1'$}
    \noLine
    \UnaryInfC{$\typctx_1 \uplus \typctxtwo_1, \vartwo \hastype \ptypethree_1 \vdash \tmtwo\isub\var\val \hastype \ptype_1$}
    \AxiomC{$\overset{n}{\dots}$}
    \AxiomC{$\ \vdots\,\tdertwo_n'$}
    \noLine
    \UnaryInfC{$\typctx_n \uplus \typctxtwo_n, \vartwo \hastype \ptypethree_n \vdash \tmtwo\isub\var\val \hastype \ptype_n$}
    \RightLabel{\footnotesize$\lambda$}
    \TrinaryInfC{$\typctx \uplus \typctxtwo \vdash \tm\isub\var\val \hastype \ptype$}
    \DisplayProof
  \end{equation*}
  where $\size{\tderthree} = \sum_{i=1}^n \size{\tdertwo_i'} = \sum_{i=1}^n \size{\tder_i} + \sum_{i=1}^n \size{\tdertwo_i} = \size{\tder} + \size{\tdertwo'} = \size{\tder} + \size{\tdertwo}$.
  \qedhere
  \end{enumerate}
\end{proof}

%% file: subject-reduction.tex
\begin{proof}
Let $\prog = (\tm,\genv)$. By induction on the length $n$ of the environment $\genv$ (note that the induction is on the \emph{length} and not on the \emph{structure}). Cases:
\begin{itemize}
  \item \emph{Empty environment}, \ie $n = 0$. This case is itself by induction on the evaluation context $\evctx$ in the step $\prog \tof \progp$. Cases of $\evctx$:
\begin{enumerate}
\item \emph{Step at the root}, \ie $\evctx = \ctxhole$. Then $\tm = (\la\var\tmtwo)\tmthree$ and the typing derivation $\tder$ of $\prog$ is:
      \begin{equation*}
        \tder =
        \AxiomC{$\ \vdots\, \tder_1$}
        \noLine
        \UnaryInfC{$\typctx_1, \var \hastype \ptypetwo \vdash \tmtwo \hastype \ptype$}
        \RightLabel{\footnotesize$\lambda$}
        \UnaryInfC{$\typctx_1 \vdash \la\var\tmtwo \hastype [\Pair{\ptypetwo}{\ptype}]$}
        \AxiomC{$\ \vdots\, \tder_2$}
        \noLine
        \UnaryInfC{$\typctx_2 \vdash \tmthree \hastype \ptypetwo$}
        \RightLabel{\footnotesize$@$}
        \BinaryInfC{$\typctx \vdash (\la\var\tmtwo)\tmthree \hastype \ptype$}
        \RightLabel{\footnotesize$\EsCoerc$}
        \UnaryInfC{$\typctx \vdash ((\la\var\tmtwo)\tmthree,\emptyenv) \hastype \ptype$}
        \DisplayProof
      \end{equation*}
      where $\typctx = \typctx_1 \uplus \typctx_2$ and $\size{\tder} = \size{\tder_1} + \size{\tder_2} + 1$. Two subcases depending on the shape of $\tmthree$:
\begin{enumerate}
\item \emph{$\tmthree$ is a value}, and $\prog = (\tm,\emptyenv) = ((\la\var\tmtwo)\val,\emptyenv) \tobv (\tmtwo\isub\var\val,\emptyenv) = \progp$. Applying the substitution lemma (\reflemma{substitution}) to $\tder_1$ and $\tder_2$, we obtain a derivation $\concl{\tder''}{\typctx}{\tmtwo\isub\var\val}{\ptype}$ such that $\size{\tder''} = \size{\tder_1} + \size{\tder_2} = \size{\tder} - 1$. Applying rule $\EsCoerc$ to $\tder''$ we coerce it to a derivation $\tder'$ for the program $(\tmtwo\isub\var\val,\emptyenv) = \progp$ such that $\size{\tder'} = \size{\tder''} = \size{\tder} - 1$.

\item \emph{$\tmthree$ is a inert term},  and $\prog = (\tm,\emptyenv) = ((\la\var\tmtwo)\itm),\emptyenv) \toin (\tmtwo, \esub\var\itm) = \progp$. Then, the following derivation $\tder'$:
      \begin{equation*}
        \tder' \defeq
        \AxiomC{$\ \vdots\, \tder_1$}
        \noLine
        \UnaryInfC{$\typctx_1, \var \hastype \ptypetwo \vdash \tmtwo \hastype \ptype$}
        \RightLabel{\footnotesize$\EsCoerc$}
        \UnaryInfC{$\typctx_1, \var \hastype \ptypetwo  \vdash (\tmtwo,\emptyenv) \hastype \ptype$}
        \AxiomC{$\ \vdots\, \tder_2$}
        \noLine
        \UnaryInfC{$\typctx_2 \vdash \itm \hastype \ptypetwo$}
        \RightLabel{\footnotesize$\EsAppend$}
        \BinaryInfC{$\typctx \vdash (\tmtwo, \esub\var\itm) \hastype \ptype$}
        \DisplayProof
      \end{equation*}
is such that $\size{\tder'} = \size{\tder_1} + \size{\tder_2} = \size{\tder} - 1$.

\end{enumerate}

 \item \emph{Application Left}, \ie~$\prog = (\tmtwo\fire,\emptyenv) \tof (\tmtwop\fire,\genvtwo) = \tmp$ with $(\tmtwo,\emptyenv) \tof (\tmtwop,\genvtwo)$. The typing derivation $\tder$ of $\prog$ then is:
      \begin{equation*}
        \tder =
        \AxiomC{$\ \vdots\, \tder_1$}
        \noLine
	\UnaryInfC{$\typctx_1 \vdash \tmtwo \hastype [\Pair{\ptypetwo}{\ptype}]$}
        \AxiomC{$\ \vdots\, \tder_2$}
        \noLine
        \UnaryInfC{$\typctx_2 \vdash \fire \hastype \ptypetwo$}
        \RightLabel{\footnotesize$@$}
        \BinaryInfC{$\typctx \vdash \tmtwo \fire \hastype \ptype$}
        \RightLabel{\footnotesize$\EsCoerc$}
        \UnaryInfC{$\typctx \vdash (\tmtwo \fire,\emptyenv) \hastype \ptype$}
        \DisplayProof
      \end{equation*}
      where $\typctx = \typctx_1 \uplus \typctx_2$ and $\size{\tder} = \size{\tder_1} + \size{\tder_2} + 1$. 
      Consider the derivation 
      \begin{equation*}
        \tdertwo =
        \AxiomC{$\ \vdots\, \tder_1$}
        \noLine
	\UnaryInfC{$\typctx_1 \vdash \tmtwo \hastype \mset{\Pair{\ptypetwo}{\ptype}}$}
         \RightLabel{\footnotesize$\EsCoerc$}
         \UnaryInfC{$\typctx_1 \vdash (\tmtwo ,\emptyenv) \hastype \mset{\Pair{\ptypetwo}{\ptype}}$}
        \DisplayProof
      \end{equation*}
      By hypothesis, $(\tmtwo,\emptyenv) \tof (\tmtwop,\genvtwo)$, and by \ih there exists a derivation $\concl{\tdertwo'}{\typctx_1}{(\tmtwop,\genvtwo)}{\mset{\Pair{\ptypetwo}{\ptype}}}$ such that $\size{\tdertwo'} = \size{\tdertwo} - 1 = \size{\tder_1}-1$. Now, two cases, depending on the nature of the rewriting step:
      \begin{enumerate}
      \item \emph{$\tobv$ step}. Then $\genvtwo = \emptyenv$ and so $\tdertwo'$ has the form:
      \begin{equation*}
        \tdertwo' =
        \AxiomC{$\ \vdots\, \tdertwo''$}
        \noLine
	\UnaryInfC{$\typctx_1 \vdash \tmtwop \hastype \mset{\Pair{\ptypetwo}{\ptype}}$}
         \RightLabel{\footnotesize$\EsCoerc$}
         \UnaryInfC{$\typctx_1 \vdash (\tmtwop ,\emptyenv) \hastype \mset{\Pair{\ptypetwo}{\ptype}}$}
        \DisplayProof
      \end{equation*}
      with $\size{\tdertwo''} = \size{\tdertwo'} = \size{\tder_1}-1$. Therefore, the following derivation:      
      \begin{equation*}
        \tder' =
        \AxiomC{$\ \vdots\, \tdertwo''$}
        \noLine
	\UnaryInfC{$\typctx_1 \vdash \tmtwop \hastype \mset{\Pair{\ptypetwo}{\ptype}}$}
        \AxiomC{$\ \vdots\, \tder_2$}
        \noLine
        \UnaryInfC{$\typctx_2 \vdash \fire \hastype \ptypetwo$}
        \RightLabel{\footnotesize$@$}
        \BinaryInfC{$\typctx \vdash \tmtwop \fire \hastype \ptype$}
        \RightLabel{\footnotesize$\EsCoerc$}
        \UnaryInfC{$\typctx \vdash (\tmtwop \fire,\emptyenv) \hastype \ptype$}
        \DisplayProof
      \end{equation*}
      satysfies the statement because 
      $$\size{\tder'} = \size{\tdertwo''} + \size{\tder_2} + 1 = \size{\tder_1} -1 + \size{\tder_2} + 1 = \underbrace{\size{\tder_1} + \size{\tder_2} + 1}_{\size{\tder}} -1 = \size{\tder} - 1$$.
      
       \item \emph{$\toin$ step}. Then $\prog = (\tmtwo \fire, \emptyenv) \toin (\tmtwop \fire, \esub\var\itm)$ for some $\var$ and inert term $\itm$ and with $\var \notin \fv\fire$, because $\var$ comes from an abstraction inside $\tmtwo$. Now, $\tdertwo'$ has the form:
      \begin{equation*}
        \tdertwo' =
        \AxiomC{$\ \vdots\, \tdertwo''$}
        \noLine
	\UnaryInfC{$\typctxtwo, \var\hastype \ptypethree \vdash \tmtwop \hastype \mset{\Pair{\ptypetwo}{\ptype}}$}
         \RightLabel{\footnotesize$\EsCoerc$}
         \UnaryInfC{$\typctxtwo, \var\hastype \ptypethree  \vdash (\tmtwop ,\emptyenv) \hastype \mset{\Pair{\ptypetwo}{\ptype}}$}
         \AxiomC{$\ \vdots\, \tderthree$}
        \noLine
        \UnaryInfC{$\typctxthree \vdash \itm \hastype \ptypethree$}
         \RightLabel{\footnotesize$\EsAppend$}
         \BinaryInfC{$\typctx_1 \vdash (\tmtwop,\esub\var\itm) \hastype \mset{\Pair{\ptypetwo}{\ptype}}$}
        \DisplayProof
      \end{equation*}
      with $\typctx_1 = \typctxtwo \uplus \typctxthree$. Note that from $\size{\tdertwo'} = \size{\tder_1}-1$ it follows that $ \size{\tdertwo''} + \size{\tderthree} =  \size{\tder_1}-1$. Now, for proving the statement we have to reintroduce the application to $\fire$ via a $@$ typing rule, but this cannot be done at the end, the typing system forces it to happen before the coercion of $\tmtwop$ to a program. To this aim, note that if $\var$ appears in the typing context $\typctx_2$ of $\fire$ then by 
      \reflemma{free} it is associated to $\emptymset$, because $\var \notin \fv\fire$. Therefore, the following derivation: 
      \begin{equation*}
        \tder' =
        \AxiomC{$\ \vdots\, \tdertwo''$}
        \noLine
	\UnaryInfC{$\typctxtwo, \var\hastype \ptypethree \vdash \tmtwop \hastype \mset{\Pair{\ptypetwo}{\ptype}}$}
	\AxiomC{$\ \vdots\, \tder_2$}
        \noLine
        \UnaryInfC{$\typctx_2 \vdash \fire \hastype \ptypetwo$}
        \RightLabel{\footnotesize$@$}
        \BinaryInfC{$\typctxtwo \uplus \typctx_2, \var\hastype \ptypethree \vdash \tmtwop \fire \hastype \ptype$}
         \RightLabel{\footnotesize$\EsCoerc$}
         \UnaryInfC{$\typctxtwo \uplus \typctx_2, \var\hastype \ptypethree  \vdash (\tmtwop \fire,\emptyenv) \hastype \ptype$}
         \AxiomC{$\ \vdots\, \tderthree$}
        \noLine
        \UnaryInfC{$\typctxthree \vdash \itm \hastype \ptypethree$}
         \RightLabel{\footnotesize$\EsAppend$}
         \BinaryInfC{$\typctxtwo \uplus \typctx_2 \uplus \typctxthree \vdash (\tmtwop \fire,\esub\var\itm) \hastype \ptype$}
        \DisplayProof
      \end{equation*}
      satisfies the statement because
      \begin{itemize}
        \item \emph{Typing context}: $\typctxtwo \uplus \typctx_2 \uplus \typctxthree = \underbrace{\typctxtwo  \uplus \typctxthree}_{\typctx_1}\uplus \typctx_2 = \typctx$.
        \item \emph{Size}: $\size{\tder'} = \underbrace{\size{\tdertwo''} + \size\tderthree}_{= \size{\tder_1}-1} + \size{\tder_2} + 1 = \size{\tder_1} + \size{\tder_2} = \size\tder -1$.
      \end{itemize}
\end{enumerate}

     \item \emph{Application right}, \ie~$\prog = (\tmthree\tmtwo,\emptyenv) \tof (\tmthree\tmtwop,\genvtwo) = \tmp$ with $(\tmtwo,\emptyenv) \tof (\tmtwop,\genvtwo)$. It is essentially identical to the previous case, the only difference is the switch of the left and right subterms. Note indeed that in the previous case the fact that the right subterm $\fire$ is a fireball is never used. 
\end{enumerate}

\item \emph{Non-empty environment}, \ie $n> 0$. Then $\genv = \overline\genv \appendOp \esub\var\itm$. Note that the existence of a redex does not depend on the environment, so that if $\prog = (\tm, \overline\genv \appendOp \esub\var\itm) \tof (\tmp, \genvtwo)$ then there exists $\overline\genv'$ such that $(\tm, \overline\genv) \tof (\tmp, \overline\genv' )$ and $\genvtwo = \overline\genv' \appendOp \esub\var\itm$. The type derivation $\tder$ for $\prog = (\tm, \overline\genv \appendOp \esub\var\itm)$ has necessarily the following form:
    \begin{equation*}
        \tder =
	\AxiomC{$\ \vdots\, \tdertwo$}
        \noLine
        \UnaryInfC{$\typctxtwo, \var \hastype \ptypetwo \vdash (\tm, \overline\genv) \hastype \ptype$}
        \AxiomC{$\ \vdots\, \tderthree$}
        \noLine
        \UnaryInfC{$\typctxthree\vdash \itm \hastype \ptypetwo$}
         \RightLabel{\footnotesize$\EsAppend$}
         \BinaryInfC{$\typctxtwo \uplus \typctxthree \vdash (\tm, \overline\genv \appendOp \esub\var\itm) \hastype \ptype$}
        \DisplayProof
      \end{equation*}
    with $\typctx = \typctxtwo \uplus \typctxthree$ and $\size\tder = \size\tdertwo + \size\tderthree$. By \ih applied to the step $(\tm, \overline\genv) \tof (\tmp, \overline\genv')$, there exists a typing derivation $\concl {\tdertwo'} {\typctxtwo, \var \hastype \ptypetwo} {(\tmp, \overline\genv')} \ptype$ such that $\size{\tdertwo'} = \size\tdertwo -1$. Then the typing derivation:
    \begin{equation*}
        \tder' =
	\AxiomC{$\ \vdots\, \tdertwo'$}
        \noLine
        \UnaryInfC{$\typctxtwo, \var \hastype \ptypetwo \vdash (\tmp, \overline\genv) \hastype \ptype$}
        \AxiomC{$\ \vdots\, \tderthree$}
        \noLine
        \UnaryInfC{$\typctxthree\vdash \itm \hastype \ptypetwo$}
         \RightLabel{\footnotesize$\EsAppend$}
         \BinaryInfC{$\typctxtwo \uplus \typctxthree \vdash (\tmp, \overline\genv \appendOp \esub\var\itm) \hastype \ptype$}
        \DisplayProof
      \end{equation*}
      satisfies the statement, because $\size{\tder'} = \size{\tdertwo'} + \size\tderthree = \size{\tdertwo} - 1 + \size\tderthree = \size\tder - 1$.
\end{itemize}
\end{proof}

%% file: normal-forms-typable.tex
\begin{proof}
We prove both statements simultaneously.
There are two cases, either $\expr$ is a fireball or a normal program.
\begin{itemize}
\item \emph{$\expr$ is a fireball $\fire$}. By induction on the definition of fireballs. 
Note that if $\fire$ is a value the statement is given by \reflemmap{value-typing}{judg}. 
Therefore, suppose that $\fire$ is a inert term of the form $\var \fire_1 \ldots \fire_n$ ($n \geq 0$). 
By \ih, $\fire_i$ is typable with a type derivation $\concl{\tdertwo_i}{\typctxtwo_i}{\fire_i}{\ptype_i}$, for some multi type $\ptype_i$. 
The type derivation $\tder$ is 
constructed as follows. Let $\ptypetwo_1 \defeq \ptypetwo$ and $\ptypetwo_{i+1} \defeq \mset{\Pair{\ptype_{n-i+1}} {\ptypetwo_i}}$ ($i \leq n$). 
  \begin{equation*}
    \tder \defeq 
    \AxiomC{}
    \RightLabel{\footnotesize$\ruleAx$}
    \UnaryInfC{$\var \hastype \ptypetwo_{n+1}  \vdash \var \hastype \ptypetwo_{n+1}$}
    \AxiomC{$\ \vdots\,\tdertwo_1$}
    \noLine
    \UnaryInfC{$\typctxtwo_1 \vdash \fire_1 \hastype \ptype_1$}
    \RightLabel{\footnotesize$\ruleAp$}
    \BinaryInfC{$\var \hastype  \ptypetwo_{n+1}  \uplus \typctxtwo_1  \vdash \var \fire_1 \hastype  \ptypetwo_n$}
    \noLine
    \UnaryInfC{$\vdots$}
        \AxiomC{$\ \vdots\,\tdertwo_n$}
    \noLine
    \UnaryInfC{$\typctxtwo_n \vdash \fire_n \hastype \ptype_n$}
    \RightLabel{\footnotesize$@$}
    \BinaryInfC{$\var \hastype \ptypetwo_{n+1}  \uplus (\biguplus_{i=1}^n\typctxtwo_i ) \vdash \var \fire_1 \ldots \fire_n \hastype \ptypetwo_1$}
    \DisplayProof
  \end{equation*}

	\item \emph{$\expr$ is a program $\prog$}. 
	By induction on the length $n$ of the environment in $\prog$. 
	If $n=0$ then $\prog = (\fire, \emptyenv)$ and we apply the case for fireballs (adding the rule $\EsCoerc$ at the end of the type derivation). 
	Otherwise $n>0$ and $\prog = (\fire, \genv\appendOp\esub\var\itm)$. 
	By \ih, there is a type derivation $\concl{\tderthree}{\typctxthree, \var\hastype \ptypetwo}{(\fire, \genv)}{\ptype}$ for some multi type $\ptypetwo$.
    By \ih for inert terms, there is a type derivation $\concl{\tdertwo}{\typctxtwo}{\itm}{\ptypetwo}$. 
    The type derivation $\tder$ is then constructed as follows:
	    \begin{equation*}
        \tder \defeq
	\AxiomC{$\ \vdots\, \tderthree$}
        \noLine
        \UnaryInfC{$\typctxthree, \var \hastype \ptypetwo \vdash (\fire, \genv) \hastype \ptype$}
        \AxiomC{$\ \vdots\, \tdertwo$}
        \noLine
        \UnaryInfC{$\typctxtwo\vdash \itm \hastype \ptypetwo$}
         \RightLabel{\footnotesize$\EsAppend$}
         \BinaryInfC{$\typctxthree \uplus \typctxtwo \vdash (\fire, \genv \appendOp \esub\var\itm) \hastype \ptype$}
        \DisplayProof
      \end{equation*}
      \qedhere
\end{itemize}
\end{proof}

%% file: proofs-completeness.tex
\subsection{Omitted proofs and lemmas of \refsect{completeness}}

\setcounter{lemmaAppendix}{\value{l:anti-substitution}}
\begin{lemmaAppendix}[Anti-substitution]
\label{lappendix:anti-substitution}
\NoteState{l:anti-substitution}
  Let $\tm$ be a term, $\val$ be a value, and $\concl{\tder}{\typctx}{\tm\isub\var\val}{\ptype}$ be a type derivation. 
  Then there exist two type derivations $\concl{\tdertwo}{\typctxtwo, \var \hastype \ptypetwo}{\tm}{\ptype}$ and $\concl{\tderthree}{\typctxthree}{\val}{\ptypetwo}$ such that $\typctx = \typctxtwo \uplus \typctxthree$ and $\size{\tder} = \size{\tdertwo} + \size{\tderthree}$.
\end{lemmaAppendix}

\input{\proofspath/anti-substitution}

\setcounter{propositionAppendix}{\value{prop:quant-subject-expansion}}
\begin{propositionAppendix}[Quantitative subject expansion]
\label{propappendix:quant-subject-expansion}
\NoteState{prop:quant-subject-expansion}
  Let $\prog$ and $\progp$ be programs and $\concl{\tder'}{\typctx}{\progp}{\ptype}$ be a type derivation for $\progp$.
If $\prog \tof \progp$ then there exists a derivation $\concl {\tder} \typctx \progp \ptype$ for $\progp$ such that $\size{\tder'} = \size{\tder} +1$.
\end{propositionAppendix}

\input{\proofspath/subject-expansion}

\setcounter{theoremAppendix}{\value{thm:completeness}}
\begin{theoremAppendix}[Completeness]
\label{thmappendix:completeness}
\NoteState{thm:completeness}
Let $\deriv: \prog \tof^* \progtwo$ be a normalizing evaluation. 
Then there is a type derivation $\concl{\tder}{\typctx}{\prog}{\ptype}$, and it satisfies $\size\deriv + \sizeap\progtwo \leq \sizeap\tder$.
\end{theoremAppendix}

\begin{proof}
  By induction on $\size\deriv$. If $\size\deriv = 0$ then $\prog = \progtwo$ and the existence of $\tder$ is given by the fact that normal programs are typable (\refprop{nfs-are-typable}); the inequality on sizes is given by \refprop{size-normal}. 
  If $\size\deriv  =k > 0$ then $\prog \tof \progthree \tof^k \progtwo$ and by \ih we obtain a type derivation $\concl{\tdertwo}{\typctx}{\progthree}{\ptype}$ satisfying $k + \sizeap\progtwo \leq \sizeap\tdertwo$. By quantitative subject expansion (\refprop{quant-subject-expansion}) there exists $\concl{\tder}{\typctx}{\prog}{\ptype}$, satisfying $\sizeap\tder = \sizeap\tdertwo +1$. Then $\size\deriv + \sizeap\progtwo = k + 1 + \sizeap\progtwo  \leq_{\ih} \sizeap\tdertwo + 1 = \sizeap\tder$.
\end{proof}

%% file: anti-substitution.tex

\begin{proof}
  By induction on $\tm \in \Lambda$.
  \begin{enumerate}
  \item \emph{Variable}, two sub-cases:
  	\begin{enumerate}
		\item $\tm = \var$, then $\tm\isub\var\val = \var\isub\var\val = \val$. Then $\tder$ is  a typing derivation for $\val$ and $\ptype = \ptypetwo$. We can set $\tderthree \defeq \tder$ and $\tdertwo$ as the $\ruleAx$-rule on $\var$ with type $\ptypetwo = \ptype$ and empty typing context ($\typctxtwo$ is empty). Since $\size \tdertwo = 0$ we obtain $\size{\tder} = \size{\tdertwo} + \size{\tderthree}$.
  
		\item $\tm = \vartwo \neq \var$, then $\tm\isub\var\val = \vartwo$. Then $\tder$ is a typing derivation for $\tm$ and $\ptypetwo = \emptymset$. We can set $\tdertwo \defeq \tder$ and by \reflemmap{value-typing}{judg} there is a derivation $\concl{\tderthree}{}{\val}{\emptymset}$, and by \reflemmap{value-typing}{empty} this derivation has size $0$. Therefore, $\size{\tder} = \size{\tdertwo} + \size{\tderthree}$.
	 \end{enumerate}
  
  \item \emph{Application}, \ie $\tm = \tmtwo\tmthree$. Then $\tm\isub\var\val = \tmtwo\isub\var\val\tmthree\isub\var\val$ and $\tder$ has the following form

  \begin{equation*}
      \tder =
      \AxiomC{$\ \vdots\,\tder_\tmtwo$}
      \noLine
      \UnaryInfC{$\typctx_\tmtwo \vdash \tmtwo\isub\var\val \hastype \mset{\Pair{\ptypethree}{\ptype}}$}
      \AxiomC{$\ \vdots\,\tder_\tmthree$}
      \noLine
      \UnaryInfC{$\typctx_\tmthree \vdash \tmthree\isub\var\val \hastype \ptypethree$}
      \RightLabel{\footnotesize$@$}
      \BinaryInfC{$\typctx_\tmtwo \uplus \typctx_\tmthree \vdash \tm\isub\var\val \hastype \ptype$}
      \DisplayProof
  \end{equation*}
  where $\size{\tder} = \size{\tder_\tmtwo} + \size{\tder_\tmthree} + 1$. By \ih applied to the two premises, there are four derivations 
  \begin{enumerate}
  
  \item $\concl{\tdertwo_\tmtwo}{\typctxtwo_\tmtwo}{\tmtwo}{\mset{\Pair{\ptypethree}{\ptype}}}$ and  $\concl{\tderthree_\tmtwo}{\typctxthree_\tmtwo}{\val}{\ptypetwo_\tmtwo}$ such that $\typctx_\tmtwo =  \typctxtwo_\tmtwo \uplus \typctxthree_\tmtwo$ and $\size{\tder_\tmtwo} = \size{\tdertwo_\tmtwo} + \size{\tderthree_\tmtwo}$,
  
  \item $\concl{\tdertwo_\tmthree}{\typctxtwo_\tmthree}{\tmthree}{\ptypethree}$ and  $\concl{\tderthree_\tmthree}{\typctxthree_\tmthree}{\val}{\ptypetwo_\tmthree}$ such that $\typctx_\tmthree =  \typctxtwo_\tmthree \uplus \typctxthree_\tmthree$ and $\size{\tder_\tmthree} = \size{\tdertwo_\tmthree} + \size{\tderthree_\tmthree}$. 
  \end{enumerate}
The derivation $\tdertwo$ is then constructed as follows:
  \begin{equation*}
      \tdertwo \defeq 
      \AxiomC{$\ \vdots\,\tdertwo_\tmtwo$}
      \noLine
      \UnaryInfC{$\typctxtwo_\tmtwo, \var \hastype \ptypetwo_\tmtwo \vdash \tmtwo \hastype \mset{ \Pair{\ptypethree}{\ptype} }$}
      \AxiomC{$\ \vdots\,\tdertwo_\tmthree$}
      \noLine
      \UnaryInfC{$\typctxtwo_\tmthree, \var \hastype \ptypetwo_\tmthree \vdash \tmthree \hastype \ptypethree$}
      \RightLabel{\footnotesize$@$}
      \BinaryInfC{$\typctxtwo_\tmtwo \uplus \typctxtwo_\tmthree, \var \hastype \ptypetwo_\tmtwo \uplus \ptypetwo_\tmthree \vdash \tm \hastype \ptype$}
      \DisplayProof
  \end{equation*}
  with $\size{\tdertwo} = \size{\tdertwo_\tmtwo} + \size{\tdertwo_\tmthree} + 1$. The derivation $\tderthree$ for $\val$ is instead obtained by applying \reflemmap{value-typing}{merg} to $\tderthree_\tmtwo$ and $\tderthree_\tmthree$. Namely, we obtain $\concl{\tderthree}{\typctxthree_\tmtwo \uplus \typctxthree_\tmthree}{\val}{\ptypetwo_\tmtwo \uplus \ptypetwo_\tmthree}$ such that $\size \tderthree = \size{\tderthree_\tmtwo} + \size{\tderthree_\tmthree}$. The statement is then satisfied because:
  \begin{itemize}
  \item \emph{Typing contexts}: $\typctx = \typctx_\tmtwo \uplus \typctx_\tmthree = \typctxtwo_\tmtwo \uplus  \typctxthree_\tmtwo \uplus \typctxtwo_\tmthree \uplus \typctxthree_\tmthree$, and

  \item \emph{Sizes}: 
  $$\size \tder = \size{\tder_\tmtwo} + \size{\tder_\tmthree} +1 = \size{\tdertwo_\tmtwo} + \size{\tderthree_\tmtwo} + \size{\tdertwo_\tmthree} + \size{\tderthree_\tmthree} +1 = \underbrace{\size{\tdertwo_\tmtwo} + \size{\tdertwo_\tmthree} +1}_{\size \tdertwo}+ \underbrace{\size{\tderthree_\tmtwo} + \size{\tderthree_\tmthree}}_{ \size \tderthree} $$
  \end{itemize}
  
  \item \emph{Abstraction}, \ie $\tm = \la\vartwo\tmtwo$. Then we can suppose without loss of generality that $\vartwo \notin \Fv{\val} \cup \{\var\} \cup \Dom{\typctx}$, therefore $\tm\isub\var\val = \la\vartwo\tmtwo\isub\var\val$. The derivation $\tder$ then has necessarily the following form:
   \begin{equation*}
    \tder = 
    \AxiomC{$\ \vdots\,\tder_1$}
    \noLine
    \UnaryInfC{$\typctx_1, \vartwo \hastype \ptypethree_1 \vdash \tmtwo\isub\var\val \hastype \ptype_1$}
    \AxiomC{$\overset{n}{\dots}$}
    \AxiomC{$\ \vdots\,\tder_n$}
    \noLine
    \UnaryInfC{$\typctx_n, \vartwo \hastype \ptypethree_n \vdash \tmtwo\isub\var\val \hastype \ptype_n$}
    \RightLabel{\footnotesize$\lambda$}
    \TrinaryInfC{$\biguplus_{i=1}^n \typctx_i \vdash \la\vartwo \tmtwo\isub\var\val \hastype \mset{ \Pair{ \biguplus_{i=1}^n \ptypethree_i }{ \biguplus_{i=1}^n \ptype_i } }$}
    \DisplayProof
  \end{equation*}
  where $\typctx = \biguplus_{i=1}^n \typctx_i$,  $\ptype = \biguplus_{i=1}^n \ptype_i$, $\ptypethree = \biguplus_{i=1}^n \ptypethree_i$, and $\size{\tder} = \biguplus_{i=1}^n \size{\tder_i}$ for some $n \in \nat$.

  By \ih, for each $\concl{\tder_i}{\typctx_i, \vartwo \hastype \ptypethree_i}{ \tmtwo\isub\var\val }{\ptype_i}$ there exist two derivations $\concl{\tdertwo_i}{\typctxtwo_i, \vartwo \hastype \ptypethree_i, \var \hastype \ptypetwo_i}{ \tmtwo }{\ptype_i}$ and $\concl{\tderthree_i}{\typctxthree_i}{ \val }{\ptypetwo_i}$ such that $\typctx_i = \typctxtwo_i \uplus \typctxthree_i$ and  $\size{ \tder_i } = \size{ \tdertwo_i } + \size{ \tderthree_i }$. 
  Since by hypothesis $\vartwo \notin \fv\val$, then $\vartwo \notin \Dom{\typctxthree_i}$ by \reflemma{free}, so that it is correct to avoid splitting $\ptypethree_i$ between the two derivations.

  The derivation $\tdertwo$ is then constructed as follows:
  \begin{equation*}
    \tdertwo \defeq 
    \AxiomC{$\ \vdots\,\tdertwo_1$}
    \noLine
    \UnaryInfC{$\typctxtwo_1, \vartwo \hastype \ptypethree_1, \var \hastype \ptypetwo_1 \vdash \tmtwo \hastype \ptype_1$}
    \AxiomC{$\overset{n}{\dots}$}
    \AxiomC{$\ \vdots\,\tdertwo_n$}
    \noLine
    \UnaryInfC{$\typctxtwo_n, \vartwo \hastype \ptypethree_n, \var \hastype \ptypetwo_n \vdash \tmtwo \hastype \ptype_n$}
    \RightLabel{\footnotesize$\lambda$}
    \TrinaryInfC{$\biguplus_{i=1}^n \typctxtwo_i, \var \hastype \biguplus_{i=1}^n \ptypetwo_i \vdash \la\vartwo\tmtwo \hastype \mset{ \Pair{ \biguplus_{i=1}^n \ptypethree_i }{ \biguplus_{i=1}^n \ptype_i } }$}
    \DisplayProof
  \end{equation*}
with $\size{\tdertwo} = \sum_{i=1}^n \size{\tdertwo_i}$.
  
  The derivation $\tderthree$ is constructed by repeatedly applying \reflemmap{value-typing}{merg} to all the derivations $\concl{\tderthree_i}{\typctxthree_i}{ \val }{\ptypetwo_i}$, obtaining $\concl{\tderthree}{\biguplus_{i=1}^n\typctxthree_i}{ \val }{\biguplus_{i=1}^n\ptypetwo_i}$, for which $\size{\tderthree} = \sum_{i=1}^n \size{\tderthree_i}$.

The statement is satisfied because:
\begin{itemize}
	\item \emph{Typing contexts}: $\typctx = \biguplus_{i=1}^n \typctx_i = \biguplus_{i=1}^n (\typctxtwo_i \uplus \typctxthree_i) = \biguplus_{i=1}^n \typctxtwo_i \uplus \biguplus_{i=1}^n \typctxthree_i$, and similarly
	
	\item \emph{Sizes}: $\size{\tder} = \sum_{i=1}^n \size{\tder_i} = \sum_{i=1}^n (\size{\tdertwo_i} + \size{\tderthree_i}) = \sum_{i=1}^n \size{\tdertwo_i} + \sum_{i=1}^n  \size{\tderthree_i} = \size \tdertwo + \size \tderthree$.
	\qedhere
\end{itemize}
  \end{enumerate}
\end{proof}

%% file: subject-expansion.tex
\begin{proof}
We prove the cases for the two rewriting rules separately.
\begin{itemize}
	\item \emph{Value}, \ie $\prog = (\tm, \genv) \tobv (\tmp, \genv) = \progp$. By induction on the length $n$ of the environment $\genv$. Cases:
\begin{itemize}
  \item \emph{Empty environment}, \ie $n = 0$. This case is itself by induction on the evaluation context $\evctx$ in the step $\prog \tobv \progp$. Cases of $\evctx$:
\begin{enumerate}
\item \emph{Step at the root}, \ie $\evctx = \ctxhole$. Then $\tm = (\la\var\tmtwo)\val$ and $\tmp = \tmtwo \isub\var\val$. By applying the anti-substitution lemma \reflemma{anti-substitution} to the type derivation $\concl{\tder'}{\typctx}{\progp}{\ptype}$ we obtain two derivations $\concl{\tdertwo}{\typctxtwo, \var \hastype \ptypetwo}{\tmtwo}{\ptype}$ and $\concl{\tderthree}{\typctxthree}{\val}{\ptypetwo}$ such that $\typctx = \typctxtwo \uplus \typctxthree$ and $\size{\tder} = \size{\tdertwo} + \size{\tderthree}$.  Then the derivation $\tder$ of the statement is given by:
      \begin{equation*}
        \tder \defeq
        \AxiomC{$\ \vdots\, \tdertwo$}
        \noLine
        \UnaryInfC{$\typctxtwo, \var \hastype \ptypetwo \vdash \tmtwo \hastype \ptype$}
        \RightLabel{\footnotesize$\lambda$}
        \UnaryInfC{$\typctxtwo \vdash \la\var\tmtwo \hastype [\Pair{\ptypetwo}{\ptype}]$}
        \AxiomC{$\ \vdots\, \tderthree$}
        \noLine
        \UnaryInfC{$\typctxthree \vdash \val \hastype \ptypetwo$}
        \RightLabel{\footnotesize$@$}
        \BinaryInfC{$\typctx \vdash (\la\var\tmtwo)\val \hastype \ptype$}
        \RightLabel{\footnotesize$\EsCoerc$}
        \UnaryInfC{$\typctxtwo \uplus \typctxthree \vdash ((\la\var\tmtwo)\val,\emptyenv) \hastype \ptype$}
        \DisplayProof
      \end{equation*}
      where $\size{\tder} = \size{\tdertwo} + \size{\tderthree} + 1 = \size{\tder'} +1$. 
      
      \item \emph{Application left}, \ie~$\prog = (\tmtwo\fire,\emptyenv) \tobv (\tmtwop\fire,\emptyenv) = \tmp$ with $(\tmtwo,\emptyenv) \tof (\tmtwop,\emptyenv)$. The type derivation $\tder'$ of $\progp$ then necessarily has the form:      
      \begin{equation*}
        \tder' =
        \AxiomC{$\ \vdots\, \tdertwo'$}
        \noLine
	\UnaryInfC{$\typctxtwo \vdash \tmtwop \hastype \mset{\Pair{\ptype}{\ptypetwo}}$}
        \AxiomC{$\ \vdots\, \tderthree$}
        \noLine
        \UnaryInfC{$\typctxthree \vdash \fire \hastype \ptype$}
        \RightLabel{\footnotesize$@$}
        \BinaryInfC{$\typctxtwo \uplus \typctxthree \vdash \tmtwop \fire \hastype \ptypetwo$}
        \RightLabel{\footnotesize$\EsCoerc$}
        \UnaryInfC{$\typctxtwo \uplus \typctxthree \vdash (\tmtwop \fire,\emptyenv) \hastype \ptypetwo$}
        \DisplayProof
      \end{equation*}
 	with for some $\tdertwo'$ and $\tderthree$ such that $\typctx = \typctxtwo \uplus \typctxthree$ and $\size{\tder'} = \size{\tdertwo'} + \size{\tderthree} + 1$. Consider the derivation:
	            \begin{equation*}
        \tdertwo'' =
        \AxiomC{$\ \vdots\, \tdertwo'$}
        \noLine
	\UnaryInfC{$\typctxtwo \vdash \tmtwop \hastype \mset{\Pair{\ptypetwo}{\ptype}}$}
         \RightLabel{\footnotesize$\EsCoerc$}
         \UnaryInfC{$\typctxtwo \vdash (\tmtwop ,\emptyenv) \hastype \mset{\Pair{\ptypetwo}{\ptype}}$}
        \DisplayProof
      \end{equation*}
      By hypothesis, $(\tmtwo,\emptyenv) \tobv (\tmtwop,\emptyenv)$, and by \ih there exists a derivation $\concl{\tdertwo}{\typctxtwo}{(\tmtwo,\emptyenv)}{\mset{\Pair{\ptypetwo}{\ptype}}}$ such that $\size{\tdertwo} = \size{\tdertwo''} + 1 = \size{\tdertwo'} + 1$ that necessarily has the following form:
            \begin{equation*}
        \tdertwo =
        \AxiomC{$\ \vdots\, \tdertwo^*$}
        \noLine
	\UnaryInfC{$\typctxtwo \vdash \tmtwo \hastype \mset{\Pair{\ptypetwo}{\ptype}}$}
         \RightLabel{\footnotesize$\EsCoerc$}
         \UnaryInfC{$\typctxtwo \vdash (\tmtwo ,\emptyenv) \hastype \mset{\Pair{\ptypetwo}{\ptype}}$}
        \DisplayProof
      \end{equation*}
	with $\size{\tdertwo^*} = \size{\tdertwo} = \size{\tdertwo'} + 1$. Therefore, the following derivation:      
      \begin{equation*}
        \tder \defeq
        \AxiomC{$\ \vdots\, \tdertwo^*$}
        \noLine
	\UnaryInfC{$\typctxtwo \vdash \tmtwo \hastype [\Pair{\ptypetwo}{\ptype}]$}
        \AxiomC{$\ \vdots\, \tderthree$}
        \noLine
        \UnaryInfC{$\typctxthree \vdash \fire \hastype \ptypetwo$}
        \RightLabel{\footnotesize$@$}
        \BinaryInfC{$\typctxtwo \uplus \typctxthree \vdash \tmtwo \fire \hastype \ptype$}
        \RightLabel{\footnotesize$\EsCoerc$}
        \UnaryInfC{$\typctxtwo \uplus \typctxthree \vdash (\tmtwo \fire,\emptyenv) \hastype \ptype$}
        \DisplayProof
      \end{equation*}
      satysfies the statement because 
      $$\size{\tder} = \size{\tdertwo^*} + \size{\tderthree} + 1 = \size{\tdertwo'} +1 + \size{\tderthree} + 1 = \underbrace{\size{\tdertwo'} + \size{\tderthree} + 1}_{\size{\tder'}} +1 = \size{\tder'} + 1$$.

     \item \emph{Application right}, \ie~$\prog = (\tmthree\tmtwo,\emptyenv) \tobv (\tmthree\tmtwop,\emptyenv) = \tmp$ with $(\tmtwo,\emptyenv) \tobv (\tmtwop,\emptyenv)$. It is essentially identical to the previous case, the only difference is the switch of the left and right subterms. Note indeed that in the previous case the fact that the right subterm $\fire$ is a fireball is never used. 

\end{enumerate}

\item \emph{Non-empty environment}, \ie $n> 0$. Then $\genv = \overline\genv \appendOp \esub\var\itm$. Note that the existence of a redex does not depend on the environment, so that if $\prog = (\tm, \overline\genv \appendOp \esub\var\itm) \tobv (\tmp, \overline\genv \appendOp \esub\var\itm)$ then $(\tm, \overline\genv) \tobv (\tmp, \overline\genv )$. The type derivation $\tder'$ for $\progp = (\tmp, \overline\genv \appendOp \esub\var\itm)$ has necessarily the following form:
    \begin{equation*}
        \tder' =
	\AxiomC{$\ \vdots\, \tdertwo'$}
        \noLine
        \UnaryInfC{$\typctxtwo, \var \hastype \ptypetwo \vdash (\tmp, \overline\genv) \hastype \ptype$}
        \AxiomC{$\ \vdots\, \tderthree$}
        \noLine
        \UnaryInfC{$\typctxthree\vdash \itm \hastype \ptypetwo$}
         \RightLabel{\footnotesize$\EsAppend$}
         \BinaryInfC{$\typctxtwo \uplus \typctxthree \vdash (\tmp, \overline\genv \appendOp \esub\var\itm) \hastype \ptype$}
        \DisplayProof
      \end{equation*}
    with $\typctx = \typctxtwo \uplus \typctxthree$ and $\size{\tder'} = \size{\tdertwo'} + \size\tderthree$. By \ih applied to the step $(\tm, \overline\genv) \tobv (\tmp, \overline\genv)$, there exists a type derivation $\concl {\tdertwo} {\typctxtwo, \var \hastype \ptypetwo} {(\tm, \overline\genv)} \ptype$ such that $\size{\tdertwo} = \size{\tdertwo'} +1$. Then the type derivation:
    \begin{equation*}
        \tder =
	\AxiomC{$\ \vdots\, \tdertwo$}
        \noLine
        \UnaryInfC{$\typctxtwo, \var \hastype \ptypetwo \vdash (\tmp, \overline\genv) \hastype \ptype$}
        \AxiomC{$\ \vdots\, \tderthree$}
        \noLine
        \UnaryInfC{$\typctxthree\vdash \itm \hastype \ptypetwo$}
         \RightLabel{\footnotesize$\EsAppend$}
         \BinaryInfC{$\typctxtwo \uplus \typctxthree \vdash (\tm, \overline\genv \appendOp \esub\var\itm) \hastype \ptype$}
        \DisplayProof
      \end{equation*}
      satisfies the statement, because $\size{\tder} = \size{\tdertwo} + \size\tderthree = \size{\tdertwo'} + 1 + \size\tderthree = \size{\tder'} + 1$.
\end{itemize}

\item \emph{Inert}, \ie $\prog = (\tm, \genv) \toin (\tmp, \genvtwo) = \progp$. By induction on the length $n$ of the environment $\genv$. Cases:
\begin{itemize}
  \item \emph{Empty environment}, \ie $n = 0$. This case is itself by induction on the evaluation context $\evctx$ in the step $\prog  = (\ctxp{(\la\var\tmtwo)\itm}, \emptyenv) \toin (\ctxp\tmtwo, \esub\var\itm) = \progp$. Cases of $\evctx$:
\begin{enumerate}
\item \emph{Step at the root}, \ie $\evctx = \ctxhole$ and $\prog  = ((\la\var\tmtwo)\itm, \emptyenv) \toin (\tmtwo, \esub\var\itm) = \progp$. Then, $\tder'$ has necessarily the following form:
      \begin{equation*}
        \tder' =
        \AxiomC{$\ \vdots\, \tdertwo$}
        \noLine
        \UnaryInfC{$\typctxtwo, \var \hastype \ptypetwo \vdash \tmtwo \hastype \ptype$}
        \RightLabel{\footnotesize$\EsCoerc$}
        \UnaryInfC{$\typctxtwo, \var \hastype \ptypetwo  \vdash (\tmtwo,\emptyenv) \hastype \ptype$}
        \AxiomC{$\ \vdots\, \tderthree$}
        \noLine
        \UnaryInfC{$\typctxthree \vdash \itm \hastype \ptypetwo$}
        \RightLabel{\footnotesize$\EsAppend$}
        \BinaryInfC{$\typctxtwo \uplus \typctxthree \vdash (\tmtwo, \esub\var\itm) \hastype \ptype$}
        \DisplayProof
      \end{equation*}
          with $\typctx = \typctxtwo \uplus \typctxthree$ and $\size{\tder'} = \size{\tdertwo} + \size\tderthree$. Then the following derivation:
      \begin{equation*}
        \tder \defeq
        \AxiomC{$\ \vdots\, \tdertwo$}
        \noLine
        \UnaryInfC{$\typctxtwo, \var \hastype \ptypetwo \vdash \tmtwo \hastype \ptype$}
        \RightLabel{\footnotesize$\lambda$}
        \UnaryInfC{$\typctxtwo \vdash \la\var\tmtwo \hastype [\Pair{\ptypetwo}{\ptype}]$}
        \AxiomC{$\ \vdots\, \tderthree$}
        \noLine
        \UnaryInfC{$\typctxthree \vdash \itm \hastype \ptypetwo$}
        \RightLabel{\footnotesize$@$}
        \BinaryInfC{$\typctx \vdash (\la\var\tmtwo)\itm \hastype \ptype$}
        \RightLabel{\footnotesize$\EsCoerc$}
        \UnaryInfC{$\typctxtwo \uplus \typctxthree \vdash ((\la\var\tmtwo)\itm,\emptyenv) \hastype \ptype$}
        \DisplayProof
      \end{equation*}
      where $\size{\tder} = \underbrace{\size{\tdertwo} + \size{\tderthree}}_{\tder'} + 1 = \size{\tder'} +1$. 
      
       \item \emph{Application Left}, \ie $\prog = (\tmtwo \fire, \emptyenv) \toin (\tmtwop \fire, \esub\var\itm)$ with $(\tmtwo, \emptyenv) \toin (\tmtwop, \esub\var\itm)$ for some $\var$ and inert term $\itm$ and with $\var \notin \fv\fire$, because $\var$ comes from an abstraction inside $\tmtwo$. Now, $\tdertwo'$ has necessarily the following form:
            \begin{equation*}
        \tder' =
        \AxiomC{$\ \vdots\, \tder'_1$}
        \noLine
	\UnaryInfC{$\typctx_1, \var\hastype \ptypethree \vdash \tmtwop \hastype \mset{\Pair{\ptypetwo}{\ptype}}$}
	\AxiomC{$\ \vdots\, \tder'_2$}
        \noLine
        \UnaryInfC{$\typctx_2 \vdash \fire \hastype \ptypetwo$}
        \RightLabel{\footnotesize$@$}
        \BinaryInfC{$\typctxtwo \uplus \typctx_2, \var\hastype \ptypethree \vdash \tmtwop \fire \hastype \ptype$}
         \RightLabel{\footnotesize$\EsCoerc$}
         \UnaryInfC{$\typctx_1 \uplus \typctx_2, \var\hastype \ptypethree  \vdash (\tmtwop \fire,\emptyenv) \hastype \ptype$}
         \AxiomC{$\ \vdots\, \tder'_3$}
        \noLine
        \UnaryInfC{$\typctx_3 \vdash \itm \hastype \ptypethree$}
         \RightLabel{\footnotesize$\EsAppend$}
         \BinaryInfC{$\typctx_1 \uplus \typctx_2 \uplus \typctx_3 \vdash (\tmtwop \fire,\esub\var\itm) \hastype \ptype$}
        \DisplayProof
      \end{equation*}
      with $\typctx = \typctx_1 \uplus \typctx_2 \uplus \typctx_3$ and $\size{\tder'} = \size{\tder'_1} + \size{\tder'_2} + \size{\tder'_3}$. Now, to apply the \ih, we have to first remove the application to $\fire$. Therefore, consider the derivation
            \begin{equation*}
        \tdertwo' =
        \AxiomC{$\ \vdots\, \tder'_1$}
        \noLine
	\UnaryInfC{$\typctx_1, \var\hastype \ptypethree \vdash \tmtwop \hastype \mset{\Pair \ptypetwo \ptype}$}
         \RightLabel{\footnotesize$\EsCoerc$}
         \UnaryInfC{$\typctx_1, \var\hastype \ptypethree  \vdash (\tmtwop,\emptyenv) \hastype \mset{\Pair \ptypetwo \ptype}$}
         \AxiomC{$\ \vdots\, \tder'_3$}
        \noLine
        \UnaryInfC{$\typctx_3 \vdash \itm \hastype \ptypethree$}
         \RightLabel{\footnotesize$\EsAppend$}
         \BinaryInfC{$\typctx_1  \uplus \typctx_3 \vdash (\tmtwo,\esub\var\itm) \hastype \mset{\Pair \ptypetwo \ptype}$}
        \DisplayProof
      \end{equation*}
      for which $\size{\tdertwo'} = \size{\tder'} - \size{\tder'_2} -1$. By hypothesis, $(\tmtwo, \emptyenv) \toin (\tmtwop, \esub\var\itm)$, and so we can apply the \ih ($\evctx$ is smaller), and obtain $\concl{\tdertwo}{\typctx_1  \uplus \typctx_3}{(\tmtwop, \esub\var\itm)}{\mset{\Pair \ptypetwo \ptype}}$ with $\size \tdertwo = \size {\tdertwo'} +1 = \size{\tder'} - \size{\tder'_2}$. Then the following derivation satisfies the statement:      
            \begin{equation*}
        \tder =
        \AxiomC{$\ \vdots\, \tdertwo$}
        \noLine
	\UnaryInfC{$\typctx_1 \uplus \typctx_3 \vdash \tmtwo \hastype [\Pair{\ptypetwo}{\ptype}]$}
        \AxiomC{$\ \vdots\, \tder'_2$}
        \noLine
        \UnaryInfC{$\typctx_2 \vdash \fire \hastype \ptypetwo$}
        \RightLabel{\footnotesize$@$}
        \BinaryInfC{$\typctx_1 \uplus \typctx_2 \uplus \typctx_3 \vdash \tmtwo \fire \hastype \ptype$}
        \RightLabel{\footnotesize$\EsCoerc$}
        \UnaryInfC{$\typctx_1 \uplus \typctx_2 \uplus \typctx_3 \vdash (\tmtwo \fire,\emptyenv) \hastype \ptype$}
        \DisplayProof
      \end{equation*}
      because $\size \tder = \size\tdertwo + \size{\tder_2'} +1 = \size{\tder'} - \size{\tder'_2} + \size{\tder_2'} +1 = \size{\tder'} + 1$.
      
\item \emph{Application right}, \ie $\prog = (\tmthree \tmtwo, \emptyenv) \toin (\tmthree \tmtwop, \esub\var\itm)$ with $(\tmtwo, \emptyenv) \toin (\tmtwop, \esub\var\itm)$ for some $\var$ and inert term $\itm$ and with $\var \notin \fv\tmthree$, because $\var$ comes from an abstraction inside $\tmtwo$. It is essentially identical to the previous case, the only difference is the switch of the left and right subterms. Note indeed that in the previous case the fact that the right subterm $\fire$ is a fireball is never used. 

\end{enumerate}

\item \emph{Non-empty environment}, \ie $n> 0$. Then $\genv = \overline\genv \appendOp \esub\var\itm$. Note that the existence of a redex does not depend on the environment, so that if $\prog = (\tm, \overline\genv \appendOp \esub\var\itm) \tof (\tmp, \genvtwo)$ then there exists $\overline\genv'$ such that $(\tm, \overline\genv) \tof (\tmp, \overline\genv' )$ and $\genvtwo = \overline\genv' \appendOp \esub\var\itm$. The type derivation $\tder'$ for $\progp = (\tmp, \overline\genv' \appendOp \esub\var\itm)$ has necessarily the following form:
    \begin{equation*}
        \tder' =
	\AxiomC{$\ \vdots\, \tdertwo'$}
        \noLine
        \UnaryInfC{$\typctxtwo, \var \hastype \ptypetwo \vdash (\tmp, \overline\genv') \hastype \ptype$}
        \AxiomC{$\ \vdots\, \tderthree$}
        \noLine
        \UnaryInfC{$\typctxthree\vdash \itm \hastype \ptypetwo$}
         \RightLabel{\footnotesize$\EsAppend$}
         \BinaryInfC{$\typctxtwo \uplus \typctxthree \vdash (\tmp, \overline\genv' \appendOp \esub\var\itm) \hastype \ptype$}
        \DisplayProof
      \end{equation*}
    with $\typctx = \typctxtwo \uplus \typctxthree$ and $\size{\tder'} = \size{\tdertwo'} + \size\tderthree$. By \ih applied to the step $(\tm, \overline\genv) \tobv (\tmp, \overline\genv')$, there exists a type derivation $\concl {\tdertwo} {\typctxtwo, \var \hastype \ptypetwo} {(\tm, \overline\genv)} \ptype$ such that $\size{\tdertwo} = \size{\tdertwo'} +1$. Then the type derivation:
    \begin{equation*}
        \tder =
	\AxiomC{$\ \vdots\, \tdertwo$}
        \noLine
        \UnaryInfC{$\typctxtwo, \var \hastype \ptypetwo \vdash (\tmp, \overline\genv) \hastype \ptype$}
        \AxiomC{$\ \vdots\, \tderthree$}
        \noLine
        \UnaryInfC{$\typctxthree\vdash \itm \hastype \ptypetwo$}
         \RightLabel{\footnotesize$\EsAppend$}
         \BinaryInfC{$\typctxtwo \uplus \typctxthree \vdash (\tm, \overline\genv \appendOp \esub\var\itm) \hastype \ptype$}
        \DisplayProof
      \end{equation*}
      satisfies the statement, because $\size{\tder} = \size{\tdertwo} + \size\tderthree = \size{\tdertwo'} + 1 + \size\tderthree = \size{\tder'} + 1$.
      \qedhere
\end{itemize}      
\end{itemize}
\end{proof}

%% file: proofs-normal-forms-tightNew.tex
\subsection{Omitted proofs and lemmas of \refsect{tight}}

\setcounter{lemmaAppendix}{\value{l:type-characterization-inert}}
\begin{lemmaAppendix}[Inert typing of inert terms]
	\label{lappendix:type-characterization-inert}
	\NoteState{l:type-characterization-inert}
	Let $\itm$ be a inert term. 
	For any inert multi type $\iptype$ there exists a \nonempty inert type derivation $\concl{\tder}{\typctx}{\itm}{\iptype}$.  
\end{lemmaAppendix}

\begin{proof}
	By induction on the definition of $\itm = \var \fire_1 \ldots \fire_n$, where the $\fire_i$'s are fireballs (for some $n > 0$).
	The base case is given when $n = 1$ and $\fire_1$ is a value.
	
	For all $1 \leq j \leq n$, if $\fire_j$ is a value then there is a type derivation $\concl{\tdertwo_j}{\typctxtwo_j}{\fire_j}{\emptymset}$ where $\Dom{\typctxtwo_j} = \emptyset$ (\ie $\typctxtwo_j$ is a inert type context), according to \reflemmap{value-typing}{judg}; 
	otherwise, $\fire_j$ is a inert term and hence, by \ih (since $\emptymset$ is a inert multi type), there is a type derivation $\concl{\tdertwo_j}{\typctxtwo_j}{\fire_j}{\zero}$ for some inert type context $ \typctxtwo_j$.
	
	For any $k \in \nat$, let 
	
	\vspace{-\baselineskip}
	\begin{equation*}
	\zero^{\iptype}_k \defeq \mset{\overbrace{\zero \multimap [\ldots \multimap [\emptymset}^{k \text{ times } \emptymset} \multimap \iptype]]},
	\end{equation*}
	which is a inert multi type.
	Then, there exists the following type derivation $\tder$:
		\begin{equation*}
		\tder \defeq 
		\AxiomC{}
		\RightLabel{\footnotesize$\ruleAx$}
		\UnaryInfC{$\var \hastype \zero^{\iptype}_n  \vdash \var \hastype \zero^{\iptype}_n$}
		\AxiomC{$\ \vdots\,\tdertwo_1$}
		\noLine
		\UnaryInfC{$\typctxtwo_1 \vdash \fire_1 \hastype \zero$}
		\RightLabel{\footnotesize$\ruleAp$}
		\BinaryInfC{$\var \hastype  \zero^{\iptype}_n,  \typctxtwo_1  \vdash \var \fire_1 \hastype  \zero^{\iptype}_{n-1}$}
		\noLine
		\UnaryInfC{$\vdots$}
		\AxiomC{$\ \vdots\,\tdertwo_n$}
		\noLine
		\UnaryInfC{$\typctxtwo_n \vdash \fire_n \hastype \emptymset$}
		\RightLabel{\footnotesize$@$}
		\BinaryInfC{$\var \hastype \zero^{\iptype}_n, \, \biguplus_{j=1}^n\typctxtwo_j  \vdash \var \fire_1 \ldots \fire_n \hastype \iptype$}
		\DisplayProof
		\end{equation*}
	
	Since inert type contexts are closed under summation, $\typctx \defeq \var \hastype \zero^{\iptype}_n, \, \biguplus_{j=1}^n\typctxtwo_i $ is a inert type context; $\typctx$ is non-empty because $n > 0$ and so $\zero^{\iptype}_n \neq \emptymset$.
	Therefore, $\tder$ is a \nonempty inert type derivation.
\end{proof}

\begin{lemma}[Normal programs are tightly typable]
	\label{l:tight-characterisation-nfs}
	Let $\prog$ be a normal program. 
	Then there exists a tight derivation $\concl{\tder}{\typctx}{\prog}{\emptymset}$. 
	Moreover, $\prog$ is a coerced value if and only if $\tder$ is \empt.
\end{lemma}

\begin{proof}
	By harmony (\refprop{harmony-split}), $\prog = (\fire,\genv)$ for some fireball $\fire$ and some environment $\genv \defeq \esub{\var_1}{\itm_1}\dots \esub{\var_n}{\itm_n}$, with $n \geq 0$.
	Since fireballs are tightly typable (\refcoro{tight-characterisation-nfs}), there is a tight derivation $\concl{\tdertwo}{\typctxtwo, \var_1 \hastype \iptype_1, \dots, \var_n \hastype \iptype_n}{\fire}{\emptymset}$ and moreover $\tdertwo$ is \empt if and only if $\fire$ is a value.
	
	If $n = 0$ then $\genv = \emptyenv$ and $\concl{\tdertwo}{\typctxtwo}{\fire}{\emptymset}$, so we have the tight derivation
	\begin{equation*}
		\tder \defeq 
		\AxiomC{$\ \vdots\,\tdertwo$}
		\noLine
		\UnaryInfC{$ \typctxtwo \vdash \fire \hastype \emptymset$}
		\RightLabel{\footnotesize$\EsCoerc$}
		\UnaryInfC{$ \typctxtwo \vdash (\fire, \emptyenv) \hastype \emptymset$}
		\DisplayProof
	\end{equation*}
	which is empty if and only if $\tdertwo$ is so if and only if $\prog = (\fire, \emptyenv)$ is a coerced value.
	
	Otherwise $n > 0$ and $p = (\fire, \genv)$ is not a coerced value.
	By \reflemma{type-characterization-inert}, for all $1 \leq j \leq n$, for any inert  multi type $\iptype_j \uplus \biguplus_{k=1}^{j-1} \iptypetwo_{j,k}$ there is a \nonempty inert derivation $\concl{\tdertwo_j}{\typctxtwo_j, \var_{j+1} \hastype \iptypetwo_{j+1,j}, \dots, \var_n \hastype \iptypetwo_{n,j}}{\itm_j}{\iptype_j \uplus \biguplus_{k=1}^{j-1} \iptypetwo_{j,k}} $.  
	Thus, we have the following type derivation $\tder$
	
	{\footnotesize
		\begin{equation*}
		\AxiomC{$\ \vdots\,\tdertwo$}
		\noLine
		\UnaryInfC{$\typctxtwo, \var_1 \hastype \iptype_1, \dots, \var_n \hastype \iptype_n \vdash \fire \hastype \emptymset$}
		\RightLabel{\footnotesize$\EsCoerc$}
		\UnaryInfC{$\typctxtwo, \var_1 \hastype \iptype_1, \dots, \var_n \hastype \iptype_n  \vdash (\fire,\emptyenv) \hastype \emptymset$}
		\AxiomC{$\ \vdots\,\tdertwo_1$}
		\noLine
		\UnaryInfC{$\typctxtwo_1, \var_2 \hastype \iptypetwo_{2,1}, \dots, \var_n \hastype \iptypetwo_{n,1} \vdash \itm_1 \hastype \iptype_1$}
		\RightLabel{\footnotesize$\EsAppend$}
		\BinaryInfC{$\typctxtwo \uplus \typctxtwo_1, \var_2 \hastype \iptype_2 \uplus \iptypetwo_{2,1}, \dots, \var_n \hastype \iptype_n \uplus \iptypetwo_{n,1}  \vdash (\fire,\esub{\var_1}{\itm_1}) \hastype \emptymset$}
		\noLine
		\UnaryInfC{$\vdots$}
		\AxiomC{$\ \vdots\,\tdertwo_n$}
		\noLine
		\UnaryInfC{$\typctxtwo_n \vdash \itm_n \hastype \iptype_n \uplus \biguplus_{k=1}^{n-1} \iptypetwo_{n,k}$}
		\RightLabel{\footnotesize$\EsAppend$}
		\BinaryInfC{$\typctxtwo \uplus \, \biguplus_{j=1}^n\typctxtwo_j  \vdash (\fire,\esub{\var_1}{\itm_1} \ldots \esub{\var_n}{\itm_n}) \hastype \emptymset$}
		\DisplayProof
		\end{equation*}
	}

	\noindent	which is tight because inert type contexts are closed under summation.
	Moreover, $\tder$ is \nonempty because $n > 0$ and so $\typctxtwo \uplus \, \biguplus_{j=1}^n\typctxtwo_j$ is a non-empty type context.
\end{proof}

%
%
%

\setcounter{propositionAppendix}{\value{prop:tight-characterization-nfs}}
\begin{propositionAppendix}[Normal expressions are tightly typable]
	\label{propappendix:tight-characterization-nfs}
	\NoteState{prop:tight-characterization-nfs}
	Let $\expr$ be a normal expression. 
	Then there exists a tight derivation $\concl{\tder}{\typctx}{\expr}{\zero}$. 
	Moreover, 
	$\expr$ is a value or a coerced value if and only if $\tder$ is \empt.
\end{propositionAppendix}

\begin{proof}
	By harmony (\refprop{harmony-split}), the normal expression $\expr$ is either a fireball $\fire$ or a program of the form $(\fire,\genv)$.
	In the first case, we conclude by \refcor{tight-characterisation-nfs}.
	In the second case, we are done by \reflemma{tight-characterisation-nfs}.
\end{proof}

\setcounter{lemmaAppendix}{\value{l:inert-ctx-imply-inert-type}}
\begin{lemmaAppendix}[Inert spreading on inert terms]
	\label{lappendix:inert-ctx-imply-inert-type}
	\NoteState{l:inert-ctx-imply-inert-type}
	Let $\concl{\tder}{\typctx}{\itm}{\ptype}$ be a type derivation for a inert term $\itm$. If $\typctx$ is a inert type context then $\ptype$ 
	and $\tder$ are inert.
\end{lemmaAppendix}

\begin{proof}
	In order to get the right \ih, we prove the same statement with a more general hypothesis: $\itm$ is either a inert term or a variable. 
	Note that terms that are inert or variables can be defined inductively by the following grammar:
	\begin{align*}
		\itm_\textup{var} &\grameq \var \mid \itm_\textup{var} \fire \qquad \textup{(where $\fire$ is a usual fireball)}
	\end{align*}
	 
	The proof of the statement is by induction on the structure of $\itm_\textup{var}$.
	Note that the fact $\ptype$ is a inert multi type implies that $\tder$ is a inert type derivation.
	
	If $\itm_\textup{var} = \var$ then $\tder$ has the form
	\begin{equation*}
		\AxiomC{}
		\RightLabel{\footnotesize$\ruleAx$}
		\UnaryInfC{$\var \hastype \ptype \vdash \var \hastype \ptype$}
		\DisplayProof
	\end{equation*}
	thus $\typctx = \var \hastype \ptype$ and so $\ptype$ is a inert multi type because $\typctx$ is a inert type context. 
	
	If $\itm_\textup{var} = \itmtwo_\textup{var} \fire$ then $\tder$ has the form:
	\begin{equation*}
	\AxiomC{$\ \vdots\, \tdertwo$}
	\noLine
	\UnaryInfC{$\typctxthree \vdash \itmtwo_\textup{var} \hastype \mset{\Pair{\ptypetwo}{\ptype}}$}
	\AxiomC{$\ \vdots\, \tderthree$}
	\noLine
	\UnaryInfC{$\typctxtwo\vdash \fire \hastype \ptypetwo$}
	\RightLabel{\footnotesize$@$}
	\BinaryInfC{$\typctxthree \uplus \typctxtwo \vdash \itmtwo_\textup{var}\fire \hastype \ptype$}
	\DisplayProof
	\end{equation*}
	with $\typctx = \typctxthree \uplus \typctxtwo$. 
	Then $\typctxthree$ is inert because $\typctx$ is inert. 
	By applying the \ih to $\tdertwo$ we obtain that $\mset{\Pair{\ptypetwo}{\ptype}}$ is inert, that implies that $\ptypetwo = \emptymset$ and $\ptype$ is inert.
\end{proof}

\setcounter{lemmaAppendix}{\value{l:size-normal-inert}}
\begin{lemmaAppendix}[Inert derivations are minimal and provide the exact size of inert terms]
	\label{lappendix:size-normal-inert} 
	\NoteState{l:size-normal-inert}
	Let $\concl{\tder}{\typctx}{\itm}{\iptype}$ be a inert type derivation for a inert term $\itm$. 
	Then $\sizeap\itm = \sizeap\tder$ and $\size{\tder}$ is minimal among the type derivations of $\itm$.
\end{lemmaAppendix}

\begin{proof}
	Minimality of $\size{\tder}$ follows from the fact that $\sizeap{\itm} = \sizeap{\tder}$ (to be proved) and \refprop{size-normal}, since $\itm$ is a normal expression by harmony (\refprop{harmony-split}).
	We prove that $\sizeap{\itm} = \sizeap{\tder}$ by induction on the structure of $\itm = \var \fire_1 \ldots \fire_n$, where the $\fire_i$'s are fireballs (for some $n > 0$).
	The base case is given when $n = 1$ and $\fire_1$ is a value.
	
	Since $\typctx$ is a inert type context, by necessity $\tder$ has the following form
	\begin{equation*}
	\tder \defeq 
	\AxiomC{}
	\RightLabel{\footnotesize$\ruleAx$}
	\UnaryInfC{$\var \hastype \zero^{\iptype}_n  \vdash \var \hastype \zero^{\iptype}_n$}
	\AxiomC{$\ \vdots\,\tdertwo_1$}
	\noLine
	\UnaryInfC{$\var \hastype \iptypetwo_1, \typctxtwo_1 \vdash \fire_1 \hastype \zero$}
	\RightLabel{\footnotesize$\ruleAp$}
	\BinaryInfC{$\var \hastype  \zero^{\iptype}_n  \uplus \iptypetwo_1,  \typctxtwo_1  \vdash \var \fire_1 \hastype  \zero^{\iptype}_{n-1}$}
	\noLine
	\UnaryInfC{$\vdots$}
	\AxiomC{$\ \vdots\,\tdertwo_n$}
	\noLine
	\UnaryInfC{$\var \hastype \iptypetwo_n, \typctxtwo_n \vdash \fire_n \hastype \emptymset$}
	\RightLabel{\footnotesize$@$}
	\BinaryInfC{$\var \hastype \zero^{\iptype}_n  \uplus \biguplus_{j=1}^n \iptypetwo_j, \, \biguplus_{j=1}^n\typctxtwo_j  \vdash \var \fire_1 \ldots \fire_n \hastype \iptype$}
	\DisplayProof
	\end{equation*}
	
	\noindent where $\typctx \defeq \var \hastype \zero^{\iptype}_n  \uplus \biguplus_{j=1}^n \iptypetwo_j, \, \biguplus_{j=1}^n\typctxtwo_j$ (with $\zero^{\iptype}_n$ and $\iptypetwo_j$ inert multi types and $\typctx_j$ inert type contexts, for all $1 \leq j \leq n$) and, for any $k \in \nat$, 
		
	\vspace{-.5\baselineskip}
	\begin{equation*}
		\zero^{\iptype}_k \defeq \mset{\overbrace{\zero \multimap [\ldots \multimap [\emptymset}^{k \text{ times } \emptymset} \multimap \iptype]]}.
	\end{equation*}
	
	For all $1 \leq j \leq n$, if $\fire_j$ is a value then  $\size{\tdertwo_j} = 0 = \size{\fire_j}$, according to \reflemmap{value-typing}{empty}; 
	otherwise, $\fire_j$ is a inert term and hence $\size{\tdertwo_j} = \size{\fire_j}$ by \ih (since $\emptymset$ is a inert multi type and $\var \hastype \iptypetwo_j, \typctxtwo_j$ is a inert type context).
	
	Therefore, $\sizeap{\itm} = n + \sum_{j=1}^n \sizeap{\fire_j} = n + \sum_{j=1}^n \sizeap{\tdertwo_j} = \sizeap{\tder}$.
\end{proof}

\setcounter{lemmaAppendix}{\value{l:size-normal-tight}}
\begin{lemmaAppendix}[Tight derivations are minimal and provide the exact size of normal forms]
	\label{lappendix:size-normal-tight} 
	\NoteState{l:size-normal-tight}
	Let $\concl{\tder}{\typctx}{\expr}{\emptymset}$ be a tight derivation and $\expr$ be a normal 
	expression. 
	Then $\sizeap\expr = \sizeap\tder$ and $\size{\tder}$ is minimal among the type derivations of $\expr$.
\end{lemmaAppendix}

\begin{proof}
	Minimality of $\size{\tder}$ follows from the fact that $\sizeap{\expr} = \sizeap{\tder}$ (to be proved) and \refprop{size-normal}.
	We prove that $\sizeap{\expr} = \sizeap{\tder}$ according to the cases for the normal expression $\expr$.
	By harmony (\refprop{harmony-split}), $\expr$ is either a fireball (\ie a value or a inert term) or a program of the form $(\fire,\genv)$.
	\begin{itemize}
		\item \emph{$\expr$ is a value}. 
		Then the statement follows from \reflemmap{value-typing}{empty}.
		
		\item \emph{$\expr$ is a inert term}. Then the statement follows from \reflemma{size-normal-inert}, since $\tder$ is a inert type derivation.
		
		\item \emph{$\expr$ is a program of the form $\prog = (\fire,\genv)$}. By induction on the length $n$ of the environment $\genv$. 
		If $n=0$ then $\prog = (\fire, \emptyenv)$ and we apply the case for fireballs. Otherwise $n>0$ and $\prog = (\fire, \genv\appendOp\esub\var\itm)$. Then $\tder$ has the following form:
		\begin{equation*}
		\AxiomC{$\ \vdots\, \tdertwo$}
		\noLine
		\UnaryInfC{$\typctxthree, \var \hastype \ptypetwo \vdash (\fire, \genv) \hastype \emptymset$}
		\AxiomC{$\ \vdots\, \tderthree$}
		\noLine
		\UnaryInfC{$\typctxtwo\vdash \itm \hastype \ptypetwo$}
		\RightLabel{\footnotesize$\EsAppend$}
		\BinaryInfC{$\typctxthree \uplus \typctxtwo \vdash (\fire, \genv \appendOp \esub\var\itm) \hastype \emptymset$}
		\DisplayProof
		\end{equation*}
		with $\typctx = \typctxthree \uplus \typctxtwo$. 
		Note that $\typctxtwo$ and $\typctxthree$ are inert because $\typctx$ is inert. By \reflemma{inert-ctx-imply-inert-type}, $\ptypetwo$ is inert, and so $\tdertwo$ is tight and $\tderthree$ is inert. 
		Then $\sizeap{(\fire, \genv)} = \sizeap\tdertwo$ by \ih and $\sizeap \itm = \sizeap\tderthree$ by \reflemma{size-normal-inert}, and hence $\sizeap \expr = \sizeap{(\fire, \genv \appendOp \esub\var\itm)} = \sizeap{(\fire, \genv)} + \sizeap \itm = \sizeap\tdertwo + \sizeap\tderthree = \sizeap\tder$.
		\qedhere
	\end{itemize}
\end{proof}

\setcounter{propositionAppendix}{\value{prop:tight-types-bound-nfs}}
\begin{propositionAppendix}[Tight derivations give the exact size of normal forms]
	\label{propappendix:tight-types-bound-nfs}
	\NoteState{prop:tight-types-bound-nfs}
	Let $\expr$ be a normal expression and $\concl{\tder}{\typctx}{\expr}{\zero}$ be a tight derivation. 
	Then $\sizeap{\expr} = \size{\typelist\typctx}$.
\end{propositionAppendix}

\begin{proof}
	In order to have the right \ih, we prove a stronger statement:
	Let $\expr$ be a normal expression and $\concl{\tder}{\typctx}{\expr}{\ptype}$ be a inert type derivation.
	\begin{itemize}
		\item If $\ptype = \emptymset$ (\ie $\tder$ is a tight type derivation), then $\sizeap{\expr} = \size{\typelist\typctx}$.
		\item If $\expr$ is an inert term, then $\sizeap{\expr} + \size{\ptype} = \size{\typelist\typctx}$.
	\end{itemize}
	
	In the proof of this stronger statement there are only two cases by harmony (\refprop{harmony-split}), since $\expr$ is a normal expression: either $\expr$ is a fireball or $\expr$ is a program of the form $(\fire,\genv)$.
	
	\begin{itemize}
		\item \emph{$\expr$ is a fireball $\fire$}. By induction on the definition of fireballs. 
		If $\fire$ is a value then $\domain{\typctx} = \emptyset$ according to 
		\reflemmap{value-typing}{empty}, 
		hence $\sizeap{\expr} = 0 = \size{\typelist\typctx}$. 
		Otherwise $\fire$ is a inert term of the form $\var \fire_1 \ldots \fire_n$ ($n > 0$). 
		Any tight type derivation $\concl{\tder}{\typctx}{\var \fire_1 \ldots \fire_n}{\ptype}$ is constructed as follows, where $\ptypetwo_1 \defeq \ptype$, $\ptypetwo_{i+1} \defeq \mset{\Pair{\emptymset}{\ptypetwo_i}}$ ($1 \leq i \leq n$) and $\typctx = \var \hastype \ptypetwo_{n+1}  \uplus (\biguplus_{i=1}^n\typctxtwo_i)$ (note that the tightness of $\typctx$ has many implications!):
		\begin{equation*}
		\tder \defeq 
		\AxiomC{}
		\RightLabel{\footnotesize$\ruleAx$}
		\UnaryInfC{$\var \hastype \ptypetwo_{n+1}  \vdash \var \hastype \ptypetwo_{n+1}$}
		\AxiomC{$\ \vdots\,\tdertwo_1$}
		\noLine
		\UnaryInfC{$\typctxtwo_1 \vdash \fire_1 \hastype \emptymset$}
		\RightLabel{\footnotesize$\ruleAp$}
		\BinaryInfC{$\var \hastype  \ptypetwo_{n+1}  \uplus \typctxtwo_1  \vdash \var \fire_1 \hastype  \ptypetwo_n$}
		\noLine
		\UnaryInfC{$\vdots$}
		\AxiomC{$\ \vdots\,\tdertwo_n$}
		\noLine
		\UnaryInfC{$\typctxtwo_n \vdash \fire_n \hastype \emptymset$}
		\RightLabel{\footnotesize$@$}
		\BinaryInfC{$\var \hastype \ptypetwo_{n+1}  \uplus (\biguplus_{i=1}^n\typctxtwo_i ) \vdash \var \fire_1 \ldots \fire_n \hastype \ptypetwo_1$}
		\DisplayProof
		\end{equation*}
		By \ih, $\sizeap{\fire_i} = \size{\typelist\typctxtwo_i}$.
		Note that $\size{\ptypetwo_{n+1}} = n + \size{\ptype}$. 
		Therefore, $\sizeap{\var \fire_1\dots\fire_n} + \size{\ptypetwo_1} = n + \sum_i^n\sizeap{\fire_i} + \size{\ptypetwo_1} = n + \sum_i^n \size{\typelist\typctxtwo_i} + \size{\ptypetwo_1} =\size{\typelist{\var \hastype \ptypetwo_{n+1}  \uplus (\biguplus_{i=1}^n\typctxtwo_i )}}$.
		
		\item \emph{$\expr$ is a program of the form $\prog = (\fire,\genv)$}. 
		By induction on the length $n$ of the environment $\genv$. 
		If $n=0$ then $\prog = (\fire, \emptyenv)$ and the last rule of any type derivation for $\prog$ is $\EsCoerc$ (which preserves types), thus we can apply the case for fireballs, since $\sizeap{(\fire, \emptyenv)} = \sizeap{\fire}$. 
		Otherwise $n>0$ and $\prog = (\fire, \genv\appendOp\esub\var\itm)$. 
		Any tight type derivation $\concl{\tder}{\typctx}{\prog}{\emptymset}$ is then constructed as follows, where $\typctx = \typctxthree \uplus \typctxtwo$ is a inert type context (in particular, $\typctxtwo$ and $\typctxthree$ are so):
		\begin{equation*}
		\tder \defeq
		\AxiomC{$\ \vdots\, \tderthree$}
		\noLine
		\UnaryInfC{$\typctxthree, \var \hastype \ptypetwo \vdash (\fire, \genv) \hastype \emptymset$}
		\AxiomC{$\ \vdots\, \tdertwo$}
		\noLine
		\UnaryInfC{$\typctxtwo\vdash \itm \hastype \ptypetwo$}
		\RightLabel{\footnotesize$\EsAppend$}
		\BinaryInfC{$\typctxthree \uplus \typctxtwo \vdash (\fire, \genv \appendOp \esub\var\itm) \hastype \emptymset$}
		\DisplayProof
		\end{equation*}
		Since $\typctxtwo$ is inert, then $\ptypetwo$ is inert by \reflemma{inert-ctx-imply-inert-type} applied to $\tdertwo$, hence $\tdertwo$ is inert and $\tderthree$ is tight.
		By \ih, $\sizeap{(\fire,\genv)} = 
		\size{\typelist{\typctxthree, \var \hastype \ptypetwo}}$ and $\sizeap{\itm} + \size{\ptypetwo} = \size{\typelist\typctxtwo}$. 
		Therefore, $\sizeap{(\fire, \genv \appendOp \esub\var\itm)} = \sizeap{(\fire, \genv)} + \sizeap{\itm} = \size{\typelist{\typctxthree, \var \hastype \ptypetwo}} + \size{\typelist\typctxtwo} - \size{\ptypetwo} = \size{\typelist{\typctxthree \uplus \typctxtwo}}$.
		\qedhere
	\end{itemize}
\end{proof}

\setcounter{theoremAppendix}{\value{thm:tight-correctness}}
\begin{theoremAppendix}[Tight correctness]
	\label{thmappendix:tight-correctness}
	\NoteState{thm:tight-correctness}
	Let $\concl{\tder}{\typctx}{\prog}{\emptymset}$ be a tight type derivation. Then there 
	is a normalizing evaluation $\deriv \colon \prog \tof^* \progtwo$ with 
	$    \sizeap\tder = \size\deriv + \sizeap\progtwo = \size\deriv + \sizeap{\typelist\typctx}$.
	In particular, if $\domain{\typctx} = \emptyset$, then $\size{\tder} = \size{\deriv}$ and $\progtwo$ is a coerced value.
\end{theoremAppendix}

\begin{proof} 
	By induction on $\sizeap\tder$. 
	If $\prog$ is normal then the statement holds with $\progtwo \defeq \prog$ and $\deriv$ the empty derivation ($\size{\deriv} = 0$), because $\tder$ and hence $\typctx$ are tight, so $\sizeap{\progtwo} = \sizeap{\tder}$ by \reflemma{size-normal-tight} and $\sizeap{\progtwo} = \sizeap{\typelist\typctx}$ by \refprop{tight-types-bound-nfs}. 
	Otherwise $\prog \tof \progthree$ and by quantitative subject reduction (\refprop{quant-subject-reduction}) there is a tight derivation $\concl{\tdertwo}{\typctx}{\progthree}{\emptymset}$ such that $\sizeap\tdertwo = \sizeap\tder -1$. 
	By \ih, there exist a noral program $\progtwo$ and an evaluation $\deriv' \colon \progthree \tof^* \progtwo$ such that $\size{\deriv'} + \sizeap\progtwo = \sizeap\tdertwo = \size{\deriv'} + \sizeap{\typelist\typctx}$ (and so $\sizeap\progtwo = \size{\typelist\typctx}$).
	Therefore, the evaluation $\deriv \colon \prog  \tof^* \progtwo$ obtained by prefixing $\deriv'$ with the step $\prog \tof \progthree$ verifies $\size\deriv + \sizeap\progtwo = \size{\deriv'} + 1 + \sizeap\progtwo = \sizeap\tdertwo + 1 = \sizeap\tder$ and $\sizeap\tder  = \size{\deriv} + \size{\typelist\typctx}$.
	
	If moreover $\domain{\typctx} = \emptyset$, then $\progtwo$ is a coerced value by \refprop{tight-characterization-nfs}, so $\size{\progtwo} = 0$.
\end{proof}

\setcounter{theoremAppendix}{\value{thm:tight-completeness}}
\begin{theoremAppendix}[Tight completeness]
	\label{thmappendix:tight-completeness}
	\NoteState{thm:tight-completeness}
	Let $\deriv \colon \prog \tof^* \progtwo$ be a normalizing evaluation. 
	Then there is a tight type derivation $\concl{\tder}{\typctx}{\prog}{\emptymset}$ with 
	$\sizeap\tder = \size\deriv + \sizeap{\progtwo} = \size{\deriv} + \size{\typelist\typctx}$.
	In particular, if $\progtwo$ is a coerced value, then $\size{\tder} = \size{\deriv}$ and $\domain{\typctx} = \emptyset$. 
\end{theoremAppendix}

\begin{proof}
	By induction on the length $\size\deriv$ of the evaluation $\deriv$. 
	If $\size\deriv = 0$ then $\prog = \progtwo$, so the existence of $\concl{\tder}{\typctx}{\prog}{\emptymset}$ with $\typctx$ tight is given by the fact that normal programs are tightly typable (\refprop{tight-characterization-nfs}), and the equalities on sizes are given by \reflemma{size-normal-tight} and \refprop{tight-types-bound-nfs}. 
	If $\size\deriv  =k + 1$ then $\prog \tof \progthree \tof^k \progtwo$ and by \ih 
	there is a tight derivation $\concl{\tdertwo}{\typctx}{\progthree}{\emptymset}$ satisfying $k + \sizeap\progtwo = \sizeap\tdertwo = k + \size{\typelist\typctx}$ (and so $\sizeap\progtwo = \size{\typelist\typctx}$). 
	By quantitative subject expansion (\refprop{quant-subject-expansion}), there is $\concl{\tder}{\typctx}{\prog}{\emptymset}$ 
	with $\sizeap\tder = \sizeap\tdertwo +1$. 
	Therefore, $\size\deriv + \sizeap\progtwo = k + 1 + \sizeap\progtwo  =_{\ih} \sizeap\tdertwo + 1 = \sizeap\tder$ and $\sizeap\tder  = \size{\deriv} + \size{\typelist\typctx}$.
	
	If moreover $\progtwo$ is a coerced value then $\size{\progtwo} = 0$ , and $\domain{\typctx} = \emptyset$ by \refprop{tight-characterization-nfs}.
\end{proof}

%% file: mainLong.bbl
\begin{thebibliography}{10}
\providecommand{\url}[1]{\texttt{#1}}
\providecommand{\urlprefix}{URL }
\providecommand{\doi}[1]{https://doi.org/#1}

\bibitem{DBLP:journals/tcs/Accattoli15}
Accattoli, B.: {Proof nets and the call-by-value {\(\lambda\)}-calculus}.
  Theor. Comput. Sci.  \textbf{606},  2--24 (2015)

\bibitem{AccattoliLengrandKesner18}
Accattoli, B., Graham-Lengrand, S., Kesner, D.: Tight typings and split bounds.
  In: ICFP 2018 (2018), to appear

\bibitem{DBLP:conf/aplas/AccattoliG16}
Accattoli, B., Guerrieri, G.: {Open Call-by-Value}. In: {{APLAS} 2016}. pp.
  206--226 (2016)

\bibitem{AccattoliGuerrieri17b}
Accattoli, B., Guerrieri, G.: {Implementing Open Call-by-Value}. In: {FSEN}
  2017. pp. 1--19 (2017)

\bibitem{AccattoliGuerrieri18}
Accattoli, B., Guerrieri, G.: {Types of Fireballs (Extended Version)}. CoRR
  \textbf{abs/1808.10389} (2018)

\bibitem{DBLP:conf/flops/AccattoliP12}
Accattoli, B., Paolini, L.: {Call-by-Value Solvability, revisited}. In:
  {FLOPS}. pp. 4--16 (2012)

\bibitem{fireballs}
Accattoli, B., {Sacerdoti Coen}, C.: {On the Relative Usefulness of Fireballs}.
  In: {{LICS} 2015}. pp. 141--155 (2015)

\bibitem{Bernadet-Lengrand2013}
Bernadet, A., Graham-Lengrand, S.: Non-idempotent intersection types and strong
  normalisation. Logical Methods in Computer Science  \textbf{9}(4) (2013)

\bibitem{DBLP:journals/tcs/BonoVB08}
Bono, V., Venneri, B., Bettini, L.: A typed lambda calculus with intersection
  types. Theor. Comput. Sci.  \textbf{398}(1-3),  95--113 (2008)

\bibitem{DBLP:journals/apal/BucciarelliE01}
Bucciarelli, A., Ehrhard, T.: On phase semantics and denotational semantics:
  the exponentials. Ann. Pure Appl. Logic  \textbf{109}(3),  205--241 (2001)

\bibitem{DBLP:journals/apal/BucciarelliEM12}
Bucciarelli, A., Ehrhard, T., Manzonetto, G.: A relational semantics for
  parallelism and non-determinism in a functional setting. Ann. Pure Appl.
  Logic  \textbf{163}(7),  918--934 (2012)

\bibitem{BKV17}
Bucciarelli, A., Kesner, D., Ventura, D.: Non-idempotent intersection types for
  the lambda-calculus. Logic Journal of the IGPL  \textbf{25}(4),  431--464
  (2017)

\bibitem{DBLP:conf/fossacs/CarraroG14}
Carraro, A., Guerrieri, G.: {A Semantical and Operational Account of
  Call-by-Value Solvability}. In: {{FOSSACS} 2014}. pp. 103--118 (2014)

\bibitem{Carvalho07}
de~Carvalho, D.: S\'emantiques de la logique lin\'eaire et temps de calcul.
  {T}h\`ese de doctorat, Universit\'e Aix-Marseille II (2007)

\bibitem{DBLP:conf/csl/Carvalho16}
de~Carvalho, D.: The relational model is injective for multiplicative
  exponential linear logic. In: {CSL} 2016. pp. 41:1--41:19 (2016)

\bibitem{deCarvalho18}
de~Carvalho, D.: Execution time of {\(\lambda\)}-terms via denotational
  semantics and intersection types. Math. Str. in Comput. Sci.  \textbf{28}(7),
   1169--1203 (2018)

\bibitem{DBLP:journals/tcs/CarvalhoPF11}
de~Carvalho, D., Pagani, M., {Tortora de Falco}, L.: {A semantic measure of the
  execution time in linear logic}. Theor. Comput. Sci.  \textbf{412}(20),
  1884--1902 (2011)

\bibitem{DBLP:journals/iandc/CarvalhoF16}
de~Carvalho, D., {Tortora de Falco}, L.: A semantic account of strong
  normalization in linear logic. Inf. Comput.  \textbf{248},  104--129 (2016)

\bibitem{DBLP:journals/aml/CoppoD78}
Coppo, M., Dezani{-}Ciancaglini, M.: A new type assignment for
  {\(\lambda\)}-terms. Arch. Math. Log.  \textbf{19}(1),  139--156 (1978)

\bibitem{DBLP:journals/ndjfl/CoppoD80}
Coppo, M., Dezani{-}Ciancaglini, M.: An extension of the basic functionality
  theory for the {\(\lambda\)}-calculus. Notre Dame Journal of Formal Logic
  \textbf{21}(4),  685--693 (1980)

\bibitem{DBLP:conf/icfp/CurienH00}
Curien, P.L., Herbelin, H.: {The duality of computation}. In: {ICFP}. pp.
  233--243 (2000)

\bibitem{DBLP:conf/lfcs/Diaz-CaroMP13}
D{\'{\i}}az{-}Caro, A., Manzonetto, G., Pagani, M.: Call-by-value
  non-determinism in a linear logic type discipline. In: {LFCS} 2013. pp.
  164--178 (2013)

\bibitem{DBLP:conf/csl/Ehrhard12}
Ehrhard, T.: {Collapsing non-idempotent intersection types}. In: {CSL}. pp.
  259--273 (2012)

\bibitem{DBLP:conf/ppdp/EhrhardG16}
Ehrhard, T., Guerrieri, G.: The bang calculus: an untyped lambda-calculus
  generalizing call-by-name and call-by-value. In: PPDP 2016. pp. 174--187.
  {ACM} (2016)

\bibitem{DBLP:conf/lics/FioreP94}
Fiore, M.P., Plotkin, G.D.: An axiomatization of computationally adequate
  domain theoretic models of {FPC}. In: {LICS} '94. pp. 92--102 (1994)

\bibitem{DBLP:conf/tacs/Gardner94}
Gardner, P.: Discovering needed reductions using type theory. In: TACS '94.
  Lecture Notes in Computer Science, vol.~789, pp. 555--574. Springer (1994)

\bibitem{DBLP:journals/tcs/Girard87}
Girard, J.Y.: {Linear Logic}. Theoretical Computer Science  \textbf{50},
  1--102 (1987)

\bibitem{Girard88}
Girard, J.Y.: Normal functors, power series and the $\lambda$-calculus. Annals
  of Pure and Applied Logic  \textbf{37},  129–177 (1988)

\bibitem{DBLP:conf/icfp/GregoireL02}
Gr{\'e}goire, B., Leroy, X.: {A compiled implementation of strong reduction}.
  In: {{ICFP} '02}. pp. 235--246 (2002)

\bibitem{Guerrieri15}
Guerrieri, G.: {Head reduction and normalization in a call-by-value
  lambda-calculus}. In: {{WPTE} 2015}. pp. 3--17 (2015)

\bibitem{Guerrieri18}
Guerrieri, G.: Towards a semantic measure of the execution time in
  call-by-value lambda-calculus. Tech. rep. (2018), submitted to ITRS 2018

\bibitem{GuerrieriPR15}
Guerrieri, G., Paolini, L., {Ronchi Della Rocca}, S.: {Standardization of a
  Call-By-Value Lambda-Calculus}. In: {{TLCA} 2015}. pp. 211--225 (2015)

\bibitem{DBLP:journals/lmcs/GuerrieriPR17}
Guerrieri, G., Paolini, L., {Ronchi Della Rocca}, S.: Standardization and
  conservativity of a refined call-by-value lambda-calculus. Logical Methods in
  Computer Science  \textbf{13}(4) (2017)

\bibitem{GuerrieriPellissierTortora16}
Guerrieri, G., Pellissier, L., Tortora~de Falco, L.: {Computing Connected
  Proof(-Structure)s from their Taylor Expansion}. In: {FSCD} 2016. pp.
  20:1--20:18 (2016)

\bibitem{DBLP:journals/tcs/HondaY99}
Honda, K., Yoshida, N.: Game-theoretic analysis of call-by-value computation.
  Theor. Comput. Sci.  \textbf{221}(1-2),  393--456 (1999)

\bibitem{DBLP:conf/rta/KesnerV17}
Kesner, D., Vial, P.: Types as resources for classical natural deduction. In:
  {FSCD} 2017. LIPIcs, vol.~84, pp. 24:1--24:17 (2017)

\bibitem{DBLP:journals/logcom/Kfoury00}
Kfoury, A.J.: A linearization of the lambda-calculus and consequences. J. Log.
  Comput.  \textbf{10}(3),  411--436 (2000)

\bibitem{Kri}
Krivine, J.L.: $\lambda$-calcul, types et mod\`eles. Masson (1990)

\bibitem{DBLP:conf/lics/Lassen05}
Lassen, S.: {Eager Normal Form Bisimulation}. In: {{LICS} 2005}. pp. 345--354
  (2005)

\bibitem{DBLP:journals/pacmpl/MazzaPV18}
Mazza, D., Pellissier, L., Vial, P.: Polyadic approximations, fibrations and
  intersection types. {PACMPL}  \textbf{2},  6:1--6:28 (2018)

\bibitem{DBLP:conf/icfp/NeergaardM04}
Neergaard, P.M., Mairson, H.G.: Types, potency, and idempotency: why
  nonlinearity and amnesia make a type system work. In: {ICFP} 2004. pp.
  138--149 (2004)

\bibitem{DBLP:conf/lics/Ong17}
Ong, C.L.: Quantitative semantics of the lambda calculus: Some generalisations
  of the relational model. In: {LICS} 2017. pp. 1--12 (2017)

\bibitem{DBLP:journals/mscs/PaoliniPR17}
Paolini, L., Piccolo, M., {Ronchi Della Rocca}, S.: Essential and relational
  models. Mathematical Structures in Computer Science  \textbf{27}(5),
  626--650 (2017)

\bibitem{DBLP:journals/ita/PaoliniR99}
Paolini, L., {Ronchi Della Rocca}, S.: {Call-by-value Solvability}. ITA
  \textbf{33}(6),  507--534 (1999)

\bibitem{DBLP:journals/tcs/Plotkin75}
Plotkin, G.D.: {Call-by-Name, Call-by-Value and the lambda-Calculus}. Theor.
  Comput. Sci.  \textbf{1}(2),  125--159 (1975)

\bibitem{Pottinger80}
Pottinger, G.: A type assignment for the strongly normalizable $\lambda$-terms.
  In: To HB Curry: essays on combinatory logic, $\lambda$-calculus and
  formalism. pp. 561--577 (1980)

\bibitem{DBLP:journals/mscs/PravatoRR99}
Pravato, A., {Ronchi Della Rocca}, S., Roversi, L.: {The call-by-value
  $\l$-calculus: a semantic investigation}. Math. Str. in Comput. Sci.
  \textbf{9}(5),  617--650 (1999)

\bibitem{parametricBook}
{Ronchi Della Rocca}, S., Paolini, L.: {The Parametric $\l$-Calculus}. Springer
  (2004)

\bibitem{DBLP:conf/csl/RoccaR01}
Ronchi Della~Rocca, S., Roversi, L.: Intersection logic. In: {CSL} 2001. pp.
  414--428 (2001)

\bibitem{DBLP:conf/fsttcs/Sieber90}
Sieber, K.: Relating full abstraction results for different programming
  languages. In: {FSTTCS} 1990. pp. 373--387 (1990)

\end{thebibliography}
